\documentclass[twocolumn]{aastex63}
\usepackage{bm}
\usepackage{amsmath,amssymb}
\usepackage{booktabs}
\usepackage{graphicx}
\usepackage{float}
\usepackage{comment}
\usepackage{diagbox}
\usepackage{tabularx}
\usepackage{threeparttable}
\usepackage{xspace}
\usepackage{placeins}   
\usepackage{isotope}
\usepackage{layouts}

\def\T{\mathrm{T}} 

\usepackage{mathtools}

\def\ltsima{$\; \buildrel < \over \sim \;$}
\def\simlt{\lower.5ex\hbox{\ltsima}}
\def\gtsima{$\; \buildrel > \over \sim \;$}
\def\simgt{\lower.5ex\hbox{\gtsima}}

\usepackage{color}

\submitjournal{AAS Journals}

\shorttitle{The \posydon{} binary population synthesis code}
\shortauthors{POSYDON Collaboration}

\newcommand{\Teff}{\ifmmode {T_{\rm eff}}\else${T_{\rm eff}}$\fi}
\newcommand{\Msun}{\ensuremath{M_\odot}\xspace}

\newcommand{\Rsun}{\ensuremath{R_\odot}}
\newcommand{\Lsun}{\ensuremath{L_\odot}}
\newcommand{\Zsun}{\ensuremath{Z_\odot}\xspace}
\newcommand{\posydon}{\texttt{POSYDON}\xspace}

\newcommand{\mesa}{\texttt{MESA}\xspace}

\newcommand{\f}[1]{{\color{purple} \textbf{Potential Figure:} \textit{#1}}}

\definecolor{turquoise}{HTML}{249494}
\definecolor{forestgreen}{rgb}{0.13, 0.55, 0.13}

\definecolor{chmagenta}{rgb}{0.54, 0.17, 0.88}

\defcitealias{2023ApJS..264...45F}{F23}

\begin{document}

\title{{\tt POSYDON} Version 2: Population Synthesis with Detailed Binary-Evolution Simulations across a Cosmological Range of Metallicities}

\correspondingauthor{J.~Andrews}
\email{jeffrey.andrews@ufl.edu}

\author[0000-0001-5261-3923]{Jeff\, J.\,Andrews}
\affiliation{Department of Physics, University of Florida, 2001 Museum Rd, Gainesville, FL 32611, USA}
\affiliation{Institute for Fundamental Theory, 2001 Museum Rd, Gainesville, FL 32611, USA}

\author[0000-0002-3439-0321]{Simone\,S.\,Bavera}
\affiliation{Département d’Astronomie, Université de Genève, Chemin Pegasi 51, CH-1290 Versoix, Switzerland}
\affiliation{Gravitational Wave Science Center (GWSC), Université de Genève, CH1211 Geneva, Switzerland}

\author[0000-0002-6842-3021]{Max\,Briel}
\affiliation{Département d’Astronomie, Université de Genève, Chemin Pegasi 51, CH-1290 Versoix, Switzerland}
\affiliation{Gravitational Wave Science Center (GWSC), Université de Genève, CH1211 Geneva, Switzerland}

\author[0000-0002-6064-388X]{Abhishek\,Chattaraj}
\affiliation{Department of Physics, University of Florida, 2001 Museum Rd, Gainesville, FL 32611, USA}

\author[0000-0002-4442-5700]{Aaron\,Dotter}
\affiliation{Department of Physics and Astronomy, Dartmouth College, Hanover, NH 03755 US}

\author[0000-0003-1474-1523]{Tassos\,Fragos}
\affiliation{Département d’Astronomie, Université de Genève, Chemin Pegasi 51, CH-1290 Versoix, Switzerland}
\affiliation{Gravitational Wave Science Center (GWSC), Université de Genève, CH1211 Geneva, Switzerland}

\author[0000-0003-0648-2402]{Monica\,Gallegos-Garcia}
\affiliation{Department of Physics and Astronomy, Northwestern University, 2145 Sheridan Road, Evanston, IL 60208, USA}
\affiliation{Center for Interdisciplinary Exploration and Research in Astrophysics (CIERA), Northwestern University, 1800 Sherman Ave, Evanston, IL 60201, USA}
\affiliation{Center for Astrophysics \textbar{} Harvard \& Smithsonian, 60 Garden St. Cambridge, MA, 02138, USA}
\affiliation{Harvard Society of Fellows, 78 Mount Auburn Street, Cambridge, MA 02138}

\author[0000-0001-6692-6410]{Seth\,Gossage}
\affiliation{Center for Interdisciplinary Exploration and Research in Astrophysics (CIERA), Northwestern University, 1800 Sherman Ave, Evanston, IL 60201, USA}
\affiliation{NSF-Simons AI Institute for the Sky (SkAI),172 E. Chestnut St., Chicago, IL 60611, USA}

\author[0000-0001-9236-5469]{Vicky\,Kalogera}
\affiliation{Department of Physics and Astronomy, Northwestern University, 2145 Sheridan Road, Evanston, IL 60208, USA}
\affiliation{Center for Interdisciplinary Exploration and Research in Astrophysics (CIERA), Northwestern University, 1800 Sherman Ave, Evanston, IL 60201, USA}
\affiliation{NSF-Simons AI Institute for the Sky (SkAI),172 E. Chestnut St., Chicago, IL 60611, USA}

\author[0009-0009-1888-8785]{Eirini\,Kasdagli}
\affiliation{Department of Physics, University of Florida, 2001 Museum Rd, Gainesville, FL 32611, USA}

\author[0000-0003-4554-0070]{Aggelos\,Katsaggelos}
\affiliation{Electrical and Computer Engineering, Northwestern University, 2145 Sheridan Road, Evanston, IL 60208, USA}
\affiliation{Center for Interdisciplinary Exploration and Research in Astrophysics (CIERA), Northwestern University, 1800 Sherman Ave, Evanston, IL 60201, USA}
\affiliation{NSF-Simons AI Institute for the Sky (SkAI),172 E. Chestnut St., Chicago, IL 60611, USA}

\author[0000-0001-9879-6884]{Chase\,Kimball}
\affiliation{Center for Interdisciplinary Exploration and Research in Astrophysics (CIERA), Northwestern University, 1800 Sherman Ave, Evanston, IL 60201, USA}

\author[0000-0003-3684-964X]{Konstantinos\,Kovlakas}
\affiliation{Institute of Space Sciences (ICE, CSIC), Campus UAB, Carrer de Magrans, 08193 Barcelona, Spain}
\affiliation{Institut d'Estudis Espacials de Catalunya (IEEC),  Edifici RDIT, Campus UPC, 08860 Castelldefels (Barcelona), Spain}

\author[0000-0001-9331-0400]{Matthias\,U.\,Kruckow}
\affiliation{Département d’Astronomie, Université de Genève, Chemin Pegasi 51, CH-1290 Versoix, Switzerland}
\affiliation{Gravitational Wave Science Center (GWSC), Université de Genève, CH1211 Geneva, Switzerland}

\author[0000-0002-8883-3351]{Camille\,Liotine}
\affiliation{Department of Physics and Astronomy, Northwestern University, 2145 Sheridan Road, Evanston, IL 60208, USA}
\affiliation{Center for Interdisciplinary Exploration and Research in Astrophysics (CIERA), Northwestern University, 1800 Sherman Ave, Evanston, IL 60201, USA}

\author[0000-0003-4260-960X]{Devina\,Misra}
\affiliation{Institutt for Fysikk, Norwegian University of Science and Technology, Trondheim, Norway}

\author[0000-0003-4474-6528]{Kyle\,A.\,Rocha}
\affiliation{Department of Physics and Astronomy, Northwestern University, 2145 Sheridan Road, Evanston, IL 60208, USA}
\affiliation{Center for Interdisciplinary Exploration and Research in Astrophysics (CIERA), Northwestern University, 1800 Sherman Ave, Evanston, IL 60201, USA}
\affiliation{NSF-Simons AI Institute for the Sky (SkAI),172 E. Chestnut St., Chicago, IL 60611, USA}

\author[0000-0002-6548-5489]{Dimitris\,Souropanis}
\affiliation{Institute of Astrophysics, Foundation for Research and Technology-Hellas, GR-71110 Heraklion, Greece}

\author[0000-0003-1749-6295]{Philipp\,M.\,Srivastava}
\affiliation{Electrical and Computer Engineering, Northwestern University, 2145 Sheridan Road, Evanston, IL 60208, USA}
\affiliation{Center for Interdisciplinary Exploration and Research in Astrophysics (CIERA), Northwestern University, 1800 Sherman Ave, Evanston, IL 60201, USA}
\affiliation{NSF-Simons AI Institute for the Sky (SkAI),172 E. Chestnut St., Chicago, IL 60611, USA}

\author[0000-0001-9037-6180]{Meng\,Sun}
\affiliation{Center for Interdisciplinary Exploration and Research in Astrophysics (CIERA), Northwestern University, 1800 Sherman Ave, Evanston, IL 60201, USA}

\author[0000-0003-0420-2067]{Elizabeth\,Teng}
\affiliation{Department of Physics and Astronomy, Northwestern University, 2145 Sheridan Road, Evanston, IL 60208, USA}
\affiliation{Center for Interdisciplinary Exploration and Research in Astrophysics (CIERA), Northwestern University, 1800 Sherman Ave, Evanston, IL 60201, USA}
\affiliation{NSF-Simons AI Institute for the Sky (SkAI),172 E. Chestnut St., Chicago, IL 60611, USA}

\author[0000-0002-0031-3029]{Zepei\,Xing}
\affiliation{Département d’Astronomie, Université de Genève, Chemin Pegasi 51, CH-1290 Versoix, Switzerland}
\affiliation{Gravitational Wave Science Center (GWSC), Université de Genève, CH1211 Geneva, Switzerland}

\author[0000-0002-7464-498X]{Emmanouil\,Zapartas}
\affiliation{Institute of Astrophysics, Foundation for Research and Technology-Hellas, GR-71110 Heraklion, Greece}

\author[0000-0002-0147-0835]{Michael Zevin}
\affiliation{The Adler Planetarium, 1300 South DuSable Lake Shore Drive, Chicago, 60605, IL, USA}
\affiliation{Center for Interdisciplinary Exploration and Research in Astrophysics (CIERA), Northwestern University, 1800 Sherman Ave, Evanston, IL 60201, USA}
\affiliation{NSF-Simons AI Institute for the Sky (SkAI),172 E. Chestnut St., Chicago, IL 60611, USA}

\begin{abstract}
Whether considering rare astrophysical events on cosmological scales or unresolved stellar populations, accurate models must account for the integrated contribution from the entire history of star formation upon which that population is built. Here, we describe the second version of \posydon, an open-source binary population synthesis code based on extensive grids of detailed binary evolution models computed using the \mesa code, which follows both stars' structures as a binary system evolves through its complete evolution from the zero-age main sequence, through multiple phases of mass transfer and supernovae, to their death as compact objects. To generate synthetic binary populations, \posydon uses advanced methods to interpolate between our large, densely spaced grids of simulated binaries. In our updated version of \posydon, we account for the evolution of stellar binaries across a cosmological range of metallicities, extending from $10^{-4}\,\Zsun$ to $2\,\Zsun$, including grids specifically focused on the Small and Large Magellanic Clouds ($0.2\,\Zsun$ and $0.45\,\Zsun$). In addition to describing our model grids and detailing our methodology, we outline several improvements to \posydon. These include the incorporation of single stars in stellar populations, a treatment for stellar mergers, and a careful modeling of ``reverse-mass transferring" binaries in which a once-accreting star later becomes a donor star. Our simulations are focused on binaries with at least one high-mass component, such as those that host neutron stars and black holes, and we provide post-processing methods to account for the cosmological evolution of metallicity and star formation as well as rate calculations for transient events. 
\end{abstract}



\section{Introduction \label{sec:intro}}
Ranging from the progenitors of Type Ia supernovae (SN) that form the backbone of our modern understanding of cosmology to the merging compact objects (COs) manufacturing $r$-process elements and emitting gravitational waves (GWs), stellar binaries are relevant to nearly every aspect of astrophysics. The best estimates suggest that over half of all stars in our Galaxy are found in a binary (or higher-order system), with higher fractions at larger stellar masses \citep{2010ApJS..190....1R, 2012Sci...337..444S, 2017ApJS..230...15M}. While many of these systems are widely separated, thus they evolve as essentially isolated stars with little impact on each other's evolution, a significant fraction form in sufficiently tight orbits that they will interact at some point in their lifetimes, for example through tidal forces or mass transfer \citep[MT;][]{2017PASA...34....1D}. It is the complexity induced by these interactions that produces the broad array of astrophysical phenomena uniquely accessible to stellar binaries.

Despite their importance, our theoretical understanding of stellar binaries is limited by our ability to accurately model both the physics dictating the evolution of individual stars as well as their mutual interaction. Decades of theoretical and computational studies on binary interactions have led to a general understanding of the dominant mechanisms relevant for binary star evolution \citep[see e.g., the monograph by][]{2023pbse.book.....T}. Yet, limitations in our ability to implement these models on a population scale prevent the ideal result: synthetic populations containing the best physics fidelity that accurately reproduce observational samples. Realistic synthetic populations require an accurate description of multiple pieces of complex astrophysics along with the representative initial conditions for stellar binaries. The nonlinear dependency is particularly acute when considering CO binaries such as X-ray binaries (XRBs) and GW sources, as any inaccuracies accrued can be compounded through multiple phases of evolution.

Despite the significant challenges, major strides have been made in stellar population modeling since pioneering efforts starting in the 1980's and the 1990's by
\citet{1983SvA....27..163K, 1983SvA....27..334K},
\citet{1987ApJ...321..780D}, \citet{1989ApJ...342..917B}, \citet{1992ApJ...399..621R}, \citet{1993MNRAS.260..675T}, \citet{1995ApJS..100..217I, 1995ApJS..100..233I}, and \citet{1998ApJ...493..351K}. These initial works were limited in their treatment of how single stars---and therefore binaries---evolve. Through a series of papers, The Cambridge STARS group developed a method to rapidly evolve large binary populations by generating a series of fitting formulas to describe the evolution of single stars \citep{1995MNRAS.274..964P, 1997MNRAS.291..732T, 2000MNRAS.315..543H}. Using Monte Carlo methods, multiple binary population synthesis (BPS) codes employ these fitting formulas, wrapping them with prescriptions that capture binary interactions \citep[e.g.,][]{2001A&A...365..491N, 2002MNRAS.329..897H, 2008ApJS..174..223B}. 

Of particular note, the formulae from \citet{2000MNRAS.315..543H} not only provide a quantitative description of how stars evolve as a function of mass and time, but they also incorporate their dependence on metallicity, a feature which expanded the applicability of these binary population models from local, solar-metallicity populations to cosmological phenomena such as Type Ia supernova \citep{2009ApJ...699.2026R, 2012A&A...546A..70T}, gamma-ray bursts \citep{2006ApJ...648.1110B} and GW sources \citep{2002ApJ...572..407B}. These early works were instrumental in providing astrophysical context for GW observatories \citep[e.g.,][]{2007ApJ...662..504B}. The detection of the first GW event \citep{2016PhRvL.116f1102A} accelerated the development of BPS, with several newly developed codes being released in subsequent years such as {\tt MOBSE} \citep{2018MNRAS.480.2011G}, {\tt ComBinE} \citep{2018MNRAS.481.1908K}, {\tt SEvN} \citep{2019MNRAS.485..889S}, {\tt COSMIC} \citep{2020ApJ...898...71B}, and {\tt COMPAS} \citep{2021arXiv210910352T} to name a few. Although each of these BPS codes differ in detail, they essentially rely on the procedure outlined by \citet{2002MNRAS.329..897H}, in which individual stars in binaries are evolved by perturbing single stars, either through the fitting formulas from \citet{2000MNRAS.315..543H} or by mapping onto grids of single-star tracks. While numerous studies have illustrated the value of this approach, it has recently been demonstrated that, at least for some regions of parameter space, these traditional BPS codes may inaccurately describe the evolution of binaries \citep[e.g.,][]{2021ApJ...922..110G, 2023ApJ...954..212S}. The binary population and spectral synthesis code {\tt BPASS} stands out as a noteworthy exception \citep{2017PASA...34...58E, 2018MNRAS.479...75S} as it is based on detailed binary evolution model grids using a custom version of the {\tt Cambridge STARS} binary evolution code \citep[][]{1971MNRAS.151..351E,2009MNRAS.396.1699S}. In this case binary tracks are generated where one star is evolved in its full internal structure and population models are constructed through relative weighting of the evolutionary sequences. 

In \citet[][hereafter \citetalias{2023ApJS..264...45F}]{2023ApJS..264...45F} we describe \posydon, an open-source, general-purpose binary population synthesis code designed to address the limitations of previous methods by incorporating extensive grids of detailed binary evolution models computed with the \mesa code \citep{2011ApJS..192....3P, 2013ApJS..208....4P,2015ApJS..220...15P, 2018ApJS..234...34P,2019ApJS..243...10P,2023ApJS..265...15J}, where the structures of both stars are evolved self-consistently. Broadly, \posydon incorporates functionality to handle three different evolutionary steps. First, it provides an infrastructure to manage and compute large grids of detailed single- and binary-star models within a high-performance computing environment. Second, it includes a series of post-processing steps to interpret, collate, compress, and combine the data from these model grids, and ultimately train a series of classification and interpolation schemes. Finally, it offers routines to initialize large binary populations, apply classification and interpolation schemes to evolve a binary through each phase, and direct individual binaries through their various phases of evolution. The result is an approach to BPS modeling that fundamentally differs from previous efforts. 

\begin{deluxetable*}{ccccccccccccc}
\tabletypesize{\footnotesize}
\tablecolumns{14}
\tablewidth{0pt}
\setlength{\tabcolsep}{3pt}
\tablecaption{
Summary of the five detailed single- and binary-star model grids each run at 8 metallicities $Z/\Zsun\in\{10^{-4}, 10^{-3}, 10^{-2}, 0.1, 0.2, 0.45, 1, 2\}$. \label{tab:grid_properties}}
\tablehead{
  \colhead{} &
\multicolumn{2}{c}{Initial state}& & \multicolumn{8}{c}{Parameters' range and resolution} &\\
\cline{2-3}\cline{5-12}
  \colhead{Grid Name} &
  \colhead{Star 1} & \colhead{Star 2}  & & \colhead{$M_1$~[\Msun]} & \colhead{$\Delta\log_{10}M_1$} & \colhead{$M_2$~[\Msun]}& \colhead{$\Delta\log_{10}M_2$} & \colhead{$q$} & \colhead{$\Delta q$} & \colhead{$P_{\rm orb}\,[{\rm day}]$} & \colhead{$\Delta\log_{10}P_{\rm orb}$} & \colhead{$N$\,\tablenotemark{a}}
 }
\startdata
single--HMS & ZAMS & - && 0.1--300 & 0.0093 & - & - & - & - & - & - & 375\\
single--HeMS & ZAHeMS\tablenotemark{b} & - && 0.5--151 & 0.0093 & - & - & - & - & - & - & 268\\
HMS--HMS & ZAMS & ZAMS && 3.92--286 & 0.05\tablenotemark{c} & - & - & 0.05--0.99 & 0.05 & 0.1--5179 & 0.14 & 39,712\\
CO--HMS & H-rich\,\tablenotemark{d} & CO  && 0.5--321 & 0.12 & 1--307 & 0.077 & - & - & 0.1-9236 & 0.31 & 13,464\\
CO--HeMS & ZAHeMS & CO && 0.5--192 & 0.15 & 1--307 & 0.077 & - & - & 0.02--1147 & 0.207 & 14,256\\
\enddata
\tablenotetext{a}{Number of models in this grid per metallicity}
\tablenotetext{b}{Zero-age He main sequence (ZAHeMS) stars}
\tablenotetext{c}{For primary masses in the range 3.92--14.04~\Msun, the spacing between successive models decreases by a factor of 3 to $\Delta \log_{10} M_1 = 0.0168$. In total, this grid is comprised of 66 separate primary masses.}
\tablenotetext{d}{Although this grid is initialized with H-rich stars at ZAMS, we ignore the portion of each simulated binary's evolution prior to the onset of RLOF. The Star 1's in this grid are therefore somewhat evolved beyond their initial states. }
\end{deluxetable*}

In \citetalias{2023ApJS..264...45F} we introduced version 1 (v1) of \posydon, which describes how binaries evolve at Solar metallicity. In this work we present version 2 (v2) of \posydon, which expands our simulations to a cosmological range of metallicities, varying from $10^{-4}\,\Zsun$ to $2\,\Zsun$ and introduces improved physical treatment in multiple areas. In Section~\ref{sec:grids} we describe the details of our different single- and binary-star model grids, including a number of improvements we have made since v1. In Section~\ref{sec:machine_learning} we detail our updated classification and interpolation methods which are trained on our grids. These grids are ingested by our post-processing pipeline which is described in Section~\ref{sec:postprocessing}. Other improvements and new features since v1, such as the inclusion of single-star populations, stellar mergers, disrupted binaries, and X-ray luminosity calculations, are outlined in Section~\ref{sec:improvements}. The method by which all these improvements are incorporated into binary populations is outlined in Section~\ref{sec:populations}, including a cosmological evolution of star formation and metallicity, as well as rate calculations for GW events, gamma-ray bursts, and other transients. Finally, in Section~\ref{sec:caveats} we describe the caveats and limitations of \posydon, provide a broad comparison with other BPS codes, and consider directions for future work.

\section{Varying Metallicity in {\tt POSYDON} Binary Grids \label{sec:grids}}
Grids of binary-star mass-transfer sequences  form \posydon's backbone, with each grid point representing a separate \mesa\ {\tt binary} simulation of one particular binary system. Due to the inherent limitations of detailed stellar structure and binary evolution codes, it is not possible to evolve both stars in a binary from zero-age main sequence (ZAMS) to collapse with a single simulation; multiple grids must be generated, each representing different evolutionary stages. An individual binary will then ``move'' through our different grids as the stars evolve through multiple phases of MT and core collapse. For the meaningful synthesis of binary populations, our grids must be comprised of accurate individual simulations, encompass the entire range of relevant parameter space, and be dense enough for faithful interpolation. Our fundamental approach is described in detail in \citetalias{2023ApJS..264...45F}. 

We provide an overview of our v2 grids in Section~\ref{sec:grid_char}, and we describe the v2 changes to our binary physics in Section~\ref{sec:changes_vI} below. We follow that with a description of how we construct our models at ZAMS in Section~\ref{sec:zams}. Finally, we provide brief overviews of our resulting grids across the full range of metallicities in the subsequent subsections. These grids are available in their processed forms (see Section \ref{sec:postprocessing}) on Zenodo under an open-source Creative Commons Attribution license: \dataset[doi:10.5281/zenodo.15194708]{https://doi.org/10.5281/zenodo.15194708}.

\subsection{Basic grid characteristics}
\label{sec:grid_char}

As in v1, our simulations are comprised of five separate grids of detailed single- and binary-star evolution tracks per metallicity (see Table~\ref{tab:grid_properties} for an overview). Our single-star hydrogen main sequence (HMS) grid is comprised of 375 single H-rich star models of different initial masses, logarithmically spaced from $0.1$ to $300\,\Msun$, evolved from ZAMS until core carbon exhaustion. Likewise, our single-star He main sequence (HeMS) grid is comprised of 268, logarithmically spaced in initial mass, single star models. The latter models are initialized without any hydrogen in their structure and are used to model stars that have lost their hydrogen envelopes either through stellar winds or a prior mass-transfer phase. Our HMS--HMS grid utilizes \mesa\ {\tt binary} to evolve the structures of both stars within an orbit and, as such, is the largest in terms of number of models ($\simeq3.9\times 10^4$) and the most computationally expensive. Our additional two binary grids, CO--HMS and CO--HeMS use \mesa\ {\tt binary}, but model the CO companion star as a point mass.

Each of these binary grids are produced by generating initial binaries that are regularly, logarithmically spaced in mass (except for the HMS--HMS grid where we increase the mass resolution for lower initial primary masses and the secondary star's mass is linearly spaced in mass ratio, $q$, which we define to be the ratio of the secondary star's mass to the primary star's mass and is therefore always less than unity) and orbital period, $P_\mathrm{orb}$. Unlike in v1 where our simulations are exclusively limited to $\Zsun$, for each of these grids in v2 we have modeled the same binaries for eight different metallicities: $Z/\Zsun\in\{10^{-4}, 10^{-3}, 10^{-2}, 0.1, 0.2, 0.45, 1, 2\}$. In all, our four-dimensional grids contain over $3.2\times10^5$ separate models which required $\mathcal{O}$(15 million) CPU hours to compute. Table~\ref{tab:grid_properties} describes the properties of each of our five grids. Compared with v1, our binary grids are over four times larger than our grids in v1. However, due to the adding of a fourth grid dimension (metallicity) as well as an increased mass range, our v2 grids have a somewhat reduced resolution compared with v1. Nevertheless, as we demonstrate in Section~\ref{sec:machine_learning}, our grids still have sufficient resolution for our classification and interpolation routines to accurately evolve binary populations.

Finally, for the purpose of testing our classification and interpolation routines, we have added additional ``random'' (randomly sampled) grids, one for each of our three binary grids across each of our eight metallicities. These ``random'' grids have the same mass and orbital period ranges as listed in Table~\ref{tab:grid_properties}, but are randomly distributed in this space and are $\simeq$20\% the original grid size. These grids, comprised of $\simeq100,000$ models in total, allow us to estimate accuracy metrics for our classification and interpolation methods described in Section~\ref{sec:machine_learning}.

\subsection{Changes from \posydon Version 1}\label{sec:changes_vI}

Although our motivation in producing this work is to expand our capabilities to evolve populations across a range of metallicities, we have made several improvements to the implementation of stellar and binary evolution physics compared to v1 \citepalias{2023ApJS..264...45F}. These changes include both alterations and improvements to our \mesa simulation setup, as well as to the binary population generation methodology. For the former, we describe the three major enhancements below, while we outline the latter in Section~\ref{sec:improvements}. There are a number of additional minor changes which we list in Section~\ref{sec:minor_changes} below.

\subsubsection{Onset of Pair-Instability Supernovae}
\label{sec:PISN}

\begin{figure}
    \centering
    \includegraphics[width=\linewidth]{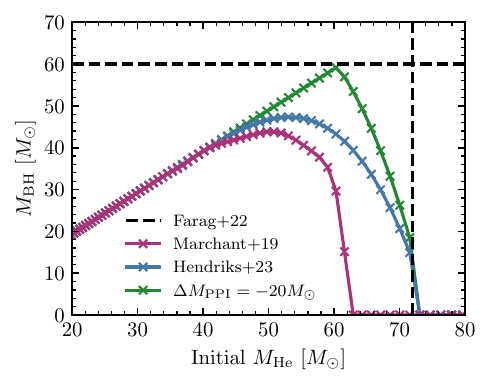}
    \caption{The pulsational pair-instability supernova prescriptions from \citet{2023MNRAS.526.4130H} without adaptations (blue) and with reduced PPI mass loss ($\Delta M_\mathrm{PPI}=-20\Msun$) (green), and the fit from \citet{2020ApJ...898...71B} to simulations from \citet{2019ApJ...882...36M} (purple) applied to the $Z=10^{-4}\Zsun$ single helium star models from \posydon with the \citet{2012ApJ...749...91F}-delayed remnant mass prescription. The maximum BH mass and PISN boundary in helium mass found by \citet{2022ApJ...937..112F} are shown as black dashed lines.}
    \label{fig:PPI}
\end{figure}

We check for the onset of pair instability \citep{1964ApJS....9..201F,1967PhRvL..18..379B} based on the star's core mass. However, in v2, we are using the approach from \citet{2023MNRAS.526.4130H} to determine the regime of (pulsational) pair-instability (PPI) from the CO-core mass at the moment of central carbon depletion. This prescription is based on the simulations of \citet{2022RNAAS...6...25R} and takes a top-down approach, where PPI removes mass before the final remnant mass is computed. If the calculated remnant mass is less than $10\Msun$, the event is assumed to be a pair-instability SN (PISN) and fully disruptive. As such, the final remnant mass is set to 0.
We provide a PISN model using the default parameters from \citet{2023MNRAS.526.4130H} and a model to reproduce the results from \citet{2022ApJ...937..112F}. The latter is the \posydon default and to achieve it, we set the additional PPI mass loss in the prescription to $\Delta M_\mathrm{PPI}=-20\Msun$, effectively reducing the mass loss from PPI. This shifts the maximum BH mass higher, while keeping the boundary of PISN at $M_\mathrm{He core} \approx  72\Msun$, in line with the findings from \citet{2022ApJ...937..112F}, as shown in Figure \ref{fig:PPI}. 
Above the pair instability gap, we assume that a direct collapse into a black hole (BH) occurs.

Depending on the central conditions, the pair-instability can happen either before or after carbon depletion \citep{2019ApJ...882...36M}. In cases where the instability occurs prior to carbon depletion, our \mesa{} models in v1 did not converge, resulting in a lack of final stellar profiles. Additionally, the final steps of those runs demonstrate rapid structural changes, violating the implicit assumption of equilibrium. This issue becomes particularly significant for models with larger ZAMS masses and lower metallicities.

\begin{figure*}
    \centering
    \includegraphics{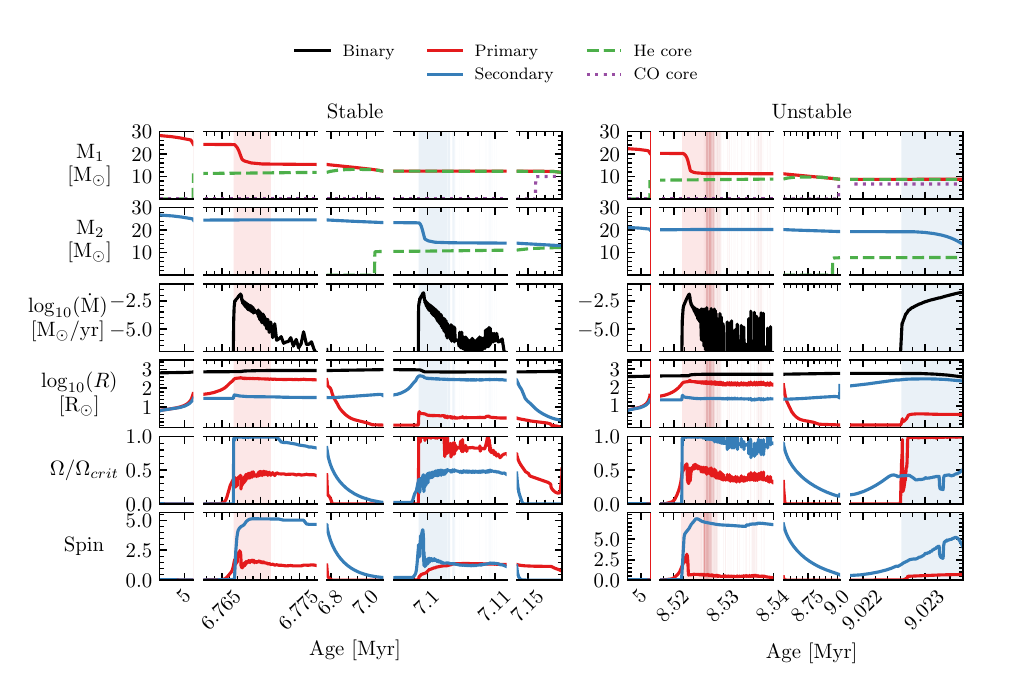}
    \caption{The evolution, as a function of age, of two binary systems experiencing reverse MT. The left panels show a stable reverse mass-transfer phase with the initial conditions: $M_\mathrm{1, ZAMS} = 28.2\,\Msun$, $q=0.95$, $P=268$ days at $Z=\Zsun$, while the right panels show a binary with initial conditions $M_\mathrm{1, ZAMS}=22.3\,\Msun$, $q=0.95$, $P=139$ days at $Z = \Zsun$ undergoing unstable reverse MT.
    The top two panels show the total mass, helium-core mass (green dashed), and carbon--oxygen--core mass (purple dotted) evolution of the primary (red) and secondary (blue) star, respectively. The mass-transfer rate is shown in the third row of panels and depending on the donor star, the Roche lobe overflow (RLOF) regime is shaded red or blue for the primary or secondary, respectively. These mass transfer sequences correspond to expansion of the donor star's radius which we show in the fourth row of panels along with the binary separation. The secondary initiates MT and reaches the $L_2$ overflow criteria for unstable MT. We also show the rotational velocity as a fraction of its critical rotation. The final row of panels indicates each star's non-dimensional spin, a measure of the angular momentum budget in units of $GM^2/c$.}
    \label{fig:reverse_MT_examples}
\end{figure*}

To address this shortcoming, in v2 we perform a check within the \mesa{} calculations to detect the emergence of pair instability, allowing us to obtain profiles and a final history uncontaminated by any dynamical-timescale effects. This check identifies the onset of pair production of neutrinos due to large central temperatures by calculating the pressure-weighted volumetric averaged adiabatic index:
\begin{equation}
    \left\langle\Gamma_1\right\rangle = \frac{\int \Gamma_1 (P/\rho) \mathrm{d}m}{\int (P/\rho) \mathrm{d}m}.
\end{equation}
If the index drops below $4/3$, we expect pair instability to ensue \citep{1999MNRAS.305..365S, 2020MNRAS.493.4333R}, as the star is expected to become dynamically unstable due to a lack of photon pressure. If this condition is triggered, we stop our \mesa{} simulations and save the final profile of the star (and its companion). As this condition supersedes our standard stopping condition for core-carbon depletion, our most massive stars may end their evolution as slightly less-evolved stars. Depending on the location of the drop of the adiabatic index, we differentiate between pulsational pair instability (off-center) and disruption of the entire star (at the center).

Because our simulations stop at central carbon depletion, we cannot automatically identify those stars that would undergo pair instability had our simulations continued. We, therefore, perform a check for pair-instability as part of the population synthesis, which uses the ranges provided by \citet{2023MNRAS.526.4130H} and can be shifted to lower or higher carbon-oxygen core masses. In rare cases, this prescription can lead to inconsistencies between the detection of (pulsational) pair instability between our \mesa{} grids and the \posydon{} simulations. Here, we disregard the information from \mesa{} because of its lower completeness in detecting pair instability during evolution past carbon depletion. 

Note that, while the prescription described previously is our default, we have included the pair-instability prescription from \citet{2019ApJ...882...36M} as an optional alternative for users.

\subsubsection{Reverse Mass Transfer}
\label{sec:reverse_MT}

In certain regions of the parameter space, particularly those where the initial mass ratio is close to unity, after a first phase of MT from the primary to the secondary star, the secondary may enter its post-main sequence (post-MS) stage and initiate a phase of ``reverse MT''. In \mesa r11701, which is used to generate \posydon grids, the treatment of reverse MT is not supported in the \texttt{Kolb} MT scheme. In v1, the secondary star would expand without being recorded as initiating MT. To address this, we made specific modifications to the \mesa r11701 code base to enable the function of switching donors when the calculated mass-transfer rate from the secondary star exceeds that from the primary star\footnote{The necessary modifications in the \mesa code base are documented here: \href{https://github.com/MESAHub/mesa/issues/545}{https://github.com/MESAHub/mesa/issues/545}. }. 
All our v2 grids include this updated MT prescription. Additionally, we provide updated versions of our v1 grids to include this treatment for reverse MT, which was introduced in \citet{2024A&A...683A.144X}, where a detailed discussion on reverse MT can be found.

In Figure~\ref{fig:reverse_MT_examples}, we show the time evolution of two example binaries at solar metallicity that experience stable and unstable reverse MT phases. In the first example (left panels), the binary initially has a primary star with a mass of $M_\mathrm{1, ZAMS} = 28.2$\,\Msun, a mass ratio of $q = 0.95$, and an orbital period of $P = 268\,\mathrm{days}$. During this first MT phase, the MT rate reaches as high as $\sim 0.01\,\Msun\,\mathrm{yr}^{-1}$, but does not reach any of our criteria indicating the onset of a common envelope \citepalias[CE; e.g., Section 4.2.4.\ in][]{2023ApJS..264...45F}. Although the secondary initially accepts all the mass lost by the primary, the accreted angular momentum quickly spins up the secondary to critical rotation at which point the accretion fraction is reduced following the procedure defined in Section 4.2.2.\ of \citetalias{2023ApJS..264...45F}. This prescription assumes that very little mass is accreted by critically rotating stars; indeed, the second row of panels in Figure~\ref{fig:reverse_MT_examples} shows that most of the mass lost by the donor during this phase is not accreted. Upon detachment and further evolution, the secondary star evolves off the MS and expands, initiating a reverse MT phase. In the bottom left panel of Figure~\ref{fig:reverse_MT_examples} we show that in this evolutionary sequence, the primary star acquires significant spin. In the second example (right panels), the binary has a slightly lower primary mass $\mathrm{M}_\mathrm{1, ZAMS} = 22.3\,\Msun$, an identical mass ratio, and a somewhat shorter orbital period of $139\,\mathrm{days}$. The binary follows, broadly, the same evolutionary path as the previous example, but in this case, the reverse MT phase reaches $L_2$ overflow, one of our CE instability criteria.

As we discuss in detail in Section~\ref{subsec:HMS-HMS grid}, we find that with decreased metallicity, reverse MT is more likely to lead to unstable MT. This occurs because the initially more massive star does not fully detach at low metallicity and can undergo stable MT until carbon exhaustion \citep[for more details, see][]{2022A&A...662A..56K}. Additionally, this star might not be fully stripped during MT and the presence of a thin hydrogen layer affects the radial response of the accreting star. Consequently, the binary system has a higher probability of coming into contact \citep[see][for more details and possible implications for long gamma-ray burst formation]{2025arXiv250209187B}.

\subsubsection{Stellar winds and their metallicity dependence}
\label{sec:winds}

\begin{figure}[t]\center
\includegraphics[width=\columnwidth]{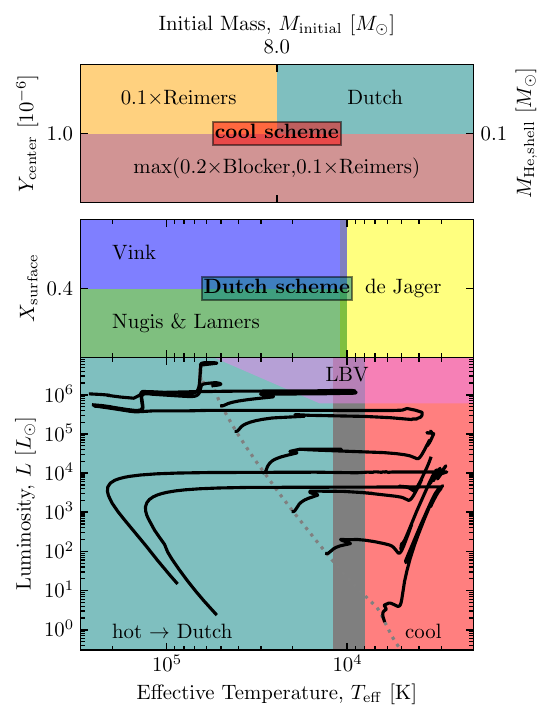}
\caption{Schematic summary of the wind prescriptions used for our grids. The bottom panel shows an Hertzsprung--Russell diagram with evolutionary tracks for $M_\mathrm{ZAMS}/\Msun\in\{1.2, 3, 6, 12, 30, 60, 120, 300\}$ at $\Zsun$ with solid, black lines and the ZAMS as a dotted, gray line. The upper panels show the criteria used to further subdivide the different schemes.}
\label{fig:schematic_wind}
\end{figure}

The suite of stellar wind prescriptions adopted in our v2 grids closely follows that of v1 (see Section~3.2.2 in \citetalias{2023ApJS..264...45F}), with the following updates: (i) a metallicity dependence of the stellar-wind strength in hot stars, (ii) a refined definition and wind prescription for thermally pulsating AGB stars, and (iii) a simple wind prescription for luminous blue variable (LBV) stars. We describe these updates in the following paragraphs.

For the wind mass loss of hot, H-rich stars \citep{Vink+2000}, we assume a power-law metallicity dependence factor, $(Z/\Zsun)^{\alpha}$ with an index $\alpha=0.68$ \citep{Vink+2001}. As the metallicity dependence of winds from cold, red (super-)giant and asymptotic giant branch (AGB) stars \citep{dejager1988, reimers1975, blocker1995} is 
uncertain and appears to be weak, both empirically \citep{Van-Loon+2005,Groenewegen2012,Goldman+2017} and theoretically \citep{Kee+2021}, we assume that these winds are metallicity-independent, following the example of MIST \citep{2016ApJ...823..102C}. In any case, cold-star winds are primarily important for single stars and for non-interacting binary systems in wide orbits, where there is sufficient space for a star to expand into a giant before Roche Lobe Overflow \citep[RLOF;][]{2024arXiv241007335Z}. For Wolf-Rayet--like winds of hot, He-rich stars, we use the prescription by \citet{nugis2000}, which includes an explicit dependence on metallicity. 

In addition, we adapt the \citet{Vink+2000,Vink+2001} stellar wind implementation in \mesa models, by adopting $ Z_{\odot,\mathrm{Vink}} = 0.0142$ and not $0.019$, which is the default in \mesa based on an older solar abundance from \citet{1998SSRv...85..161G}. This is because, despite updates to the metal content, the solar Fe abundance, which provides the dominant carriers of line-driven winds, did not change significantly, and thus any $Z$-dependence of the winds should be compared to our adopted \Zsun (de Koter, private communication). This adaptation led to a slight increase in the stellar winds during the blue phase of evolution compared to v1. 

Stars in a mass range of $\sim 0.6-8\,\Msun$ (which avoid core collapse while allowing for He fusion) can ascend the AGB and experience thermal pulses (TP) \citep{Schwarzschild.Harm1965}. These TP-AGB stars typically exhibit large mass-loss rates (perhaps as high as $10^{-4}\,\Msun\,\mathrm{yr}^{-1}$; \citealt{Wilson2000}), shedding a significant portion of the star's outer envelope, eventually leaving behind a white dwarf (WD). In v1 we considered the TP-AGB as a stopping point of the simulation, with the nascent C/O core well-formed at that point. In our v2 grids, we evolve stars through the TP-AGB to obtain fully modeled WD masses. We adopt the \cite{blocker1995} wind prescription, parameterizing TP-AGB mass loss through dust-driven winds which, following the definition in \citet{2016ApJS..222....8D} we trigger when a star has $T_\mathrm{eff}<12,000\,\mathrm{K}$, the core is He-depleted (central $Y < 10^{-6}$), and the He-burning shell becomes less massive than $0.1\,\Msun$ (in v1 we used a check built into \texttt{MESA r11701} called \texttt{have\_done\_TP}). Once these conditions are met, and thereafter, we take the maximum value of the \cite{reimers1975} red-giant branch and \cite{blocker1995} TP-AGB winds to provide a smooth transition between the two (as is similarly done with models in MIST; \citealt{2016ApJ...823..102C}). Shortly after reaching TP-AGB conditions, the \cite{blocker1995} winds overtake the \cite{reimers1975} winds.

Finally, in v2 we introduce a simple prescription for LBV-type winds, following \citet{2010ApJ...714.1217B}. Specifically, we apply an enhanced mass loss of $10^{-4}\,\Msun\,\mathrm{yr}^{-1}$ for stars that cross the Humphreys-Davidson limit \citep{1979ApJ...232..409H}, defined here as stars having simultaneously $ L > 6\times 10^{5}\,\Lsun$ and $(R/\Rsun) \times (L / \Lsun)^{1/2} > 10^5$. As discussed in \citet{2023NatAs...7.1090B}, LBV-like winds have a limited effect on the evolution of massive stars at solar metallicity, as line-driven winds can become sufficiently strong to prevent stars from crossing the Hamphreys--Davidson limit, but become increasingly important at lower metallicities. Given the considerable uncertainties, both theoretical and observational, regarding LBV winds and their metallicity dependence, as well as the simplicity of our adopted prescription, we opted not to account for a potential metallicity dependence of the LBV-like wind strength.

In Figure~\ref{fig:schematic_wind} we summarize the various wind prescriptions implemented in our stellar models and the criteria used to define the evolutionary phase in which each prescription is applied. Overlapping regions of different wind prescriptions indicate where a linear interpolation between two prescriptions is used (for LBV winds the maximum is taken instead of an interpolated value).

\subsubsection{Other Minor Changes in our \mesa simulations}
\label{sec:minor_changes}

In addition to the three major changes described in the previous sections, we have made a number of minor changes to our \mesa simulation setup in v2 which we describe here for completeness. The inlists used in our simulations are publicly available\footnote{\url{https://github.com/POSYDON-code/POSYDON-MESA-INLISTS}}. We have removed overshooting in shell burning and shell convective regions, leaving only core overshooting turned on. We made this change principally for stability reasons, but note that these processes are highly uncertain with few (if any) observational constraints. For similar reasons, we have turned off thermohaline mixing in v2; however, in our procedure for handling \mesa runs that do not converge with our default v2 inlists, we add a re-run that includes thermohaline mixing (see Section~\ref{sec:postprocessing:rerun}). We removed the stopping condition on the age of a star \citepalias[see Section~5.2 in][]{2023ApJS..264...45F} for our grids because binary interactions may produce evolved, low-mass stars that would otherwise (in the case of single, isolated evolution) require longer than the age of the Universe to produce. We disable magnetic braking to act on He stars, because they nearly always violate the assumption of being tidally synchronized. Motivated by population models within MIST2 (Dotter et al., in prep), we have additionally updated our treatment of semi-convection, altering the mixing parameter, {\tt alpha\_sc}, to 0.1. 

We have compared our results for single hydrogen-rich (H-rich) stars at \Zsun to our results from v1, and we have found that the improvements we have made do not significantly alter the final stellar masses.

\subsection{Generating our stellar models at ZAMS}
\label{sec:zams}

\begin{figure*}[t]
    \center
    \includegraphics[width=0.8\textwidth,angle=0]{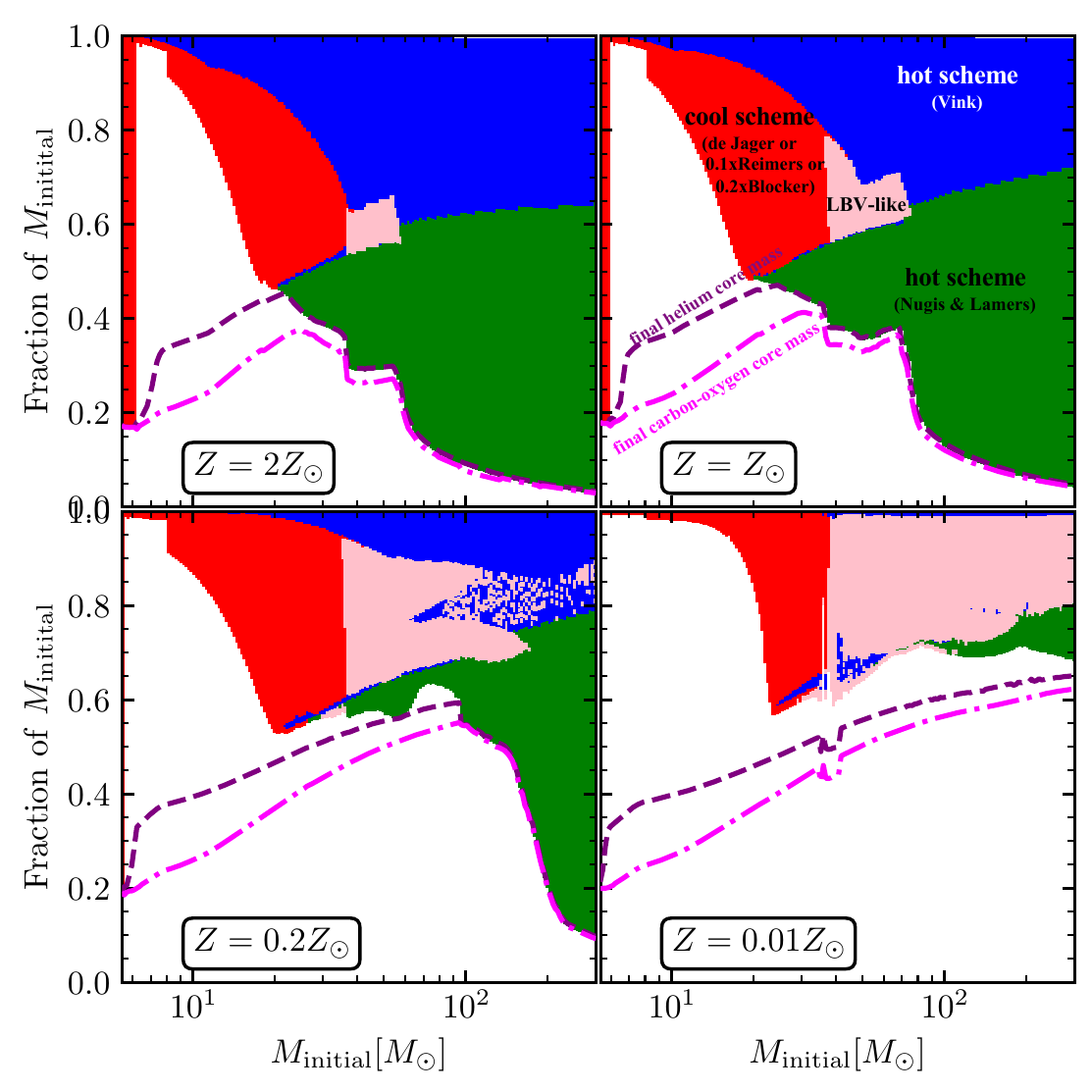}
    \caption{Depending on a star's characteristics, different mass-loss schemes will be employed (see top right panel for labeling). Colored regions indicate the fractional mass lost due to the different wind schemes as a function of ZAMS mass for single, H-rich stars for four indicative metallicities. The white region defines the star's mass at the end of the simulation which may differ from the remnant mass. We also indicate the final helium- (carbon-oxygen-)core mass as a purple, dashed (magenta, dashed-dotted) line.}
    \label{fig:mass_loss}
\end{figure*}

As in \citetalias{2023ApJS..264...45F}, our initial stellar models are drawn from our custom library of ZAMS models. As described in that work, these are generated using the \texttt{create\_zams} template provided in \mesa r11701. The primary difference regarding the present work is that we generate additional models for our expanded metallicity range which requires an extra relaxation step to adjust the elemental abundances. As in v1, we follow the \citet{Asplund+2009} definition for a solar elemental abundance distribution\footnote{We note that although we use \citet{Asplund+2009} for our abundances, we use opacity tables based on the Solar abundance reference defined by \citet{1998SSRv...85..161G}.}. However, we adjust the He abundance between the assumed primordial value of $Y_\text{Big Bang} = 0.249$ \citep{Planck2015} and its protosolar value of $Y=0.2703$ \citep{Asplund+2009}, also assuming $\Zsun=0.0142$ as in v1: 
\begin{equation}
Y = Y_\text{Big Bang} + \left(Z/\Zsun\right) \times \left(Y_{\odot} - Y_\text{Big Bang}\right).
\end{equation}
The remainder of the star is assumed to be hydrogen.

We additionally expand the mass range of our library, extending from $0.1$ to $300\,\Msun$ to support the lower masses now provided in our single star grids (see Section~\ref{sec:singles}). In total, our ZAMS library consists of eight files (one for each metallicity). For some masses, particularly at the higher mass end, the pre-MS model from which the ZAMS model is built fails to converge. In these cases, we initialize stars by choosing a similar-mass star from our ZAMS library and relaxing the model to the desired mass.

To generate our initial ZAMS models for pure He stars, we follow the same procedure described in Section~5.1 of \citetalias{2023ApJS..264...45F}. There is an additional complexity in identifying a star as initializing on the helium ZAMS (ZAHeMS), since for high-metallicity helium stars, $\alpha$-capture onto metals can contribute a non-negligible fraction to the overall luminosity. To account for this, we define the ZAHeMS as occurring when: 1) helium burning is the dominant contributor to nuclear burning and 2) the ratio of nuclear burning to overall luminosity is above the critical threshold of 0.985 as used in v1.

Finally, we must make a choice about the initial rotation rate of stars in our grids. We opt to initialize the single stars in our grids as non-rotating. However, for binaries the two obvious choices are: (1) to assume that stars are initially non rotating, and any rotation comes from tidal and accretion spin-up, and (2) to assume that all binaries have stellar spins that are already synchronized with the orbit by the time they reach ZAMS. In \posydon, we decided to go with the second option, which as we describe in \citetalias{2023ApJS..264...45F}, implies that the synchronization has happened during the pre-main sequence phase. The latter might not be true for wide binaries, but our assumption induces negligible rotation to the stellar components of those systems and does not affect their further evolution.

\subsection{Single star grids: H--rich and He-rich}\label{sec:singles}

\begin{figure*}
    \centering
    \includegraphics[width=\textwidth,angle=0]{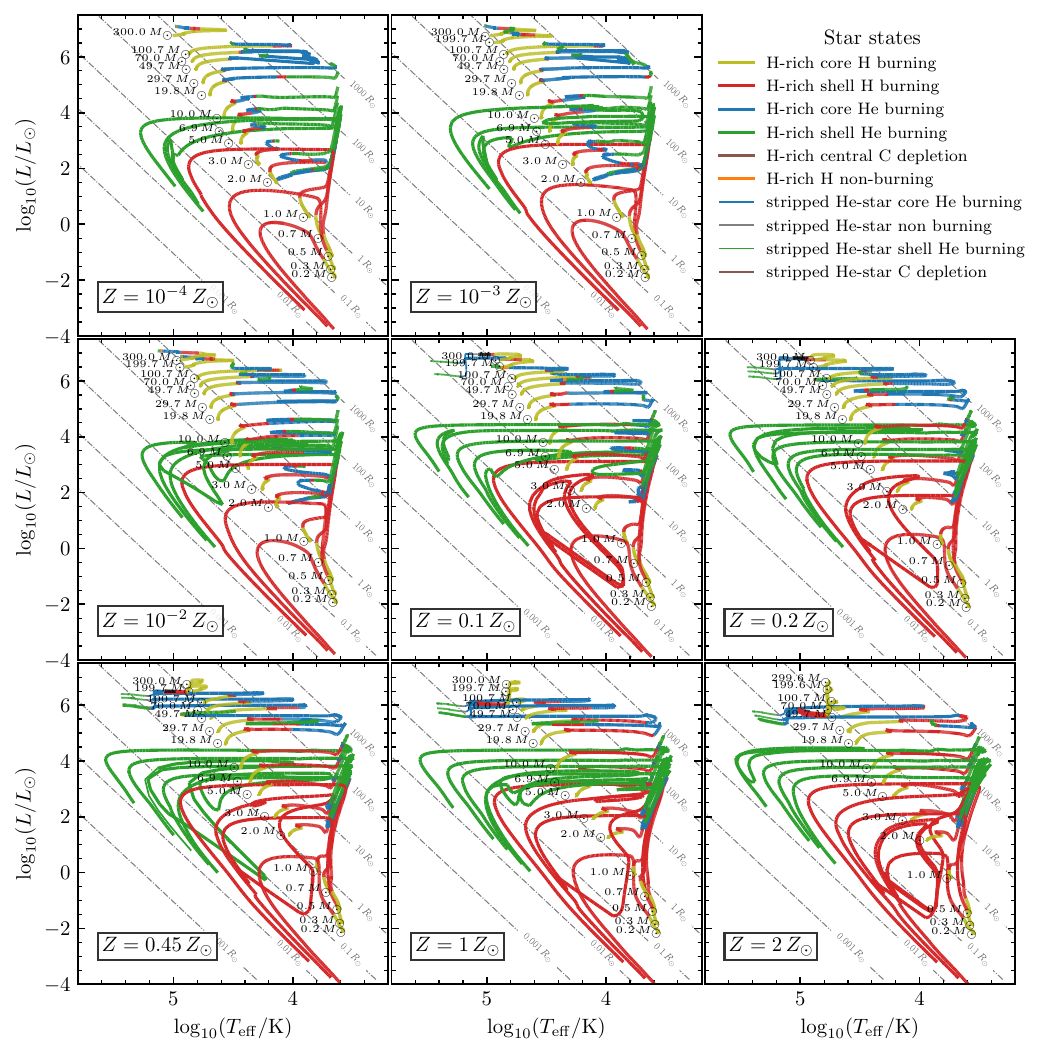}
    \caption{Hertzsprung--Russell diagrams for the \posydon grids of single star H-rich tracks, each corresponding to a different initial metallicity, displayed on bottom left of each panel. For clarity reasons, we only show 17 of our 375 tracks in each metallicity, where we state the initial mass at the beginning of each track. Line colors indicate the state of the stars. The largest differences between metallicities are seen for high-mass stars, characterized by the highest luminosity tracks in each panel.}
    \label{fig:singles_HMS_HRD}
\end{figure*}

For use in our detached step, as well as for the newly introduced features of single-star populations (see Section~\ref{sec:single_stellar_pop}), we produce grids of single H-rich stars and single He-rich stars across our eight chosen metallicities. Following v1, these grids are non-rotating, densely spaced, and extensive, ranging from very low masses ($0.1\,\Msun$ for our H-rich grid and $0.5\,\Msun$ for our He-rich grid) to very high masses ($300\,\Msun$ for our H-rich grid and $151\,\Msun$ for our He-rich grid). We provide details about the grid spacing and failure rates in Tables~\ref{tab:grid_properties} and \ref{tab:failure}. We initialize each model from ZAMS as described in Section~\ref{sec:zams}, follow it until termination (typically central carbon exhaustion, onset of PISN (see Section~\ref{sec:PISN}), or, for lower mass stars, when the central Coulomb coupling parameter $\Gamma_c$ exceeds 10 signifying the formation of a WD), and post-process the set of grids following the procedure outlined in Section~\ref{sec:postprocessing}. In addition to all the steps taken for stellar binaries, our post-processing procedure additionally resamples the evolutionary history output of single star models using the Equivalent Evolutionary Points (EEP) algorithm \citep{2016ApJS..222....8D}. As in v1, this resampling allows for accurate and robust interpolation of the evolutionary tracks across initial masses and time and is used by the detached step to evolve detached binaries after a common envelope or supernova.

\subsubsection{Effect of metallicity on single, H-rich stars}
\label{sec:single_H}

\begin{figure}[t]\center
    \includegraphics[scale=1,angle=0]{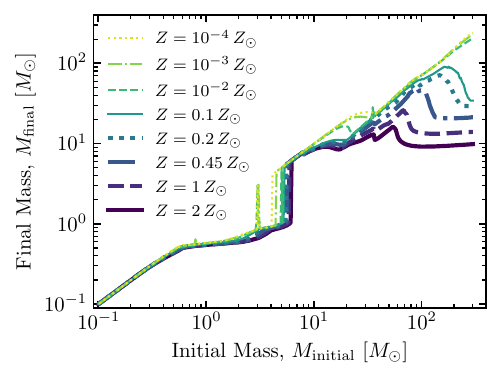}
    \caption{Initial-final mass relation for the v2 single star HMS grid for all metallicities. Note, the spike at about $3\,\Msun$ is an artifact from not finding an EEP at low metallicity for one model. }
    \label{fig:initial_final_v2_allZ}
\end{figure}

The most significant effects of metallicity on massive, single star evolution come from the impact of stellar winds \citep[for a detailed review of the effects of metallicity on single star evolution, see][]{2003ApJ...591..288H, 2014ARA&A..52..487S}. This dependence is highlighted in Figure~\ref{fig:mass_loss}, where we show, for four indicative metallicities, the fractional mass lost to different wind schemes during different evolutionary phases as a function of initial mass. High-mass, high-metallicity stars (top rows) lose most of their mass in a combination of cool- and hot-scheme winds, resulting in stars with cores containing only a fraction of their original mass at collapse. Even a slight reduction in the metallicity, down to $Z = 0.2\,\Zsun$, shows a modest impact in the final stellar mass. For $Z \lesssim 0.1\,\Zsun$ most mass lost from a massive star only occurs once it reaches the LBV phase, resulting in stars retaining at least half their initial mass immediately prior to collapse. Massive stars with $Z<0.1\,\Zsun$ exhibit behavior similar to the bottom right panel in Figure~\ref{fig:mass_loss}. 

For lower mass stars, the effects of metallicity are less pronounced, naturally resulting from our adoption of cool scheme winds that are metallicity independent. Any differences in the panels of Figure~\ref{fig:mass_loss} between stars with $M_\mathrm{initial} < 8\,\Msun$ are therefore due to indirect effects (e.g., the opacity effects of metallicity altering a star's equilibrium radius, effective temperature, and luminosity). 

\begin{figure}[t]\center
    \includegraphics[scale=1,angle=0]{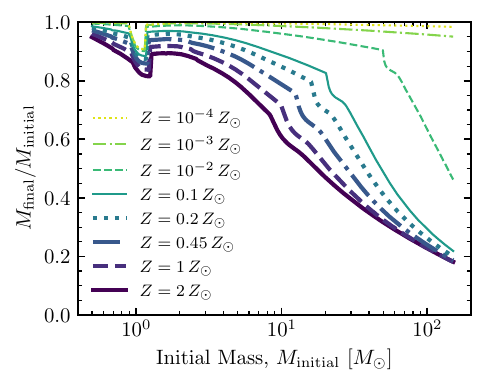}
    \caption{Mass retention rates for v2 single stars in our HeMS grid for all metallicities. }
    \label{fig:mass_retention_HeMS_v2_allZ}
\end{figure}

The metallicity dependence of stellar winds can also be seen in Figure~\ref{fig:singles_HMS_HRD} where we display the evolutionary tracks of single H-rich stars in a Hertzsprung-Russell (HR) diagram, one panel for each metallicity. As was seen in Figure~\ref{fig:mass_loss}, for stars with $M_\mathrm{initial} \lesssim 8\,\Msun$, differences between different metallicities are minor. For example, low-mass main sequence stars have slightly smaller radii at lower metallicities. Additional differences, such as hotter core temperatures and shorter lifetimes for low-metallicity stars, are not apparent in Figure~\ref{fig:singles_HMS_HRD}. At higher masses however, differences begin to appear. In our $2\,\Zsun$ grid for example, the most massive stars evolve toward lower luminosities before moving blueward on the HR diagram, while somewhat less massive stars maintain their luminosities, evolving bluewards as they lose their envelopes due to high wind mass-loss rates from the hot scheme winds. At progressively lower metallicities, the bluewards shift of massive stars occurs at higher and higher masses, until there is essentially no formation of Wolf-Rayet stars at $Z\lesssim10^{-2}\,\Zsun$. Other subtle effects on the evolution of single H stars can be seen in Figure~\ref{fig:singles_HMS_HRD}, a description of which is outside the scope of this work where we focus on stellar binaries. For a more thorough description of the impact of metallicity on single stars, we refer the reader to dedicated efforts \citep[e.g.,][]{2016ApJ...823..102C}.

\begin{figure*}
    \centering
    \includegraphics[width=\textwidth,angle=0]{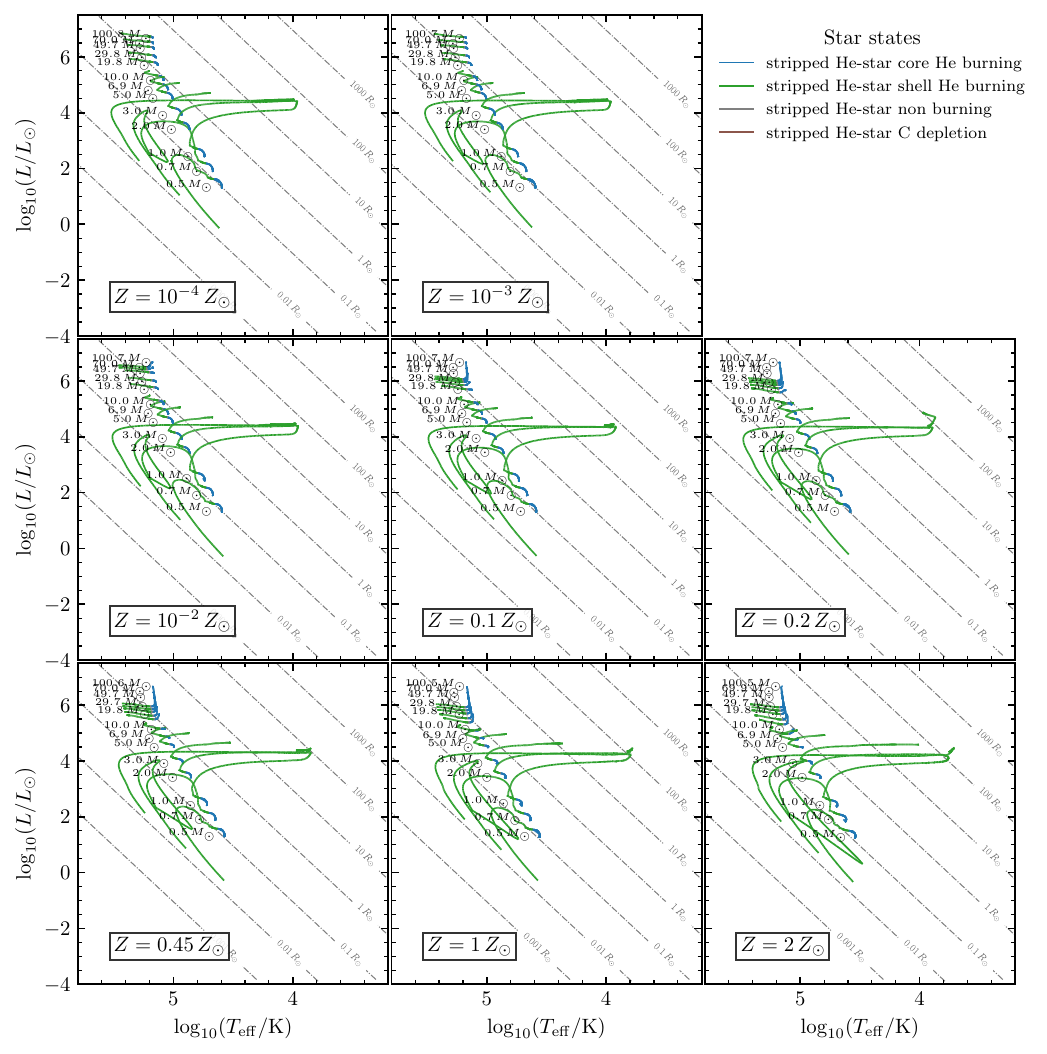}
    \caption{Hertzsprung--Russell diagrams for the \posydon grids of helium single star tracks, each corresponding to a different initial metallicity, displayed on bottom left of each panel. For clarity reasons, we only show 13 of our 268 tracks in each metallicity, where we state the initial mass at the beginning of each track. The color indicates the burning state of the stars. As with H-rich stars, the largest differences in He-star evolution across different metallicities occur at the highest masses; at these masses winds are significant for high-metallicity stars but inconsequential for low-metallicity stars.}
    \label{fig:singles_HeMS}
\end{figure*}

\begin{figure*}
    \centering
    \includegraphics[width=\textwidth,angle=0]{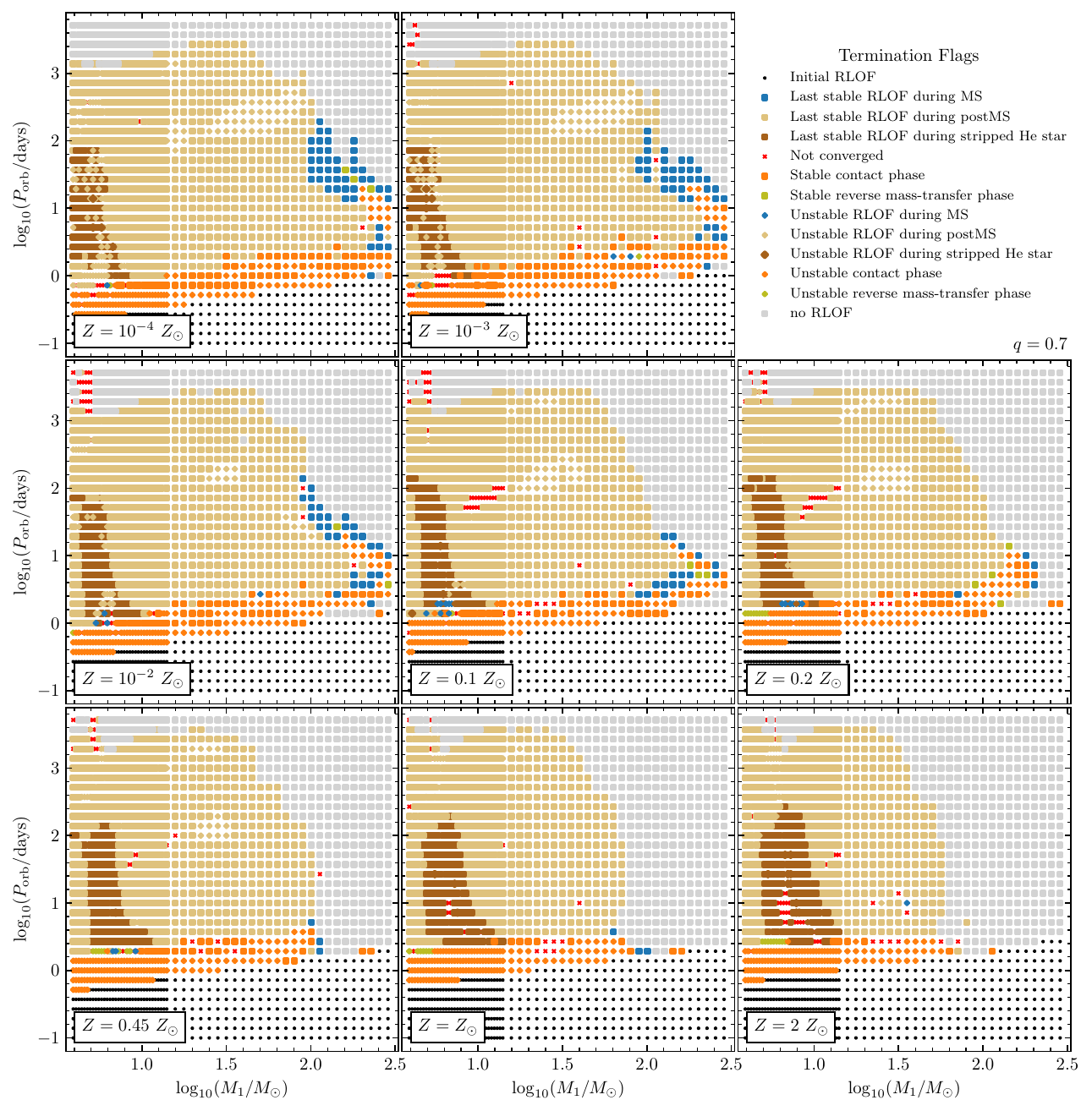}\\
    \caption{Eight slices ($q=0.7$) from our grid of two H-rich stars, one panel per metallicity. The evolution of each model is summarized using different symbols. Models that experienced stable or no MT are represented by squares, indicating that they reached the end of the life of one of the stars. Models that stopped during MT due to conditions for dynamical instability are represented by diamonds. Colors indicate the evolutionary phase of the donor star during the latest episode of MT (or no MT at all in gray). Small black dots represent systems that were in initial RLOF at birth, while red crosses represent models that stopped prematurely for numerical reasons. }
    \label{fig:HMS-HMS_MESA_grid_TF12_9panels_temp}
\end{figure*}

Due to our focus on the CO masses resulting from our model, we include one more comparison of single star evolution as a function of metallicity: the initial--final mass relation as shown in Figure~\ref{fig:initial_final_v2_allZ}. Note the final masses shown here are the stellar masses at the end of our \mesa simulations upon central carbon depletion, the onset of PISN, or WD formation, not our derived CO masses which is discussed in Section~\ref{sec:CCSN}. Therefore, these do not account for mass loss due to PISN or pulsational PISN (PPISN). In agreement with Figures~\ref{fig:mass_loss} and \ref{fig:singles_HMS_HRD}, we see in Figure~\ref{fig:initial_final_v2_allZ} that metallicity predominantly affects the final masses of the most massive stars, where stellar winds are most impactful.

\subsubsection{Effect of metallicity on single, He-rich stars}
\label{sec:single_He}

\begin{figure*}
    \centering
    \includegraphics[width=\textwidth,angle=0]{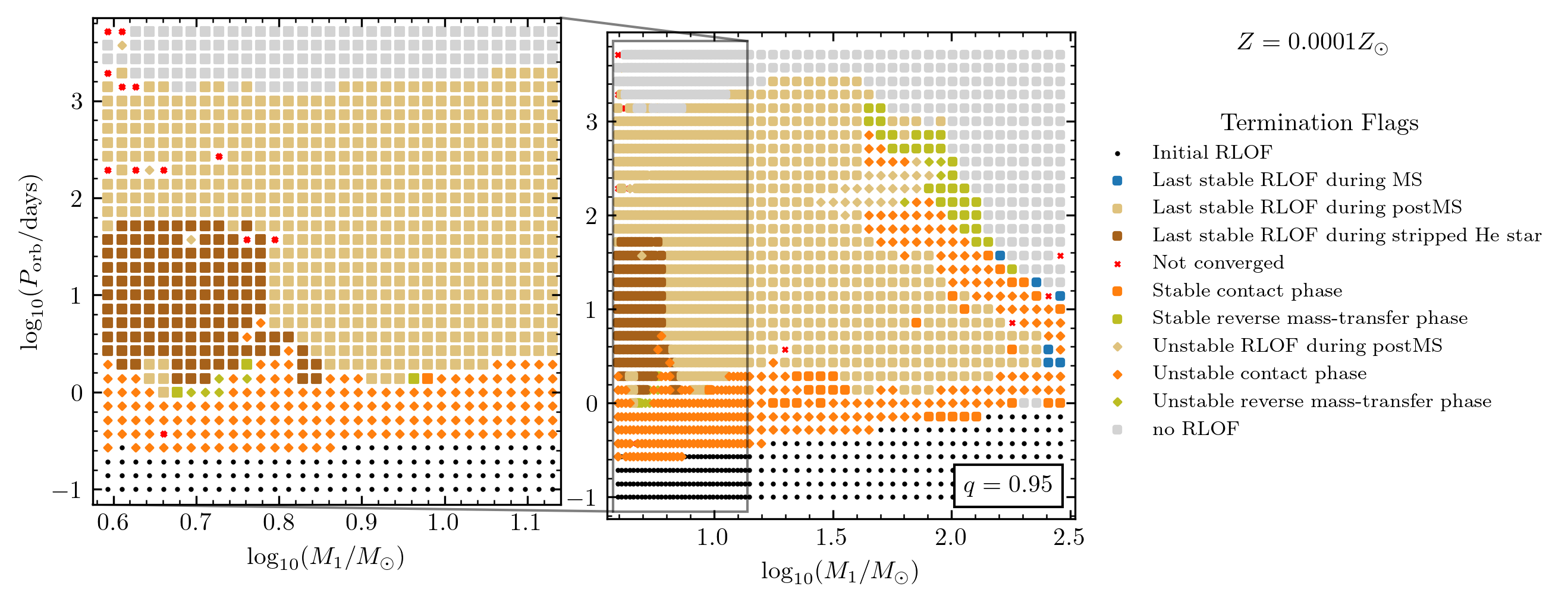}\\
    \caption{In the left panel we highlight the increased resolution of our grids for primary star masses of 3.92--14.04~\Msun for one example grid slice at $Z = 10^{-4}~\Zsun$ with $q=0.95$. The markers in this high-resolution region of the grid overlap in the right panel which shows the entire mass range in the grid slice analogous to the grid slices in Figure~\ref{fig:HMS-HMS_MESA_grid_TF12_9panels_temp}. }
    \label{fig:HMS-HMS_low_mass_extension}
\end{figure*}

Similar to v1, we generate  a grid of pure-He single stars following the procedure described in Section~\ref{sec:zams}, expanded to eight different metallicities \citep[for a detailed review of the evolution of single helium stars, see][]{2019ApJ...878...49W}. In Figure~\ref{fig:mass_retention_HeMS_v2_allZ} we show the ratio between the final  and initial mass of these single HeMS stars for our eight metallicities. Lower mass helium stars, below a few $\Msun$, do not experience significant wind mass loss, and thus their evolution shows little dependence on metallicity. The dip at low masses is caused by He stars $\gtrsim1\,\Msun$ evolving through a He giant phase. This trend can alternatively be seen in Figure~\ref{fig:singles_HeMS} where we show Hertzsprung–Russell diagrams for these grids, with one panel for each metallicity. The tracks for $1\,\Msun$ evolve redwards to large radii, whereas less massive He stars evolve first bluewards then to lower luminosities before becoming WDs. 

At higher masses, the differences in He-star evolution induced by metallicity becomes more obvious. In Figure~\ref{fig:mass_retention_HeMS_v2_allZ} the most massive He stars lose $\simeq80\%$ of their mass in winds at \Zsun, while at the lowest metallicities these stars may lose only a few per cent of their mass. The difference in their evolution can also be seen in Figure~\ref{fig:singles_HeMS}, where the highest mass He stars at high metallicities evolve towards lower luminosities before trending bluewards due to mass loss while the same stars at low metallicities never reduce their luminosities. We discuss this point further in Section~\ref{sec:CO-HeMS} where we describe the impact of metallicity on our binary grid comprised of a CO with a He-star companion.

\subsection{Binary Evolution with Two H-rich Main-Sequence Stars: HMS--HMS Grid}
\label{subsec:HMS-HMS grid}

Following our procedure outlined in Section 5.5 of \citetalias{2023ApJS..264...45F}, we simulate each binary in our HMS--HMS grid by first initializing two single stars at ZAMS as described in Section~\ref{sec:zams}. After being placed in a binary, the stars' rotation periods are increased to be initially synchronized with the orbital period, and each binary is then evolved until one of our stopping conditions is met. Using our infrastructure for evolving large grids of models within high-performance computing environments, we simulate 39,712 models at each of our eight metallicities for a total of 317,696 models in this grid type.

In Figure~\ref{fig:HMS-HMS_MESA_grid_TF12_9panels_temp} we show a two-dimensional slice of this double H-rich star grid at a fixed initial binary mass ratio $q=0.7$ for each of our metallicities. Each point in the figure corresponds to one binary simulation, with the marker's shape and color indicating information about its final state: black points indicate binaries that would overfill their Roche lobes at initialization (``Initial RLOF''), gray squares indicate binaries that never overfill their Roche lobes (``no RLOF''), colored squares indicate binaries that went through stable MT and therefore evolve to central carbon exhaustion (``Stable RLOF''), and colored diamonds evolve into a CE (``unstable RLOF''). Following \citetalias{2023ApJS..264...45F}, since the stars in Initial RLOF binaries do not fit inside their orbits, we do not consider these systems to be viable binaries and halt their evolution at ZAMS. The black points in each panel are therefore non-viable and indicate the lower orbital period bound for our binary populations. The colors of the markers indicate the evolutionary state when MT ceases (this scheme differs from what is commonly used in the literature where the evolutionary state at RLOF is indicated). Finally, red points represent binaries that failed to converge at some point during their evolution (typically a few per cent of our models).

Comparison between the different panels of Figure~\ref{fig:HMS-HMS_MESA_grid_TF12_9panels_temp} shows that MT at this mass ratio ($q=0.7$) is largely stable during post-MS and the boundaries between no RLOF, stable RLOF, contact binaries, and initial RLOF are all found at similar positions. For lower and higher mass ratios, some variations start to appear; however, in general metallicity plays a minor role in the MT outcome of low-mass binaries. Of course, as metallicity affects a star's opacity and therefore its radius, quantitative differences exist. For higher mass stars ($M\gtrsim 40\,\Msun$), differences between metallicities become substantial. Rooted in the dependence of stellar winds on metallicity, massive stars at high metallicity expand less, or even not at all, producing tightly orbiting binaries ($P_{\rm orb}\sim$ days) that never interact \citep{2024arXiv241018501K}. At metallicities $\lesssim 0.1\,\Zsun$, even the most massive stars expand, leading to stable MT on the MS.

\begin{figure}
    \centering
    \includegraphics[width=\columnwidth,angle=0]{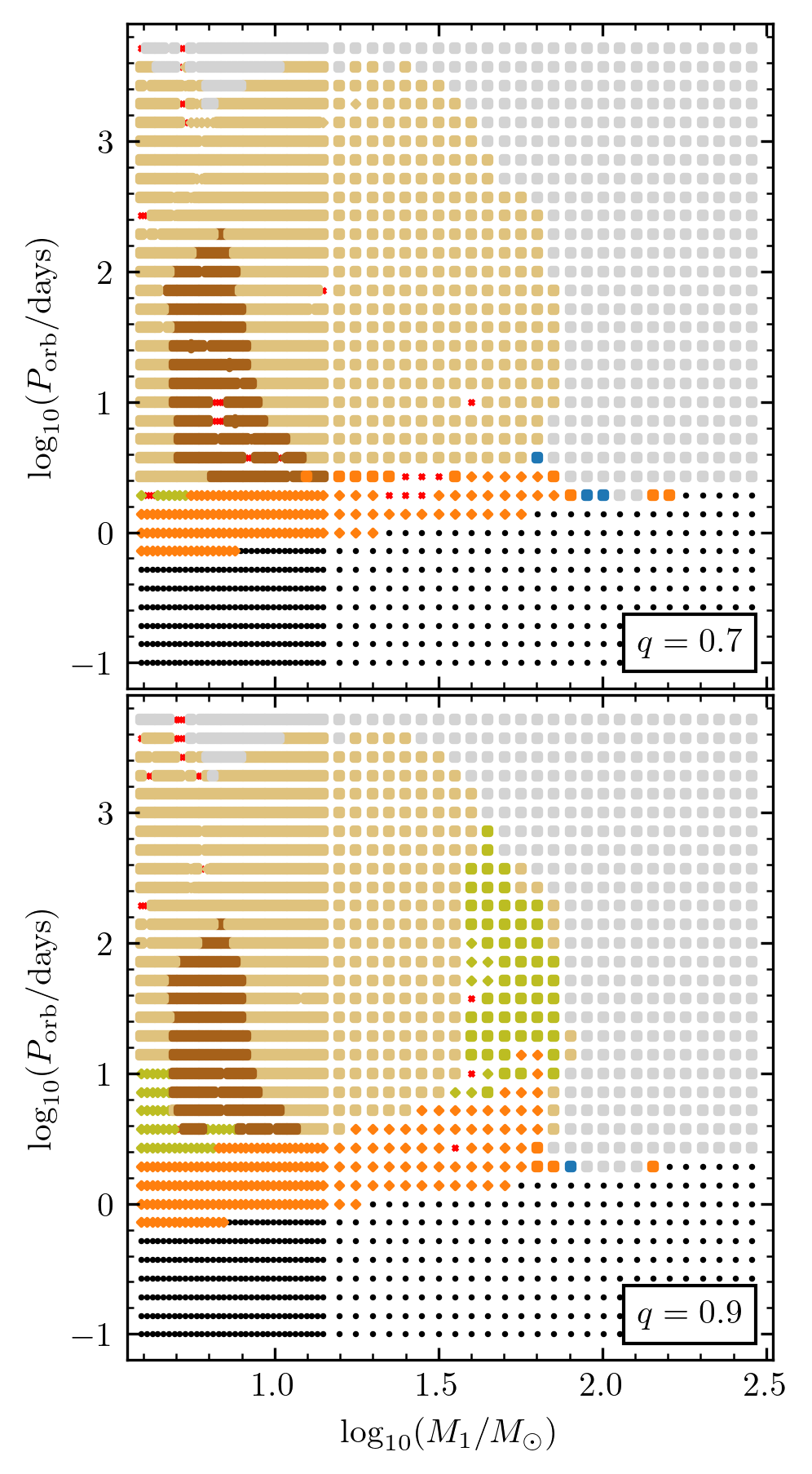}\\
    \caption{For $Z_{\odot}$ we show two slices of our HMS-HMS grids (plot markers are the same as in Figure~\ref{fig:HMS-HMS_MESA_grid_TF12_9panels_temp}). Whereas the reverse MT class only exists in a very small region of the $q=0.7$ slice, reverse MT becomes a significant class for $q\gtrsim0.9$, particularly for the most massive stars that overfill their Roche lobes. }
    \label{fig:HMS-HMS_q_trend}
\end{figure}

\begin{figure*}
    \centering
    \includegraphics[width=\textwidth,angle=0]{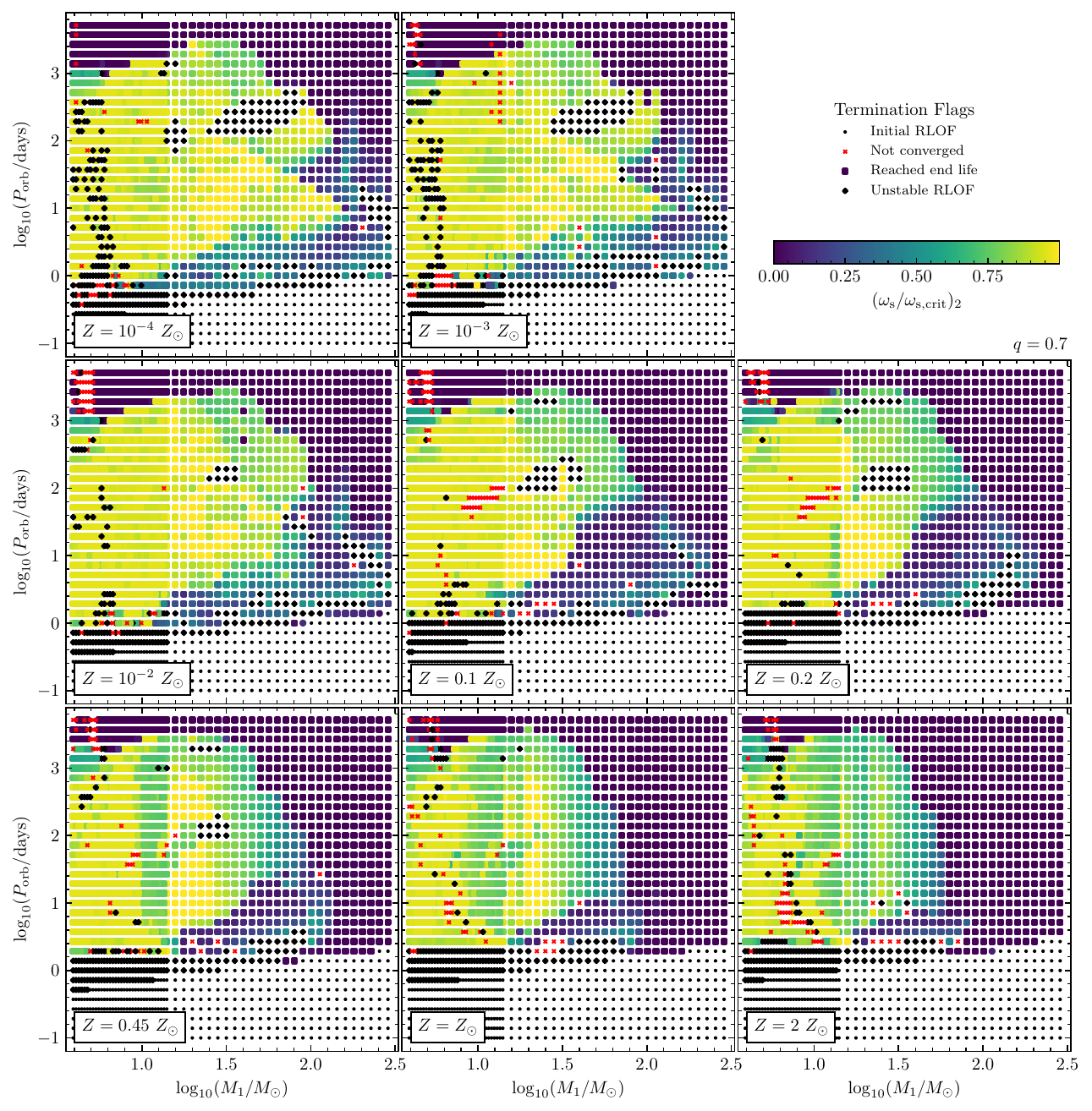}\\
    \caption{For the same grid slices shown in Figure~\ref{fig:HMS-HMS_MESA_grid_TF12_9panels_temp} with eight metallicities and a fixed initial binary mass ratio $q=0.7$, we show systems where one of the two stars reached the end of its life, represented by a color point indicating the final ratio of the angular velocity of the secondary star (initially less massive) divided by its critical rotation rate, denoted as $(\omega_\mathrm{s}/\omega_\mathrm{s,crit})_2$. In cases where MT occurred, the secondary star gained angular momentum and spun up, resulting in a highly spinning accretor star that remained in this state until the end of the simulation.}
    \label{fig:HMS-HMS_MESA_grid_omega_9panels_temp}
\end{figure*}

\begin{figure*}
    \centering
    \includegraphics[width=\textwidth,angle=0]{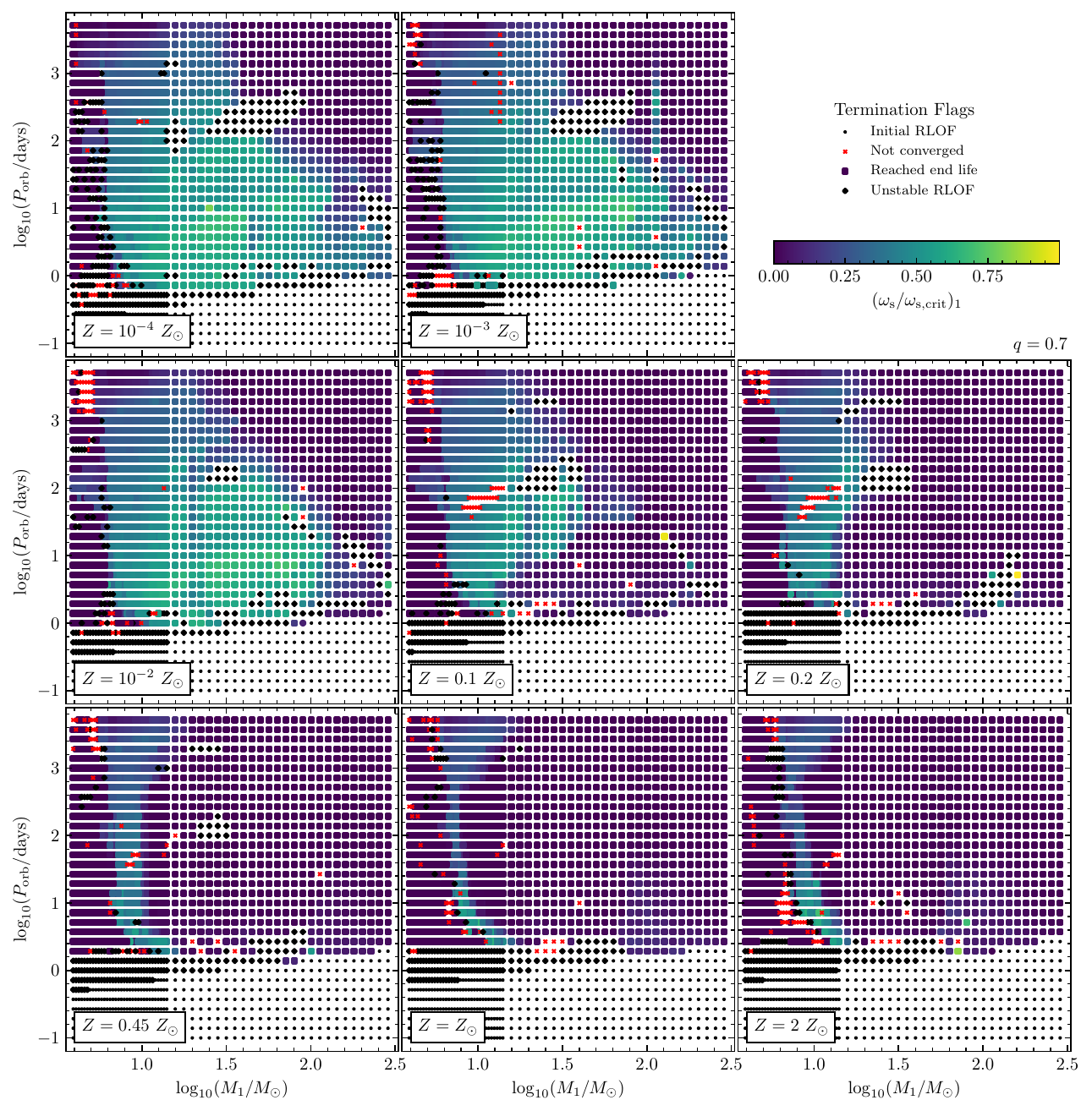}\\
    \caption{For the same grid slices shown in Figure~\ref{fig:HMS-HMS_MESA_grid_TF12_9panels_temp}, and the figure follows the description of Figure~\ref{fig:HMS-HMS_MESA_grid_omega_9panels_temp}, but the color indicates the final ratio of the angular velocity of the primary star (initially more massive) divided by its critical rotation rate, denoted as $(\omega_\mathrm{s}/\omega_\mathrm{s,crit})_1$. In cases where MT occurred, the primary star lost angular momentum and spun down, resulting in a slowly rotating donor star that remained in this state until the end of its life.}
    \label{fig:HMS-HMS_MESA_grid_omega_9panels_temp1}
\end{figure*}

While our HMS-HMS grid is regularly spaced, for simulations with primary star masses 3.92--14.04~\Msun, we triple our grid resolution to better capture model variations in this mass range. In Figure~\ref{fig:HMS-HMS_MESA_grid_TF12_9panels_temp} the markers for individual simulations in this mass range overlap. To visually obtain a sense of how this increased resolution region compares to the remainder of the HMS-HMS grid, we select one example grid slice ($Z=10^{-4}~\Zsun$ and $q=0.95$) and provide a zoomed-in view of these high-resolution models in Figure~\ref{fig:HMS-HMS_low_mass_extension}. Roughly half of all our HMS-HMS models are contained within this regime.

In Figure~\ref{fig:HMS-HMS_MESA_grid_TF12_9panels_temp} we note a few peculiarities which we discuss explicitly. At the highest masses and shortest orbital periods (that are not classified as Initial RLOF; $\sim$days), there are some models that never interact (gray squares). The stars in these models expand only slightly due to their rapid rotation rates (synchronized with the orbit) while the same mass stars at larger periods allow the stars to expand further during their evolution, leading to eventual RLOF. We further note a narrow region of models at high masses with $Z=10^{-2}-10^{-1}\,\Zsun$ that exhibit reverse MT (see Section~\ref{sec:reverse_MT}). Although small in the panels in Figure~\ref{fig:HMS-HMS_MESA_grid_TF12_9panels_temp}, the regions expand immensely for binaries with $q\gtrsim 0.9$ at higher metallicity. In Figure~\ref{fig:HMS-HMS_q_trend} the growth of the reverse MT regions  at larger $q$ can be seen explicitly; where the reverse MT class exists only as a small region at $q=0.7$, for higher $q$ values, reverse MT occupies a higher fraction of the HMS-HMS grid, particularly at larger primary masses. (all our grid slices with different $q$ values are provided in the online version of Figure~\ref{fig:HMS-HMS_MESA_grid_TF12_9panels_temp} as a figure set). 

Finally, there are several ``jagged" boundaries, for instance at high masses as well as within the contact regime, suggesting there is room for improvement within our resolution. We plan to explore this in future work. Nevertheless, our classification schemes do not consider the evolutionary state of the donor star during MT, only whether that MT was stable or unstable. We are therefore confident that our classification scheme, at least, is unaffected by such numerical issues, given our high classification accuracies (see Section~\ref{sec:machine_learning}).

Of course, we track not only the MT history of the binary, but a whole slew of parameters describing the binary's evolution as well as its component stars: mass, radius, rotation speed, bolometric luminosity, effective temperature, and stellar core mass and radius to name a few. As an example in Figure~\ref{fig:HMS-HMS_MESA_grid_omega_9panels_temp}, we provide the secondary (the initially less-massive star, in this case the accretor in a mass-transferring system) star's rotation rate as a fraction of its critical rate at the final evolutionary step. During the MT phase, angular momentum is efficiently transferred from the donor to the accretor star, resulting in rapidly spinning stars in regions of stable mass transfer, as shown in Figure~\ref{fig:HMS-HMS_MESA_grid_omega_9panels_temp}. For more massive accretors, stronger stellar winds lead to greater angular momentum loss, keeping them farther from critical rotation speeds compared to lower-mass stars.

Conversely, the rotation rates of the donor stars, as analyzed from Figure~\ref{fig:HMS-HMS_MESA_grid_omega_9panels_temp1}, are generally much lower than their critical rotation rates. For metal-poor stars, as seen in the upper left panel, the donor star shrinks in radius and spins up near the end of MT, maintaining a relatively fast rotation rate until the end of the simulation. In contrast, for metal-rich stars, observed in the lower right panel, the donor star is nearly non-rotating at the end of the simulation, even in close massive binaries. This could be attributed to stronger stellar winds that: 1) continuously reduce the star’s spin angular momentum, and 2) cause a reduction in the ratio between the star's radius and the system separation $R/a$. Since tidal strength is sensitive to this ratio, tidal synchronization may not be sufficient to spin up the star again, even in close orbits.

\subsection{Binary Evolution with a CO and a H-rich Main-Sequence Star: CO--HMS Grid}
\label{sec:CO-HMS}

\begin{figure*}[t]\center
    \includegraphics[width=\linewidth,angle=0]{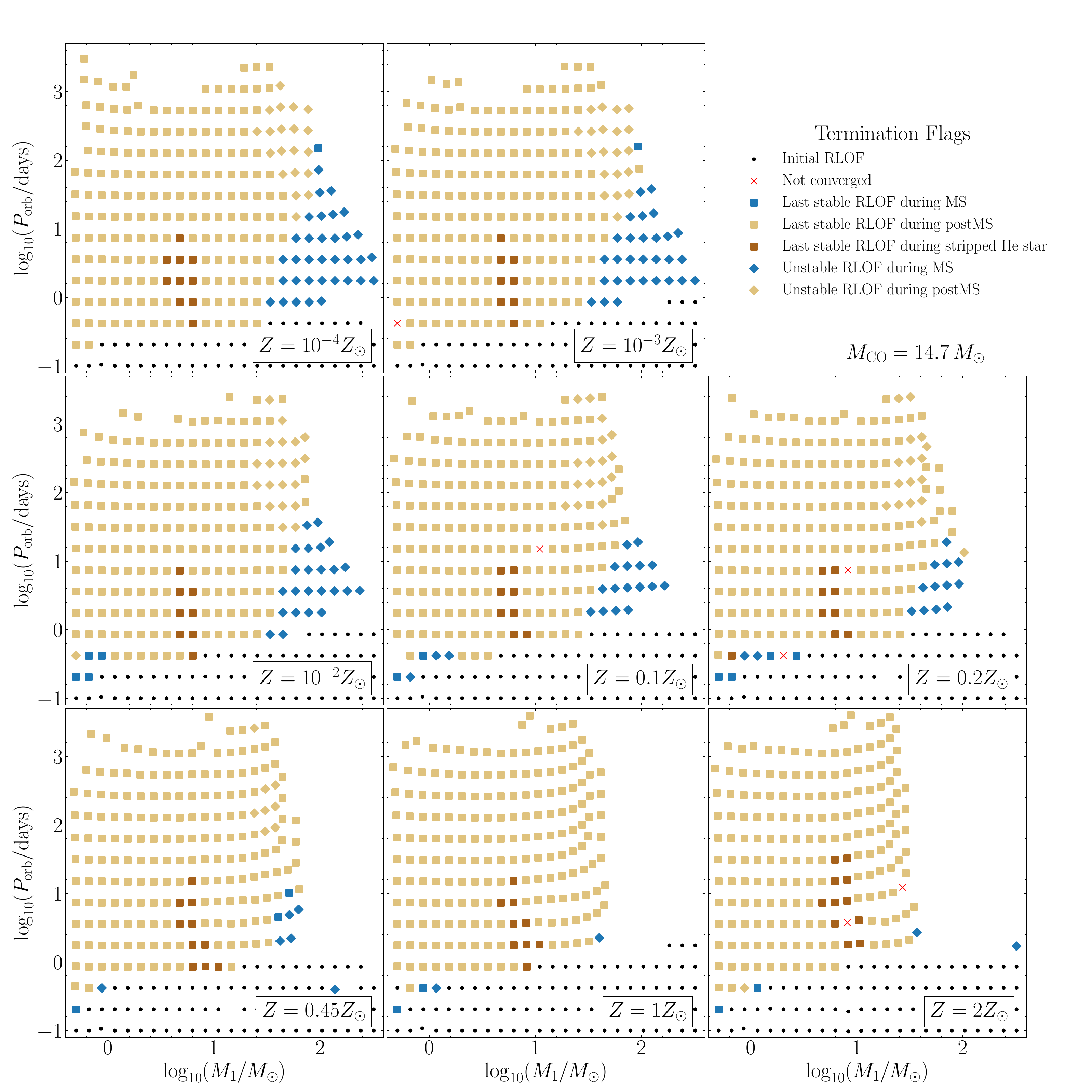}\\
    \caption{View of eight grid slices, each corresponding to a different initial metallicity, is displayed in each panel. The grid consists of binary-star models consisting of a CO and a H-rich star at the onset of RLOF and correspond to a CO mass of $14.7\,\Msun$, similar to a stellar-mass BH. The evolution of each model is summarized using different symbols and is the same as that described in Figure~\ref{fig:HMS-HMS_MESA_grid_TF12_9panels_temp}. Small black dots represent systems that were in initial RLOF at birth, while red crosses represent models that stopped prematurely for numerical reasons.}
    \label{fig:CO-HMS_RLO_grid_TF12_9panels_BH}
\end{figure*}

At the end of the HMS--HMS grid sequences, one of the two stars collapses into a CO. If the binary survives the CO formation event, then its next evolutionary phase involves a CO in orbit with a non-degenerate stellar companion. We have generated a grid of binary sequences comprised of a CO and a H-rich star to model cases when the non-degenerate star fills its Roche lobe initiating MT. The full parameters of this grid, which is comprised of over $10^5$ separate binary simulations, are listed in Table~\ref{tab:grid_properties}. Because our procedure only requires the self-consistency afforded by a detailed binary evolution simulation once MT has begun, our post-processing step removes any evolution prior to RLOF and simulations without MT; our processed grids are somewhat smaller than the $>10^5$ binaries initialized.

In Figure~\ref{fig:CO-HMS_RLO_grid_TF12_9panels_BH} we show slices from this grid for our eight metallicities and a CO mass of $14.7\,\Msun$ (corresponding to a stellar-mass BH). We provide additional figures with different mass, representing neutron star (NS) and BH companions in the online version of Figure~\ref{fig:CO-HMS_RLO_grid_TF12_9panels_BH} as a figure set. As was the case in v1, prior to RLOF we only incorporate changes to the orbital angular momentum due to wind-mass loss, resulting in the non-regular grid seen in the figure. Once RLOF begins, angular momentum terms such as tides, magnetic breaking, and gravitational radiation are turned on. The binary is initially synchronized with the orbit and synchronization via tides is calculated for the remainder of a binary's evolution. 

Binary tracks that avoid any MT are not included in the grid \citepalias[such binaries in our populations are handled by our detached evolution step, see Section~8.1 in][]{2023ApJS..264...45F}, which causes the white space surrounding the parameter space in Figure~\ref{fig:CO-HMS_RLO_grid_TF12_9panels_BH}. Similar to Figure~\ref{fig:HMS-HMS_MESA_grid_TF12_9panels_temp}, binaries with very wide orbits (around $\gtrsim 1000\,\mathrm{days}$ in Figure~\ref{fig:CO-HMS_RLO_grid_TF12_9panels_BH}) are too widely separated to interact, and stars that are massive enough to have strong stellar winds never enter the giant phase to start RLOF. Instead, these stars lose their H-rich envelopes and become Wolf--Rayet stars. The limiting mass at which this occurs depends on metallicity. 

\begin{figure*}[ht]
    \includegraphics[scale=1.0,angle=0]{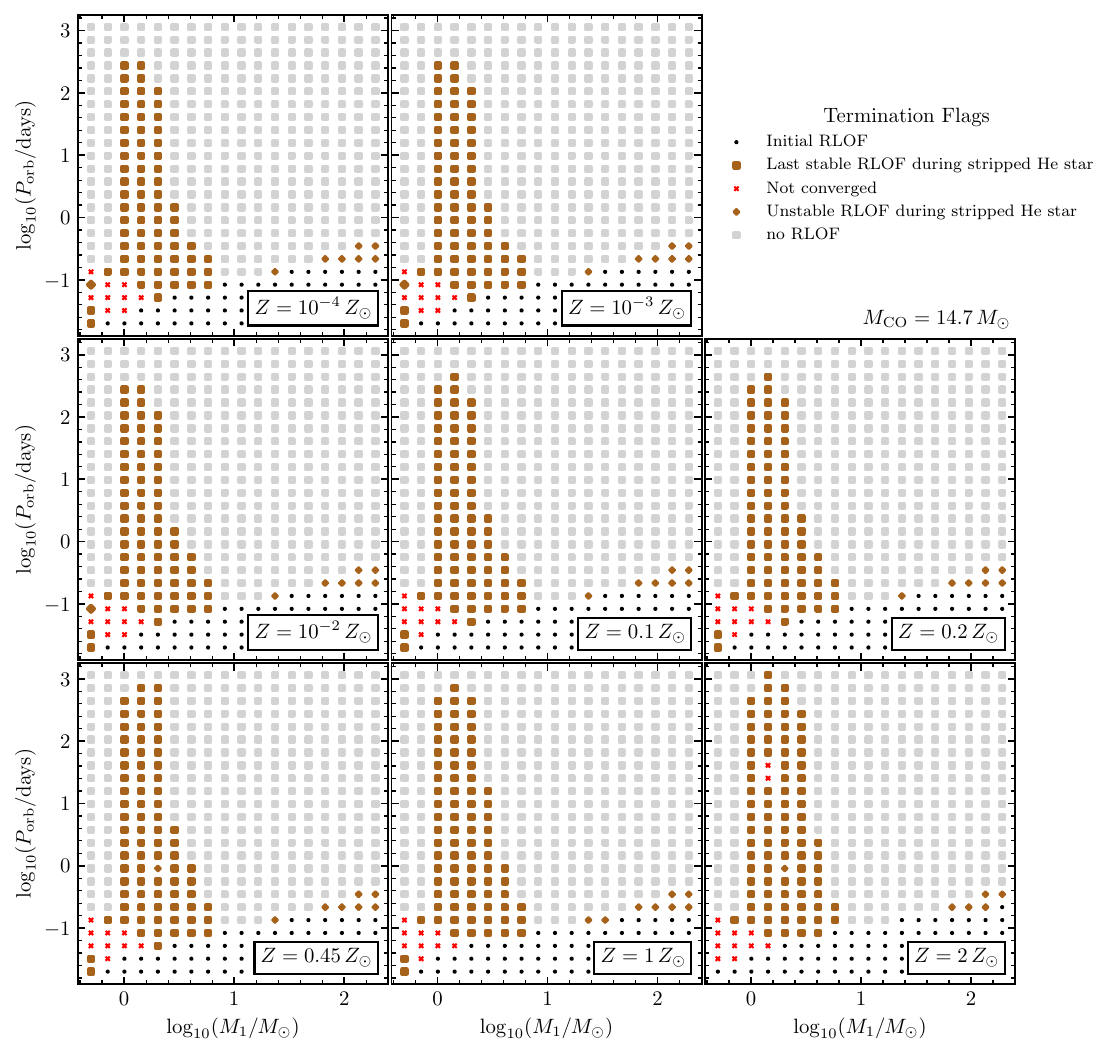}
    \caption{The panels show views of eight grid slices, each representing a different initial metallicity. The grids contain binary star models consisting of a He star and a CO with a mass of $14.7\,\Msun$, intended to represent a stellar-mass BH. The symbols summarizing the evolution of each model have been kept the same as those in Figure~\ref{fig:HMS-HMS_MESA_grid_TF12_9panels_temp}. }
\label{fig:CO-HeMS_BH}
\end{figure*}

Comparison between the panels of Figure~\ref{fig:CO-HMS_RLO_grid_TF12_9panels_BH} show that lower metallicities result in a much larger parameter space of interacting binaries, particularly in mass space. This trend is analogous to one seen in the HMS--HMS grid; massive stars at lower metallicity tend to be more compact, with weaker stellar winds, allowing them to eventually expand as giant stars. Because these more massive stars at low $Z$ are now overfilling their Roche lobes, for a constant CO mass, RLOF occurs for binaries with a more extreme mass ratio. Therefore, Figure~\ref{fig:CO-HMS_RLO_grid_TF12_9panels_BH} shows that there are now large pockets of unstable MT (diamond markers) in the newly introduced parameter regions with decreasing metallicity. For accreting BHs, initial donor masses for stable RLOF remain $\lesssim50\,\Msun$ for metallicities $\gtrsim 0.45\,\Zsun$ and $\lesssim 10^{-2}\,\Zsun$, while reaching $90\,\Msun$ for the intermediate metallicities ($0.1\,\Zsun$ and $0.2\,\Zsun$). For accreting NSs, on the other hand, most MT becomes unstable due to the more extreme mass ratios at RLOF (see online version of Figure~\ref{fig:CO-HMS_RLO_grid_TF12_9panels_BH} where we provide additional grid slices as a figure set). Therefore, metallicity does not strongly affect the region of stable RLOF. However, at lower metallicities, the unstable parameter space increases in initial donor mass from $50\,\Msun$ at $\Zsun$ to $240\,\Msun$ at $10^{-4}\,\Zsun$. 

Generally, the increased interacting parameter space (with decreasing metallicity) mainly corresponds to MS or post-MS donors at the onset of RLOF that suffer a dynamical instability and enter a CE phase \citepalias[handled by the CE step, see Section~8.2 in][]{2023ApJS..264...45F}. Since decreasing metallicity is associated with more compact and massive stellar cores and the rapid radial expansion of the star does not happen until later in their evolution when their stellar cores are more developed \citep{1992A&A...264..105M, 2003ApJ...591..288H, 2010ApJ...714.1217B}, the chances of surviving a CE are generally higher at lower metallicities, increasing the parameter space forming double CO (DCO) binaries.

\subsection{Binary Evolution with a CO and a Helium Main-Sequence Star: CO--HeMS Grid}
\label{sec:CO-HeMS}

For those binaries that survive a CE phase between a H-rich donor and its CO companion, the donor star reveals its He-rich core. To study the evolution of these systems, we run detailed binary grids consisting of 14,256 evolutionary tracks at 8 different metallicities, resulting in a grid comprised of $>100,000$ models. For a complete summary of the grid specifications, we refer to Table~\ref{tab:grid_properties}. The methodology closely follows that for CO--HMS stars described in Section~\ref{sec:CO-HMS}, but with a He-star primary, initialized as described in Section~\ref{sec:zams}. 

In Figure~\ref{fig:CO-HeMS_BH} we show an example slice from this grid at different metallicities corresponding to a BH accretor ($M_\mathrm{CO}=14.7\,\Msun$). An equivalent figure for different mass, representing NS and BH companions are provided in the online version of Figure~\ref{fig:CO-HeMS_BH} as a figure set. Models with close binary separations and initial RLOF are shown with black dots, while models that do not interact are represented by gray square markers. 

The detailed characteristics of this grid at $\Zsun$ are described in \citetalias{2023ApJS..264...45F}. Here, we only point out that we typically observe MT initiated by He stars only within a mass range of 1 to $7\,\Msun$. Donors with lower masses remain too compact, while more massive He stars expand even less and never fill their Roche lobes. Comparison between the different panels shows that at least qualitatively, the grids are nearly unchanged by metallicity. As discussed in Section~\ref{sec:single_He} He stars with masses $\lesssim 8\,\Msun$ are not strongly affected by winds, regardless of their metallicity. Picking apart the minor differences, we find that higher metallicity stars expand more when they become He giants, allowing them to fill their Roche lobes at progressively wider orbital periods. As a result, MT occurs at increasingly larger $P_\mathrm{orb}$ for both NS and BH companions. Even for high mass He stars, where the winds are strongly metallicity dependent (as seen in Figure~\ref{fig:mass_retention_HeMS_v2_allZ}), the differences between metallicities are slight. For these systems, MT quickly becomes unstable upon RLOF, mitigating the effects from wind mass loss.

Across the eight metallicities, we identify one island of non-converging models, indicated by red crosses, with short orbital periods and low He-star masses. These failed simulations predominantly go on to form WDs while becoming further stripped. We leave improvements to \posydon specifically addressing these systems for future work.

The grids shown in Figure~\ref{fig:CO-HeMS_BH} describe the evolution of a CO with a ZAHeMS companion, which serves to model systems that survive a CE with a BH companion. The resulting post-CE orbit ought to be circular, making this grid optimal for a starting point for the next phase of evolution. However, during tests of our v2 grids, we found an alternative type of evolutionary sequences for which where this grid was insufficient. For binaries with mass ratios close to unity, the primary could undergo core collapse while the second star has lost its H envelope, either through winds or MT. The result is a CO with an evolved helium star companion in an eccentric orbit. Our detached step can handle the evolution of such systems so long as the binary remains detached. However, lower mass helium stars will expand and could overfill their Roche lobes in a Case BB MT phase. In v1, we could not appropriately account for the evolution of these systems. In v2, we address these systems by including a second version of our CO--HeMS grids in which the evolution prior to RLOF is removed. This treatment is analogous to our processing of the CO--HMS grid, and allows for MT with evolved He-star donors. Since there is relatively minor evolution of helium stars prior to MT the resulting grid appears very similar to that seen in Figure~\ref{fig:CO-HeMS_BH}. This grid is discussed further in Appendix~\ref{sec:CO-HeMS_RLO}.

\section{Interpolating through \posydon Binary Grids \label{sec:machine_learning}}

\begin{figure*}[t]
    \center
    \includegraphics[width=2\columnwidth,angle=0]{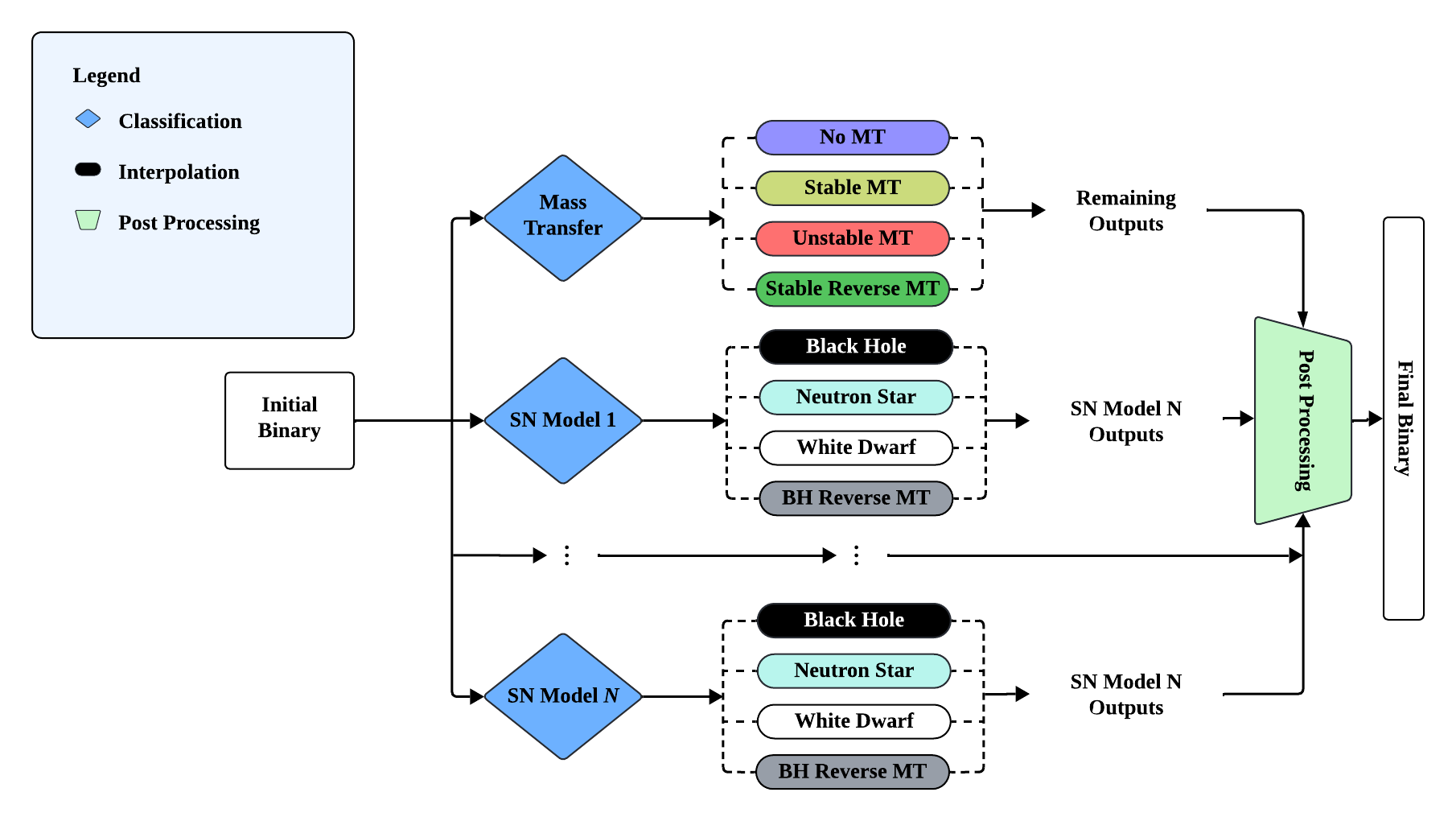}\\
    \caption{A flow chart diagram outlining the algorithm used to perform our interpolation on each of the grids. Interpolation happens homogeneously within a class and the output is combined at the end before a post-processing step.}
    \label{fig:flow-chart}
\end{figure*}

In v1 we developed a scheme for evolving binaries through each of our three binary grids using a combination of classification and three-dimensional interpolation methods. The aim is to predict the state of an arbitrary binary star system at the end of its evolution given its initial stellar masses and orbital period. For this, we require the mapping of our binary grids from their initial, input parameters to their final characteristics through each grid. Here we maintain the same goal and describe below improvements to our interpolation algorithm and its performance on the v2 grids. Elsewhere \citep{2025ApJ...984..154S} we address the more difficult problem of generating through three-dimensional interpolation the full, time-dependent evolutionary tracks of binary sequences using our pre-simulated grids. Such track generation will be included in future \posydon versions. 

Each grid is comprised of binaries with different types of evolutionary sequences, we first segment each grid into different broadly defined classes. To classify an input sample, we use a weighted $k$-nearest neighbor ($k$-NN) approach, where $k$ is optimized using a Monte Carlo cross-validation (MCCV) scheme based on the balanced accuracy (bACC) which is computed by averaging the sensitivity and specificity of each class \citep{watt_borhani_katsaggelos}. After a non-grid point input is classified, it is passed to an interpolator trained on the subset of our binary-star model grids within the same class, so a separate interpolation object is produced for each class. The interpolator uses Delaunay triangulation in the input space, forming a convex hull comprised of simplexes. To interpolate and find the end state of a system, we first determine the simplex within which the system's input parameters reside. We then calculate the final interpolated value at the hyperplane that passes through the vertices of that simplex. The metrics defining the quality of our classification and interpolation methods are computed using an independent validation set comprised of our ``random grid'' described in Section~\ref{sec:grids}.

Our scheme also includes pre- and post-processing steps: the former normalizes the data so that the input and output quantities span a comparable dynamical range, while the latter ensures that interpolated values are physical (e.g., maintaining monotonicity when physically appropriate, ensuring that stellar properties are consistent with known physical laws like the Stefan-Boltzmann Law, etc.). For a full description of the constraints we apply and our implementation, we refer the reader to Section 7.4 in \citetalias{2023ApJS..264...45F}. 

In the present work, we essentially follow the same procedure as in \citetalias{2023ApJS..264...45F}, making improvements to the classification and interpolation schemes, described in Section~\ref{sec:interpolation_improvements} below. We do not interpolate between metallicities, as this would add a fourth dimension to our input space. Instead, we generate separate, analogous classification and interpolation objects for each grid at each of our eight metallicities. This choice, which has several implications, was made predominantly for practical reasons. Furthermore, it ensures that the analysis of our classification and interpolation accuracies below is essentially identical to what was described in \citetalias{2023ApJS..264...45F}. However, we note that since we can only evolve binary populations at our eight specified metallicities, multi-metallicity populations with intrinsically smooth metallicity distributions, such as a cosmically evolving star formation history, must be mapped to a mixture of our discreet metallicity grids. We describe our procedure for modeling such populations in Section~\ref{sec:transients}.

Before proceeding to the specific improvements we make to our classification and interpolation schemes along with the accuracies we find for our v2 grids, we comment that \posydon is built to alternatively allow for binaries to evolve following a nearest neighbor approach. Under this setting, no classification or interpolation is required within our binary grids, as individual binaries are mapped to the nearest \mesa simulation. In Appendix~\ref{sec:nearest_neighbor} we compare the accuracy of the two methods. For essentially all parameters tested in each of our grids, our classification and interpolation methods are an improvement over a nearest-neighbor matching scheme. Our classification and interpolation methods are therefore used within \posydon by default, although we offer evolution using nearest-neighbor matching as a user option.

\subsection{Interpolation improvements}
\label{sec:interpolation_improvements}

In v1 each simulated binary is classified into a MT class before being passed to the corresponding interpolator. However, in v2, we have added an additional 24 classification schemes, one for each of the 24 SN models built into \posydon (see Section~\ref{sec:CCSN}). The interpolation input remains the same: the system's initial stellar masses and period. However, the resulting CO masses (along with a handful of additional parameters) are separately interpolated for each SN model, depending on whether the resulting object is a WD, NS, BH, or BH-Reverse-MT (a new class described by systems undergoing ``reverse MT'' described in Section~\ref{sec:reverse_MT}). Figure~\ref{fig:flow-chart} shows the schematic describing our procedure.
Independent classification is performed 25 times (24 SN models and one MT classification; blue diamonds). Within each classification, interpolation is conducted as per the class assigned by the classifier. This step is shown as the different colored tabs in the interpolation section of the flow chart. Finally, the output for every interpolated quantity of each classification scheme is combined into one output vector before following the same post-processing steps used in v1 (see Section 7 in \citetalias{2023ApJS..264...45F}). 

\begin{figure*}[t]
    \centering
    \includegraphics[width=\linewidth]{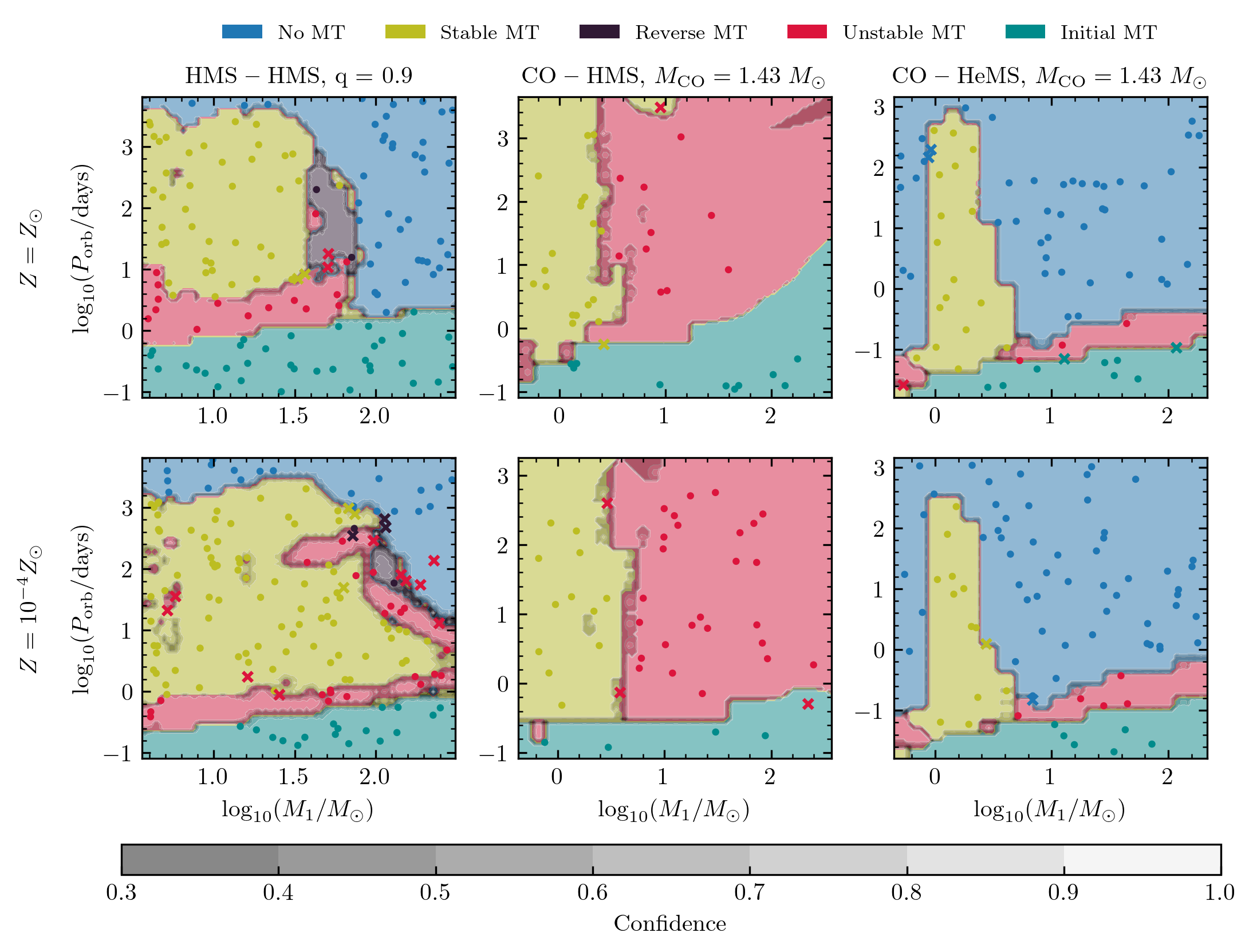}
    \caption{Decision boundaries of our classifier for two sample metallicities (\Zsun, top panels; $10^{-4}\,\Zsun$, bottom panels). Each panel shows the logarithm of the primary's initial mass on the horizontal axis and logarithm of the initial orbital period on the vertical axis. The constant (mass ratio or CO mass) at which the slice is fixed is indicated in the titles of the plots in the top row. Each decision boundary has a different color as indicated by the legend, with grayscale as a transparency layer indicating classification confidence. Reverse MT only exists in the HMS--HMS grid, predominantly when $q$ approaches unity. Colored points indicate the positions and classes of our validation set, where predicted correctly (filled circle) and incorrectly (crosses). Note that, for the CO--HMS grid, although the classification boundaries span the entire plot range, they are only applied to systems that undergo RLOF. }
    \label{fig:decision_boundaries}
\end{figure*}

We have also recently identified a fraction of systems that evolve through ``reverse MT'' \citep[see discussion in Section~\ref{sec:reverse_MT} and][]{2024A&A...683A.144X}. Such systems are distinct in that reverse MT is the only mechanism by which a primary star can increase in mass and spin magnitude through accretion, yielding evolutionary outcomes that are unique compared to other classes. Therefore, we expand the set of MT classes defining a system's evolution to include these reverse cases.

Additionally, we optimized our procedure for pre-processing. In v1, we optimized our normalization of inputs and outputs using Monte Carlo cross-validation on the entire training set for both classification and interpolation schemes. In v2, we maintained our normalization scheme for the classification step, but we implemented class-wise normalization for the interpolation step. 

Finally, when optimizing our hyper-parameter $k$ for the classification step (described in more detail in the following section), we constrain its value to be greater or equal to three to promote smooth class decision boundaries.

\subsection{Classification Results}

\begin{figure*}[t]
    \center
    \includegraphics[width=2\columnwidth,angle=0]{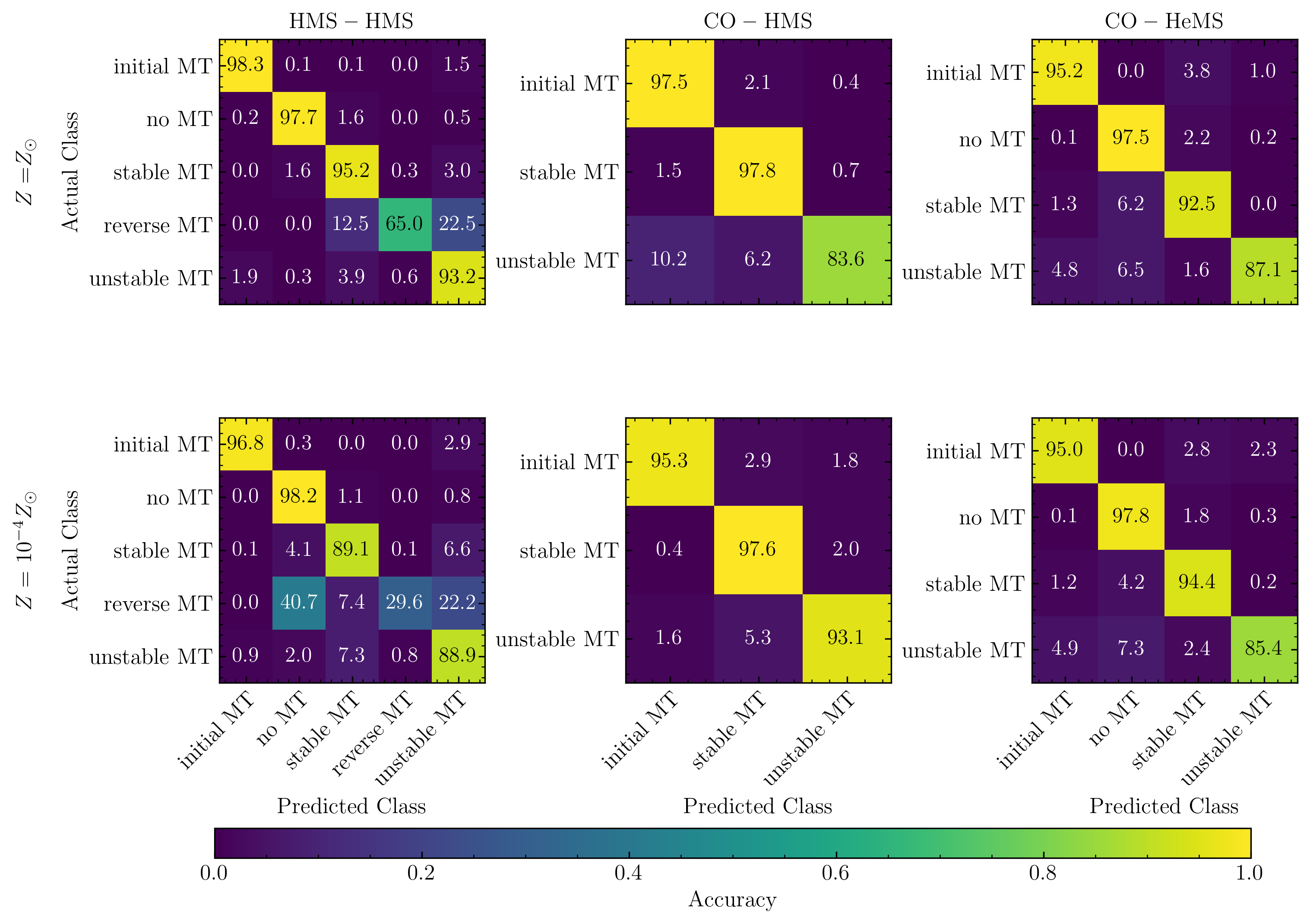}\\
    \caption{Confusion matrices showing the classification error exhibited for each of our three grids as well as two different metallicities selected. The top row of confusion matrices corresponds to $\Zsun$ while the bottom row corresponds to $Z = 10^{-4}\,\Zsun$. The horizontal axis of each matrix corresponds to the predicted class while the vertical axis corresponds to the ground truth class. Each value in cell $C_{ij}$ corresponds to the fraction of samples that belong to class $i$ (horizontal axis) while being classified as class $j$ (vertical axis). The color represents the magnitude of the value $C_{ij}$.}
    \label{fig:confusion_matrices}
\end{figure*}

Given the tendency for MT classes to form clusters within the parameter space (as can be seen in e.g., Figure~\ref{fig:HMS-HMS_MESA_grid_TF12_9panels_temp}), $k$-NN serves as a powerful and interpretable algorithm to solve the classification step of our approach in \posydon. In Figure~\ref{fig:decision_boundaries} we show the clustering properties of our classes for two different metallicities (\Zsun, top panels; $10^{-4}\,\Zsun$, bottom panels). Each panel represents a slice of the initial conditions with fixed $q$ (from our HMS--HMS grid, first column) or $M_\mathrm{CO}$ (from our CO--HMS or CO--HeMS grids, second and third columns, respectively). 

The classification boundaries are generated using our $k$-NN algorithm, with an independently chosen value of $k$ for each grid. Optimization of $k$ is carried out with Monte Carlo cross-validation, which consists of randomly partitioning the training grid into a training and validation set. Next, a different $k$ is used to classify the samples in the validation set. This process is repeated three times for each $k = 1-30$ which allows us to pick the hyperparameter value of $k$ which produces the highest classification accuracy. In Table~\ref{table:hyperparameters} we provide the list of optimal values of $k$ calculated by our $k$-NN classifier for each grid and metallicity combination. Next, we evaluate our classification models using a separate testing set of binary simulations described in Section~\ref{sec:grid_char}.

While the class boundaries result in a few classification errors, Figure~\ref{fig:decision_boundaries} shows that they are generally rather smooth. Furthermore, the top and bottom rows exhibit similar behavior, suggesting similar classification performance. This is indeed borne out by the confusion matrices calculated from our validation set (see Figure~\ref{fig:confusion_matrices}). Each row of the matrix corresponds to the predicted class, while each column corresponds to the ground truth class. Because each row is normalized to sum to unity, each cell $C_{i,j}$ represents the percentage of samples that are classified as class $i$ when their actual class is $j$. Ideally, the diagonal entries of the confusion matrix would be close to one (correct classifications), while the off-diagonal values (misclassification) would be very small, which is the case in Figure~\ref{fig:confusion_matrices}. As in Figure~\ref{fig:decision_boundaries}, the top row of matrices corresponds to the three grids with $\Zsun$ while the bottom row corresponds to the three grids with $10^{-4}\,\Zsun$.

\begin{deluxetable*}{l c c c c c c c c c}
\tabletypesize{\scriptsize}
\tablewidth{0pt}
\setlength{\tabcolsep}{1.2\tabcolsep}  
\tablecaption{Number of Neighbors Used in MT Classification
\label{table:hyperparameters}}
\tablehead{
    \colhead{Grid} &
    \colhead{2\,\Zsun} &
    \colhead{\Zsun} & 
    \colhead{0.45\,\Zsun} & 
    \colhead{0.2\,\Zsun} & 
    \colhead{0.1\,\Zsun} & 
    \colhead{$10^{-2}\,\Zsun$} & 
    \colhead{$10^{-3}\,\Zsun$} & 
    \colhead{$10^{-4}\,\Zsun$}
    }
\startdata
    HMS--HMS & 3 & 17 & 5 & 3 & 3 & 4 & 3 & 3\\
    CO--HMS & 3 & 3 & 5 & 3 & 3 & 3 & 3 & 13\\
    CO--HeMS & 3 & 3 & 13 & 3 & 3 & 3 & 3 & 5\\
\enddata 
\tablecomments{The optimal $k$ used in our $k$-NN classification scheme for our three different grids as well as our eight different metallicities. Optimization is carried out through a Monte Carlo cross-validation scheme.}
\end{deluxetable*}

\begin{figure*}[t]
    \center
    \includegraphics[width=2\columnwidth,angle=0]{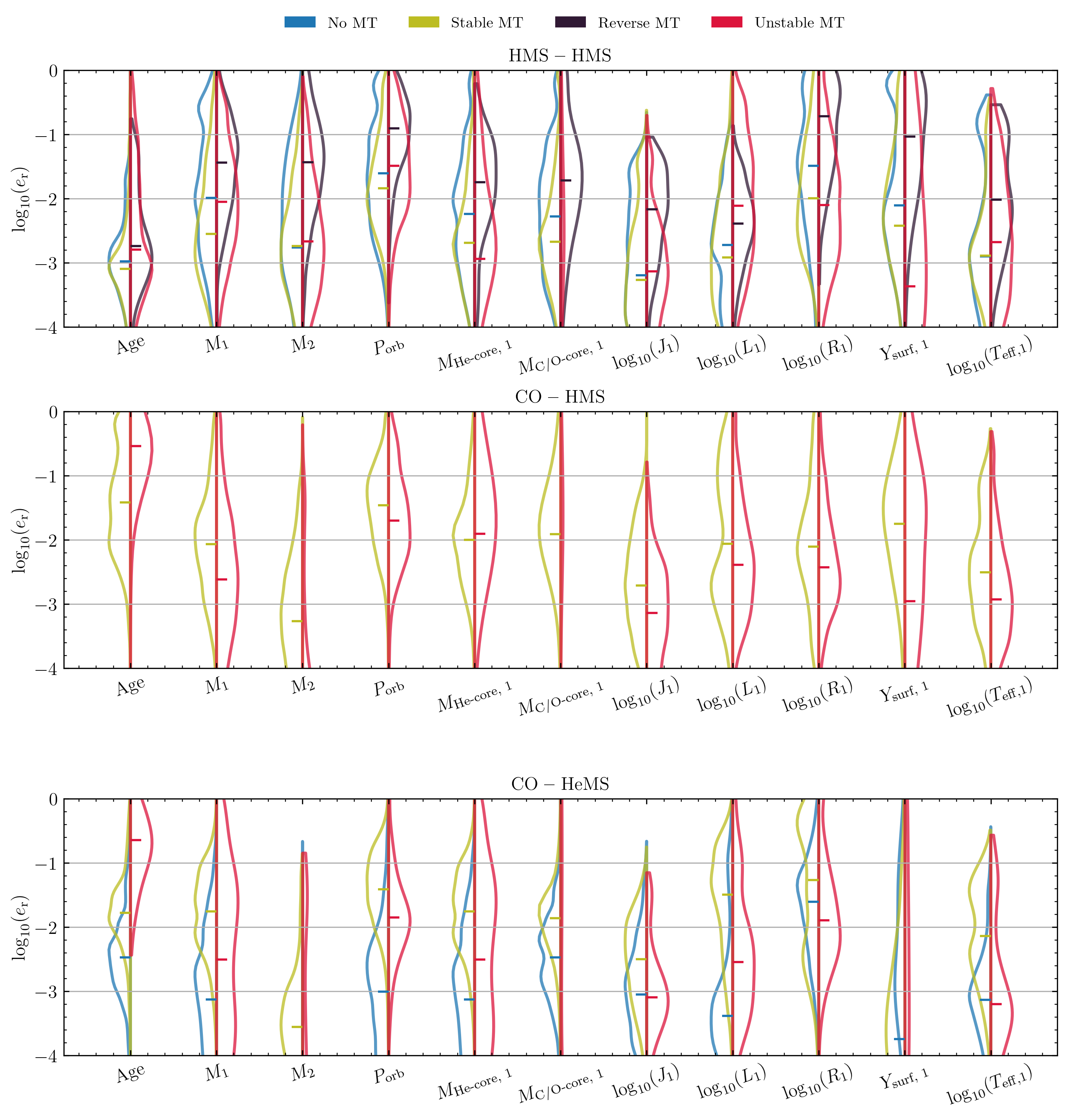}\\
    \caption{Violin plots showing relative interpolation error for our three grids for all metallicities. Each of the different colored curves corresponds to a different interpolation class. The horizontal axis shows a select 11 interpolation fields while the vertical axis shows the relative error in $\log_{10}$ scale. Tick marks indicate median values. The bulk of the distributions for most parameters are almost all within 10\%, typically under 1\%. }
    \label{fig:error_distributions}
\end{figure*}

For the HMS--HMS grid, performance is similar across metallicities and aside from the reverse MT class, accuracy is high ($>$95\%). These accuracies are similar to those characterizing our v1 grids, despite the decreased resolution of our v2 grids. The lower reverse MT accuracy is to be expected given that this class is relatively rare (typically this outcome occurs when $q$ is close to unity). Classification errors can be attributed to regions of high uncertainty near class boundaries, particularly the regions of stable MT surrounded by the unstable MT class which can be seen in both of the leftmost panels in Figure~\ref{fig:decision_boundaries}. The middle column of Figure~\ref{fig:confusion_matrices} shows the confusion matrices for the CO--HMS RLOF grid. The performance is similar aside from the unstable MT class which performs slightly better with $Z = 10^{-4}\,\Zsun$. This matches with what is seen in Figure~\ref{fig:decision_boundaries} where the middle two panels show that $Z = \Zsun$ has a greater mix between classes in the top right than when $Z = 10^{-4}\,\Zsun$. Finally, the CO--HeMS grid performs similarly across the two select metallicities. As discussed in Section~\ref{sec:CO-HeMS}, metallicity has only a minor effect on this grid, so we would not expect significant differences between the two rows.

\subsection{Interpolation Accuracies}

To evaluate a predicted final value $\hat{y}$ using its ground truth value $y$ we define the relative error as,
\begin{equation}
    e_r = \left| \frac{\hat{y} - y}{y} \right|.
\end{equation}
Class-wise distributions of $e_r$ for 11 selected parameters can be seen in Figure~\ref{fig:error_distributions}. These parameters were chosen based on their relative importance across both binary and single star parameters. Each row corresponds to a different set of grids where the interpolation errors from our validation sets in each of our eight metallicities are combined into one set of distributions. There is a great deal of variation in our interpolation accuracies depending on which parameter is under consideration. While a few parameters exhibit relatively large errors, the bulk of the distributions for most parameters are almost all within 10\%, typically under 1\% of relative error. Despite our reduced grid resolution in v2, these errors are similar to what was found in v1, although there are marginal differences depending on the exact combination of parameter, class, and grid.

In some cases, large parameter errors are not meaningful. For the unstable MT class of the CO--HMS grid starting at RLOF, the middle panel of Figure~\ref{fig:error_distributions} shows median errors in excess of 30\%. However, this is an extremely short-lived phase, and it contributes negligibly to the overall duration of the binary's evolution.

Some of the error seen in this figure is due to incorrect classification. For instance in the HMS--HMS grid, the reverse MT class has the worst performance. Misclassified binaries will use the incorrect interpolator, correspondingly leading to relatively large interpolation errors in system parameters. Likewise, the no MT distribution in the CO--HeMS grid tends to perform well as its classification accuracy is above 97\%. Clearly, improvements to our classification algorithm, or using active learning methods as described in \citet{2022ApJ...938...64R}, would improve our interpolation accuracy, as well as focus on which regions and metallicities allow for the most opportunity for optimization.

\section{Postprocessing of {\tt POSYDON} Model Grids \label{sec:postprocessing}}
All single and binary star \mesa{} simulations produce data sets that must be cleaned, collated, and analyzed before being integrated into the \posydon{} framework for simulating stellar populations. With the inclusion of multiple metallicities in v2, we have developed a pipeline to efficiently post-process the increased data volume and complexity of our detailed single- and binary-star model grids. In this section, we briefly summarize the standard post-processing procedure presented in Section 6 of \citetalias{2023ApJS..264...45F} and highlight relevant additions and changes made for v2. 

\begin{figure*}[t]
    \center
    \includegraphics[width=\textwidth,angle=0]{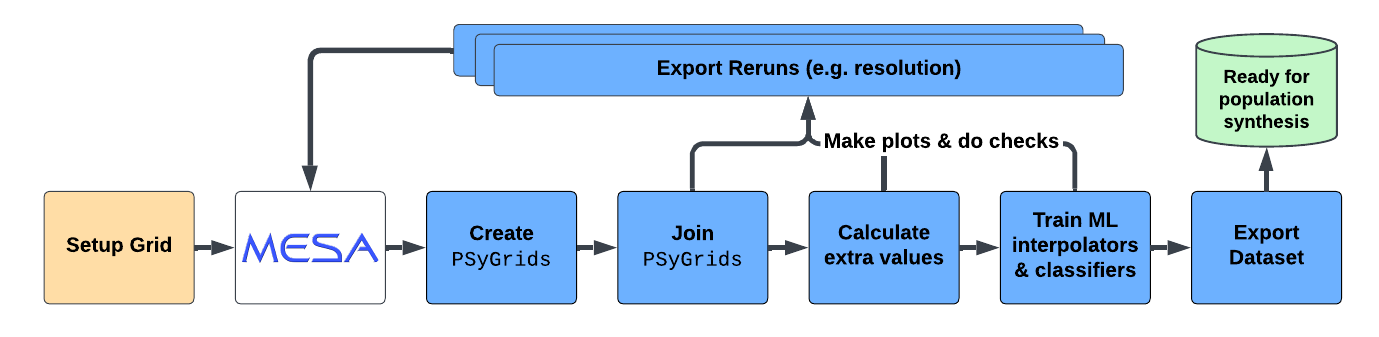}\\
    \caption{Flowchart of the post-processing pipeline (detailed in Section~\ref{sec:postprocessing}) which translates raw \mesa output (white) into data products (green) used by \posydon for binary population synthesis.
    Starting on the left, we first define the initial parameter space (orange) to run \mesa, then perform the \mesa{} simulations, followed by the main post-processing loop in blue, resulting in our final data product in green.
    During the collation of \mesa grids, calculating extra quantities from the detailed final profiles, and training our machine learning models (Section~\ref{sec:machine_learning}), we perform intermediate checks (e.g. generating plots, calculating non-convergence rates, classification and interpolation performance) to verify the quality of our \mesa models and data processing steps.
    Additionally, if we identify regions experiencing convergence issues in \mesa, we perform reruns (Section~\ref{sec:postprocessing:rerun}) to increase the convergence rate of our simulations (e.g. by adjusting the spatial or temporal resolution).
    Both the binary and single-star \mesa models broadly follow this control sequence, with each dataset having a few unique steps (e.g. identifying EEPs for single stars).
    }
    \label{fig:postprocessing_pipeline_flowchart}
\end{figure*}

\begin{figure*}[t]\center
    \includegraphics[width=\textwidth,angle=0]{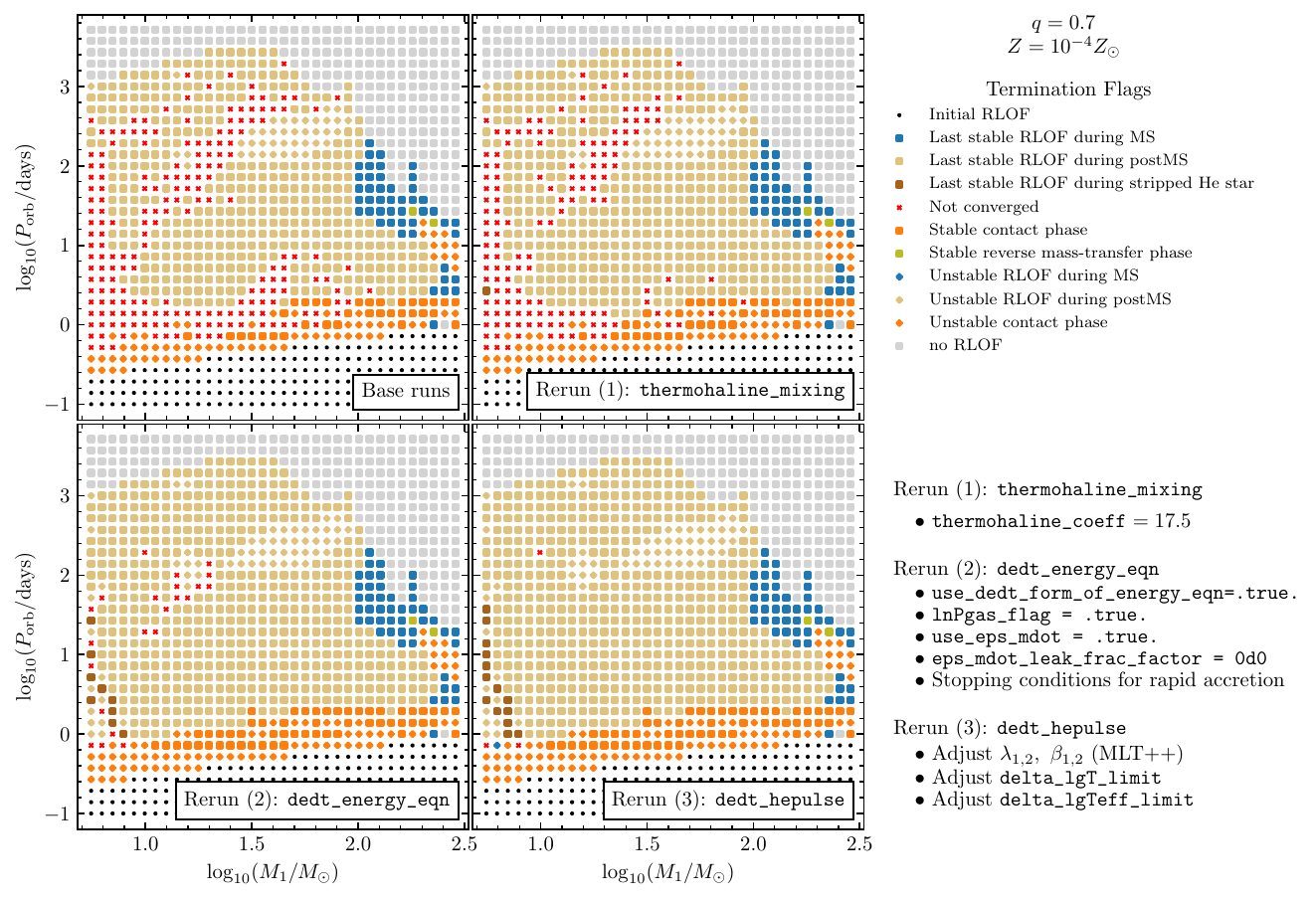}\\
    \caption{An example grid slice showing the progressive improvement of simulation success rate from our subsequent reruns (described in the text) addressing specific physical problems that the grids originally encountered at the base level. This example uses models from our HMS--HMS grid at $Z=10^{-4}\,Z_{\odot}$ and $q = 0.7$, with red crosses marking simulations that failed to converge. After the three re-runs, our non-convergence rates have dropped to $\lesssim3$\% for all of our grid slices (see Table~\ref{tab:failure}). Note that in generating this figure, we have reduced the grid resolution for $M_1 < 14 \Msun$ for visual clarity. The same series of re-runs was performed for all non-converging models in each of our binary grids. }
    \label{fig:rerun_comparison}
\end{figure*}

Our post-processing steps are as follows:
\begin{enumerate}
    \item Compute single- and binary-star models (see Sections~3-5 in \citetalias{2023ApJS..264...45F} and changes in v2 described in Section~\ref{sec:grids}).
    \item For H-rich and He-rich single-star grids, we resample the time evolution output using equivalent evolutionary points (EEPs; see Section 6.3 in \citetalias{2023ApJS..264...45F}; \citealt{2016ApJS..222....8D}). 
    \item Combine our runs into an hdf5-based file format. During this process, we classify each individual model based on the different resulting stellar and binary types as described in Section~7 in \citetalias{2023ApJS..264...45F} and changes outlined in Section~\ref{sec:machine_learning}.
    \item Downsample the time resolution of our models' output in time to reduce data size while retaining salient information \citepalias[see Section~6.4 in][]{2023ApJS..264...45F}.
    \item Join individual grid objects into a combined grid object that manages each of the different layers, in the process replacing non-converged models from earlier reruns (see Section~\ref{sec:postprocessing:rerun} below).
    \item Add additional post-processed quantities to the grids (e.g., for SNe or CE during population synthesis; see Section~\ref{sec:postprocessing:extra_quantities}).
    \item Train classifiers and interpolators on the stellar and binary parameters in each grid (see Section~\ref{sec:machine_learning}).
    \item Identify non-converged models (non-converged \mesa runs), and stage these models for a rerun (see Section~\ref{sec:postprocessing:rerun}). Our post-processing procedure is repeated until our final rerun has been completed.
    \item Collect the final combined grid and interpolator objects into the structure expected by \posydon{}.
\end{enumerate}

The post-processing pipeline is designed to be modular, with options allowing users to customize post-processing steps (e.g., starting a grid at ZAMS or RLOF), and adding or removing entire steps when necessary (e.g., different subsets of reruns). The pipeline reads a parameter file to determine the user's desired post-processing steps and can be parallelized to run on high-performance computing resources using the \texttt{slurm} job scheduler. Upon submission our software pipeline executes the aforementioned post-processing steps and produces an array of visualizations along with a status report for inspection at each step of the pipeline.  Furthermore, it checks the rate of non-converged models and automatically integrates the reruns in the final output model grid. Figure~\ref{fig:postprocessing_pipeline_flowchart} shows the primary steps of our pipeline and the control sequences relating them, while intermediate steps (e.g., compressing files) are omitted for clarity.

\subsection{Rerunning non-converged models}
\label{sec:postprocessing:rerun}

After our initial set of single- and binary-star model grids has been computed, we perform a series of reruns on \mesa models that either failed to converge or did not complete within our imposed 48-hour wall-time limit for each individual binary simulation. Non-converging models are due to numerical failures in MESA (i.e., a solution can not be found due to difficult-to-model physical processes, leading to extraordinarily small timesteps). As an example, the non-converging island in Figure~\ref{fig:CO-HeMS_BH} is due to relatively low-mass He stars losing their outer envelopes during mass transfer. These stars have several factors conspiring to cause non-convergence: their distended envelopes experience superadiabatic conditions, their mass transfer tend to reach an oscillatory state where they fluctuate between mass transferring and not, and they are also experiencing crystallization in their cores. By altering the physics assumptions (in particular dealing with energy transport) in our simulations, we aim to fill-in the gaps left by non-converging models. 

We display the evolutionary summary for our base runs alongside subsequent reruns in Figure \ref{fig:rerun_comparison}. Panels in this Figure show a selection of our HMS--HMS grid models at $10^{-4}\,Z_{\odot}$ and $q = 0.7$ for demonstration. While some non-converging models (red crosses) appear randomly distributed across parameter space in our base runs, suggesting a cause of numerical nature, in many cases they form ``islands" in the input parameter space (most visible are the islands formed in the top center and bottom left portions of the top panels of Figure~\ref{fig:rerun_comparison} or alternatively the lower left region of the panels in Figure~\ref{fig:CO-HeMS_BH}). Our interpolators can span across isolated non-converged models, but clustered groups of non-converged models pose a major challenge for interpolators. Following the approach we took in v1, having identified those runs that did not converge, we rerun them with slightly altered physics or numerical settings to facilitate their successful completion. One may see in Figure \ref{fig:rerun_comparison} that the ``Not converged'' models are mostly eliminated by our subsequent reruns. Our working assumption is that any inaccuracies accrued through inconsistent physics across a grid will be small compared with the large gaps in our grids that our classification and interpolation methods would have to manage. Indeed, we test the revised simulation parameters for each of our reruns on a handful of successful models to ensure consistency of the results. In comparing Hertzsprung-Russell diagram morphologies and key stellar evolution parameters (such as mass and radius) across comparable non-converged and neighboring runs, rerun simulations lead to similar evolutionary behavior.

In v1, we reduced our non-convergence rate by including a rerun that limited the maximum radiative opacity to mitigate the impact of resolving stellar pulsations caused by the $\kappa$-mechanism. We found this had several unintentional effects on low-mass stars which altered not only the evolution of a star, but also the shape of the main sequence. In v2, we have improved our rerunning steps, performing three sequential reruns: 
\begin{enumerate}
\item {\tt thermohaline\_mixing} rerun: we adjust the thermohaline mixing parameter $\alpha_\mathrm{th}$ from 0 (our default has no such mixing) to $\alpha_\mathrm{th}=17.5$ \citepalias[see Section 3.2.3 of][]{2023ApJS..264...45F} which helps the convergence of some models that undergo accretion.
\item {\tt dedt\_energy\_eqn} rerun: we switch to the \texttt{dedt}-form of the energy equation in \mesa's hydro solver \citep{2019ApJS..243...10P} which helps models during rapid (superthermal) MT. We additionally modify some mesh parameters responsible for determining the homologous region of the model during mass accretion \citep{2015ApJS..220...15P}. This rerun also introduces two new stopping conditions to handle situations where \mesa typically experiences numerical convergence issues during rapid MT. These conditions are described in more detail below in discussion of the \texttt{dedt\_energy\_eqn} rerun.
\item {\tt dedt\_hepulse} rerun: we slightly modify the default MLT++ \citep{2013ApJS..208....4P} parameters and timestep controls to handle superadiabatic conditions in models that are stripped of their H envelopes and continue their evolution as stripped He star models. 
\end{enumerate}

Among these three types of reruns, the largest improvement in terms of \mesa model convergence comes from the adoption of the \texttt{dedt}-form of the energy equation in our \texttt{dedt\_energy\_eqn} rerun. As described in \cite{2019ApJS..243...10P}, this is a physically equivalent reformulation of the energy equation solved in \mesa that is advantageous to the numerical conservation of energy. This form of the energy equation holds a further advantage, owing to its formalism, in that it is not reliant on a ``non-homologous'' term; a problem described by \citealt{sugimoto.nomoto.eriguchi:1981} (and see the related discussion in \citealt{2015ApJS..220...15P}) where models would be forced to take small time steps in order to solve the energy equation within tolerance during rapid accretion. In more recent versions of \mesa (as in \citealt{2023ApJS..265...15J}), the \texttt{dedt}-form of the energy equation is the default choice. Alongside this, using the \texttt{dedt}-form of the energy equation also allows for an accounting of the ingestion of accretion energy during superthermal accretion (i.e., accretion at rates faster than the accretor's thermal timescale). Many of these non-converged models accrete at such rapid rates where assuming the accretion energy is negligible becomes inappropriate.

Accounting for accretion energy is a capability introduced in \citet{2019ApJS..243...10P} (\mesa r11701) that can lead to heating and inflation of the accretor's envelope. Previous versions of \mesa assumed accretion rates were slower than the thermal timescale, and that the energy was efficiently radiated away \citep{2015ApJS..220...15P}. We find that the inclusion of this energy (via setting \texttt{use\_eps\_mdot = .true.}) is critical to solving the stellar structure of stars experiencing accretion at such high rates, which typically occurs towards the lower mass end of our parameter space. In the methodology described by \cite{2019ApJS..243...10P}, as mass is accreted into cells of the modeled structure, the rate at which accreted energy is leaked into those cells may be set by the user (done via \texttt{eps\_mdot\_leak\_frac\_factor}). We set \texttt{eps\_mdot\_leak\_frac\_factor = 0} for numerical stability, so that energy is not leaked until the final cell in which material is accreted. This essentially sets the accreted energy to be leaked adiabatically and fully ingested by the accretor. Furthermore, while \mesa normally uses the density $\rho$ as a basic variable in solving the energy equation, the gas pressure ($P_{\rm gas}$) may be used instead (as mentioned in \citealt{2018ApJS..234...34P}) by setting \texttt{lnPgas\_flag = .true.}. We find that this leads to better numerical stability for these models and therefore utilize $P_{\rm gas}$ rather than $\rho$ in this and the subsequent \texttt{dedt\_hepulse} rerun.

The \texttt{dedt\_energy\_eqn} rerun also introduces two new stopping conditions, as there are certain cases where superthermal accretion leads inevitably to numerical convergence issues in our models. We find that there are instances where binary components come into contact while undergoing superthermal accretion (i.e., both the donor and accretor fill their Roche lobe). We assume that this will result in an unstable MT episode.

Similarly, as an accretor accumulates angular momentum during MT, they may spin up to critical rates (i.e., where centrifugal forces overwhelm their own gravity). We introduce a second stopping condition in cases where accretors reach critical rotation rates while undergoing superthermal accretion. Critical rotation is expected to create a decretion disk of outflowing material. Precisely how this should manifest while such a star is simultaneously undergoing accretion is uncertain (let alone at accretion rates faster than the thermal relaxation timescale). In our \mesa simulations, critical rotation in a stellar models induces mass loss such that the model's rotation rate is made sub-critical \citep{2013ApJS..208....4P}. As the accretor is simultaneously gaining angular momentum and mass at high rates (typically exceeding $10^{-4}\,\Msun\,\mathrm{yr}^{-1}$) while also losing material to remain at sub-critical rotation rates, the models can struggle to find a solution in such scenarios.

Analytical calculations by \cite{lu.etal:2023} (although primarily focused on binaries with CO accretors) examined cases where rapid accretion inflows interacting with an accretion disk could heat the material such that it causes significant outflow from the binary's L2 point. Under such circumstances, the material could form a circumbinary outflow that can facilitate inspiral. Based on this, we treat cases where our models are rotating at critical rates while undergoing superthermal accretion as unstable MT cases. Our assumption being that material accreted at such high rates onto a critically rotating star will lead to instances where the surrounding disk of accreted/decreted material is lost through the L2 point in a similar fashion, leading to L2 overflow. In nature, critically rotating Algols (e.g., \citealt{deschamps.etal:2013}) may be examples of such a physical scenario. In our modeling, we typically see the accretors also undergoing expansion during superthermal MT (as also described in recent modeling by \citealt{lau.etal:2024}). This expansion, lowering the accretor's surface gravity, is another factor conducive to super-critical rotation that can contribute to numerical convergence issues. The final fate of these systems is uncertain, but we assume these systems will enter a CE phase. A more detailed treatment of this scenario will be a goal of future work.

While the adjustments to the energy equation and other changes described above help many models (especially, where the accretor's thermal timescale is longer than the MT timescale), a class of non-converged models remains. These tend to represent systems wherein the donor becomes a stripped star (losing most of its H-rich envelope). Some of these models experience prohibitively small time steps during the stripping phase. We found that this is due to our default requirement that temperature (both internal, cell-wise $T$ and surface $T_{\rm eff}$) cannot change more than 1 degree Kelvin between two consecutive timesteps. We relax this limit during MT to allow changes of up to roughly 3 and 10 degrees Kelvin for the internal cell-wise $T$ and $T_{\rm eff}$, respectively. This relaxation allows roughly half of these models to proceed through their evolution to become stripped helium stars (the other half already successfully evolves). 

Many of the models go on to experience additional numerical issues as they evolve off the helium MS and expand. These models have very diffuse envelopes that typically trigger a second MT episode, wherein they typically run into numerical issues once again. The outer parts of these diffuse envelopes are typically superadiabatic as well, and we find that adjusting the thresholds at which MLT++ (see details in \citealt{2013ApJS..208....4P} and an overview in e.g., Appendix A of \citealt{agrawal:2022}) reaches full efficiency helps the majority of these models succeed. Specifically, we adjust $\lambda_1 = 0.05$, $\lambda_2 = 0.01$, $\beta_1 = 0.05$, $\beta_2 = 0.01$ (from their defaults of $\lambda_1 = 1$, $\lambda_2 = 0.5$, $\beta_1 = 0.35$, $\beta_2 = 0.25$) after stars lost their H envelope. Briefly, $\lambda_{1,2}$ ($\beta_{1,2}$) describe threshold values of the ratio of radiative to Eddington luminosity (gas pressure to pressure) within a model cell, above which, MLT++ reduces superadiabatic temperature gradients. Above $\lambda_1$ and $\beta_1$, the superadiabaticity within a cell is fully suppressed; while between $\lambda_1$ and $\lambda_2$ ($\beta_1$ and $\beta_2$), the superadiabaticity is suppressed to a lesser extent, reducing to no suppression below values of $\lambda_2$ and $\beta_2$ within the model. With these changes, most of the remaining non-converged models are able to converge and successfully complete their second MT episode.

\begin{table}
    \centering
    \caption{Non-convergence rates of our binary grids after our reruns. Without the reruns, the non-convergence rates would be an order of magnitude larger on average.}
    \label{tab:failure}
    \begin{tabular}{c c c c}
         Metallicity, $Z$ & HMS--HMS & CO--HMS & CO--HeMS \\
         \hline
         $2\,\Zsun$ & $3.2\%$ & $1.1\%$ & $2.7\%$ \\
         $1\,\Zsun$ & $1.9\%$ & $0.0\%$ & $2.5\%$ \\
         $0.45\,\Zsun$ & $1.8\%$ & $0.1\%$ & $2.2\%$ \\
         $0.2\,\Zsun$ & $2.2\%$ & $0.2\%$ & $2.2\%$ \\
         $0.1\,\Zsun$ & $2.5\%$ & $0.3\%$ & $2.5\%$ \\
         $10^{-2}\,\Zsun$ & $2.1\%$ & $0.1\%$ & $2.2\%$ \\
         $10^{-3}\,\Zsun$ & $1.8\%$ & $0.2\%$ & $2.1\%$ \\
         $10^{-4}\,\Zsun$ & $1.3\%$ & $0.5\%$ & $2.2\%$ \\
    \end{tabular}
\end{table}

We apply these reruns in sequence, where only the subset of non-converged models in the previous run for each grid are rerun with the altered option. As seen in Figure~\ref{fig:rerun_comparison} our series of reruns include progressively fewer models as our set of non-converged models reduces. Importantly, the remaining non-converged models tend to be isolated rather than forming islands, allowing our machine learning methods to classify and interpolate over them. After our final rerun, we find a non-convergence rate of up to a few percent in each of our grids (see Table~\ref{tab:failure} for our binary grid failure rates).

\subsection{Calculating post-processed quantities}
\label{sec:postprocessing:extra_quantities}

From our \mesa{} simulations we aim to provide users with any data that might be required to both interpret their evolution as well as determine their next evolutionary steps within the \posydon{} population synthesis framework. These must be provided while considering memory constraints so as to still allow population synthesis with reasonable computing resources. To start, our processed grids include the single-star and binary-star evolution over time as well as the final structural profiles of the primary and (in the case of the HMS--HMS grid) the secondary stars. Additionally, we define four termination flags broadly describing the physical outcomes of our binary \mesa{} simulations following the procedure described in \citetalias{2023ApJS..264...45F}: 1) the end condition of the \mesa{} simulation; 2) the system's mass-transfer history; 3) the final state of the primary star; 4) the final state of the secondary star.

For the first flag, we define a variety of ending conditions for our \mesa{} simulations corresponding to: an imminent SN (central carbon depletion, or onset of (P)PISN; Section~\ref{sec:changes_vI}), WD formation (\texttt{gamma\_center\_limit = 10}), unstable RLOF (L2 overflow, exceeding the maximum $\dot{M}$, photon trapping), or a model which has RLOF at ZAMS (initial MT). Our designation of MT history comprising the second flag follows the approach in v1, while we now record a history of all MT cases with the canonical definitions of Case A, Case B, and Case C MT \citep[for a pedagogical description, see][]{1991ApJS...76...55I} into a single label. In v2, we modify our labeling convention such that a binary going through Case A followed by Case B MT has the label \texttt{case\_A1/B1} where the number indicates which star was overfilling its Roche lobe. Termination flags 1 and 2 are combined to infer the interpolation class (stable, unstable, initial, no, and reverse MT; see \citetalias{2023ApJS..264...45F} for more details on the first four).

The third and fourth flags designating the state of the primary and secondary stars follow our naming convention described in \citetalias[][]{2023ApJS..264...45F}, where we indicate the predominant nucleosynthetic process occurring as well as the envelope composition (e.g., BH, H-rich-core-C-depletion, stripped-He-core-He-burning). In v2, we added a new characterization for the envelope, accreted He. In addition to the standard values calculated in v1 \citepalias[see Table~4 in][]{2023ApJS..264...45F}, we calculate additional quantities including a CO interpolation class (BH, NS, WD, BH\_reverse\_MT). These quantities are directly related to the new classification and interpolation schemes described in Section \ref{sec:machine_learning}.

\subsection{Other post-processing pipeline capabilities}
\label{sec:postprocessing:other_capabilities}

The new version of the post-processing script enables users to streamline the procedure, efficiently make checks and visualizations, and enable/disable optional settings and extended functionalities. Such features are: 1) producing multipage PDFs with figures, 2) performing sanity checks based on the properties of the systems, and 3) classification and interpolation model training and accuracy metrics (classification, initial-final interpolation, profile interpolation; see Section \ref{sec:machine_learning}). For additional information we direct the reader to the code documentation at \href{https://posydon.org}{posydon.org}.

\section{Other Changes from {\tt POSYDON v1} \label{sec:improvements}}
Separate from the changes to the stellar and binary evolution physics adopted in our single- and binary-star \mesa model grids, we implement a series of improvements to the physical prescriptions throughout \posydon.

\subsection{Matching binary evolution products to single-star models}\label{sec:matching}

In the \posydon approach to binary population synthesis, binaries traverse through multiple grids during their complete evolution from ZAMS to both stars undergoing core collapse. Some errors are accrued during the transition between grids. As an example, when the primary star in the HMS--HMS grid undergoes core collapse, for further evolution of the partially evolved, non-degenerate secondary in the binary we match to the nearest star from our single star grid (see discussion in Section 8.1 of \citetalias{2023ApJS..264...45F}). Inevitably, the closest single star does not have identical structural properties to the secondary star resulting from the HMS--HMS grid. While this matching was required in v1, we have since improved our procedure for finding the nearest matching single star model.

As in v1, we aim to find the best-fit single-star model by defining a square Euclidean distance which is the weighted quadrature sum of the differences between the input stellar model and a single star model across a few specific stellar parameters, where each parameter has its own weight to account for their differences in dynamical range.  Depending on the star's evolutionary state, we define this distance using different parameters: i) For MS stars, in v2 we choose to match the total mass of the star, its central H abundance $X_{\mathrm{center}}$, the radius of the MS star, and its helium core mass (which has not been formed for the longest part of the MS); ii) For post-MS stars, we replace the central H abundance with the central He abundance ($Y_{\mathrm{center}}$) in the list above; iii) For evolved stripped stars, we use the He-core mass of the star (equal to its total mass), the radius, and its central He abundance. We keep the weight factors (used to enforce similar weighting during the minimization process) the same as in v1. However, in v2, we allow users to modify the list of parameters quantified in the matching process for each case, as well as the weight factors. 

When identifying the nearest single star model, we search  in both mass and age, calculated using the EEP approach described in Section 6.3 of \citetalias{2023ApJS..264...45F}. Efficiently finding the closest matching star in this two-dimensional space is a non-trivial process. We utilize the truncated Newton (\citealt{dembo.steihaug:1983}) minimization method in the Python library \texttt{scipy} \citep{2020SciPy-NMeth} and consider a match successful if the identified star is consistent to within 1\% for the computed square distance using each of the matched parameters listed previously. If the identified closest matching star does not reach 1\% consistency, we define a series of triage steps to keep searching for a sufficiently similar single star model: First, we adopt an alternative minimization method, a modified version of Powell's method \citep{powell:1964}. If we still cannot find a model matching to within 1\%, we remove the stellar radius as a matching parameter and try again with the truncated Newton method. If all else fails, as a final effort we try to match the (MS or post-MS) binary component(s) to the He single star grid of models, rather than the H single-star grid. For He stars the above steps are identical except the matching is initially attempted with our He single-star grid, and as a final attempt, we match to the H single-star grid. For this final attempt, we aim to find a star with matching parameters within 10\%, otherwise we consider the match to have failed and the evolution of the binary is stopped.

\subsection{Isolated star evolution without a binary companion} \label{sec:single_stellar_pop}

While v1 focused exclusively on stellar binaries, in v2 we include this capability, accounting for: 1) stars that are born single, 2) binaries that disrupt, forming two separate, single stars, and 3) the merger of a binary into a more massive single star. In \posydon, each of these cases are handled by the {\tt detached\_step} \citepalias[see Section 8.1 in][]{2023ApJS..264...45F}, which maps each of these single star cases onto our pre-calculated single star grids. We individually discuss our treatment for each of the three cases in the following sections below.

\begin{figure*}[t]
    \center
    \includegraphics[scale=0.499,angle=0]{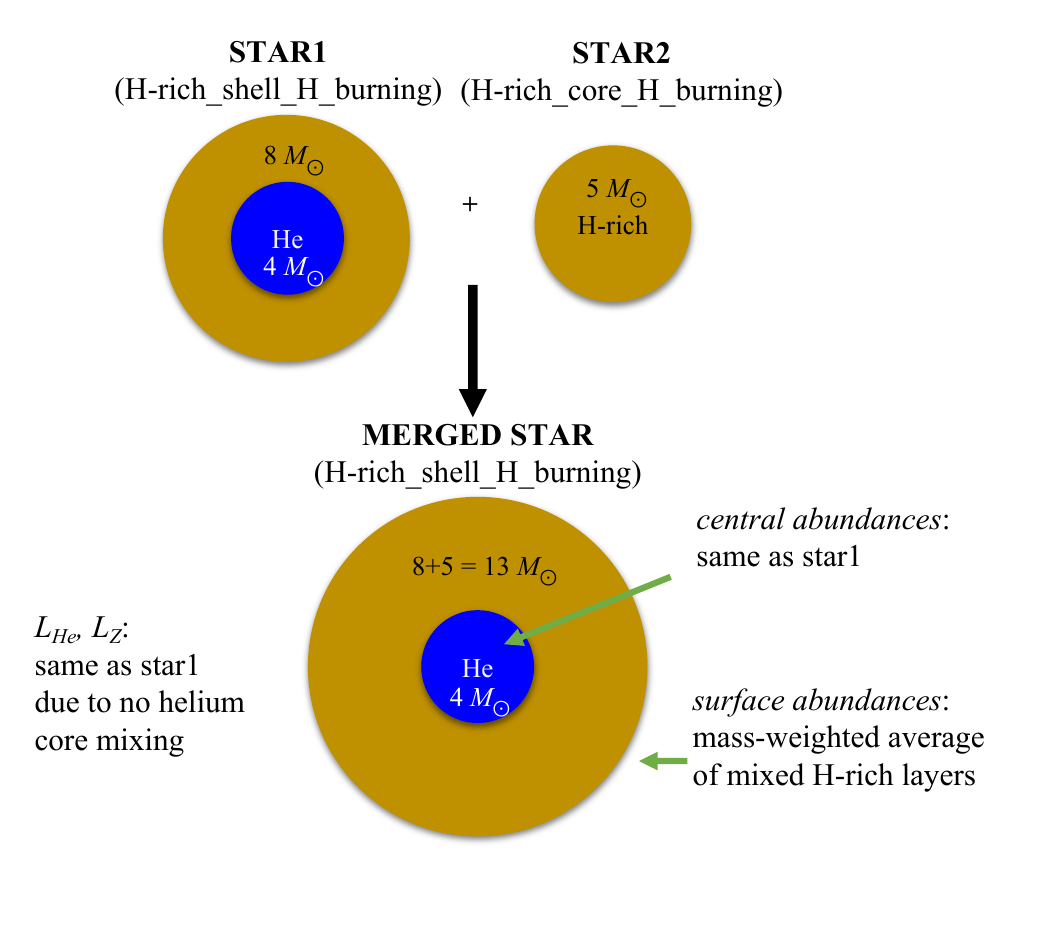}\hfill \includegraphics[scale=0.499,angle=0]{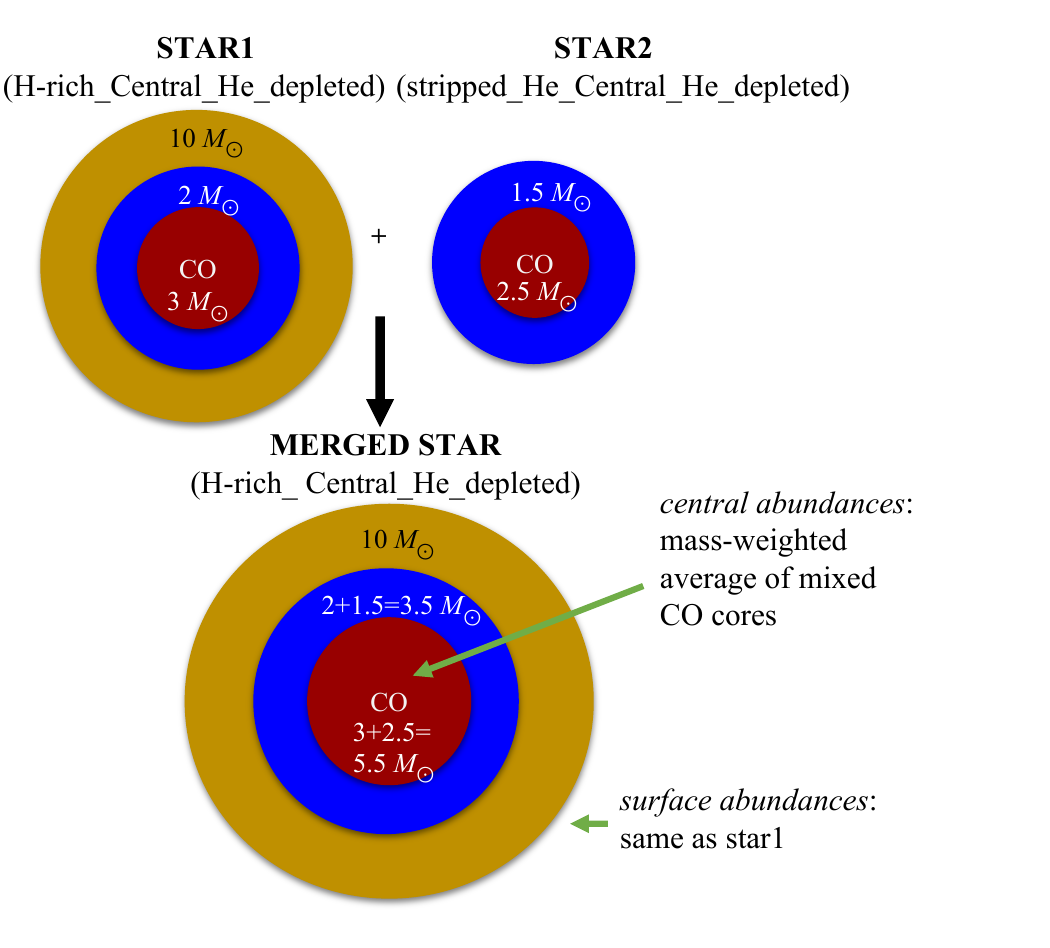}
    \caption{Schematic diagram illustrating our process for estimating the properties of the merger product for two indicative cases. The left is an example of the mixing of the surface H-rich layers, leaving the core properties constant. The right is an example of an assumed core mixing. 
    }
    \label{fig:merger_schematic}
\end{figure*}

\subsubsection{Initially single stars}
\label{sec:initially_single}

The default \posydon setting for population generation has a constant binary fraction of unity, and thus only accounts for initial stellar binaries. To model mixed populations of binaries with initially single stars, we have incorporated two different schemes for the binary fraction of our populations. We allow for a user-specified constant binary fraction as well as a mass-dependent binary fraction following \citet{2017ApJS..230...15M}. 

Our single stars follow the same initial mass function (with the same mass ranges) as is set for the masses of the primary stars in our binaries. While our binary star grids extend to primary masses down to $5.5\,\Msun$, the single star grids span a much wider range. Therefore, for use cases requiring populations with only initially single stars (a binary fraction of zero), the lower mass limit can be extended down to $0.1\,\Msun$.

After being generated, an initially single star follows the evolutionary tracks of our single star grids using EEPs. To maintain consistency within our generated populations, single stars are modeled as instances of binary objects, but have a massless companion, with all of its stellar properties equal to \texttt{None}. The stars are then evolved until the maximum simulation time.

\subsubsection{Disrupted binaries} \label{sec:disrupted}

In v1 we did not account for the evolution of a binary system that disrupted due to its SN kick; if a binary ever became unbound, we simply stopped following its evolution. In v2 we now track the evolution of the ejected companion through a new isolated step. The isolated step directs the binary into the detached step assuming an extremely large orbit and turning all options of orbital evolution off. These stars are effectively matched to the nearest single-star tracks following the approach described in Section~\ref{sec:matching}) and evolved until completion. 

\begin{table*}[]
    \centering
    \begin{tabular}{c c c c c}
          &  & & & Conserve \\
         Name & Mechanism & Engine & PPI extra mass loss & H-rich envelope \\
         \hline
         \texttt{SN\_MODEL\_v2\_01} & \citet{2012ApJ...749...91F} delayed & -- & -20 \Msun & No \\
         \texttt{SN\_MODEL\_v2\_02} & \citet{2012ApJ...749...91F} delayed & -- & -20 \Msun & Yes \\
         \texttt{SN\_MODEL\_v2\_03} & \citet{2012ApJ...749...91F} delayed & -- & 0 & No \\
         \texttt{SN\_MODEL\_v2\_04} & \citet{2012ApJ...749...91F} delayed & -- & 0 & Yes \\
         \texttt{SN\_MODEL\_v2\_05} & \citet{2012ApJ...749...91F} rapid & -- & -20 \Msun & No \\
         \texttt{SN\_MODEL\_v2\_06} & \citet{2012ApJ...749...91F} rapid & -- & -20 \Msun & Yes \\
         \texttt{SN\_MODEL\_v2\_07} & \citet{2012ApJ...749...91F} rapid & -- & 0 & No \\
         \texttt{SN\_MODEL\_v2\_08} & \citet{2012ApJ...749...91F} rapid & -- & 0 & Yes \\
         \texttt{SN\_MODEL\_v2\_09} & \citet{2016ApJ...821...38S} & N20 & -20 \Msun & No \\
         \texttt{SN\_MODEL\_v2\_10} & \citet{2016ApJ...821...38S} & N20 & -20 \Msun & Yes \\
         \texttt{SN\_MODEL\_v2\_11} & \citet{2016ApJ...821...38S} & N20 & 0 & No \\
         \texttt{SN\_MODEL\_v2\_12} & \citet{2016ApJ...821...38S} & N20 & 0 & Yes \\
         \texttt{SN\_MODEL\_v2\_13} & \citet{2020MNRAS.499.2803P} & N20 & -20 \Msun & No \\
         \texttt{SN\_MODEL\_v2\_14} & \citet{2020MNRAS.499.2803P} & N20 & -20 \Msun & Yes \\
         \texttt{SN\_MODEL\_v2\_15} & \citet{2020MNRAS.499.2803P} & N20 & 0 & No \\
         \texttt{SN\_MODEL\_v2\_16} & \citet{2020MNRAS.499.2803P} & N20 & 0 & Yes \\
         \texttt{SN\_MODEL\_v2\_17} & \citet{2016ApJ...821...38S} & W20 & -20 \Msun & No \\
         \texttt{SN\_MODEL\_v2\_18} & \citet{2016ApJ...821...38S} & W20 & -20 \Msun & Yes \\
         \texttt{SN\_MODEL\_v2\_19} & \citet{2016ApJ...821...38S} & W20 & 0 & No \\
         \texttt{SN\_MODEL\_v2\_20} & \citet{2016ApJ...821...38S} & W20 & 0 & Yes \\
         \texttt{SN\_MODEL\_v2\_21} & \citet{2020MNRAS.499.2803P} & W20 & -20 \Msun & No \\
         \texttt{SN\_MODEL\_v2\_22} & \citet{2020MNRAS.499.2803P} & W20 & -20 \Msun & Yes \\
         \texttt{SN\_MODEL\_v2\_23} & \citet{2020MNRAS.499.2803P} & W20 & 0 & No \\
         \texttt{SN\_MODEL\_v2\_24} & \citet{2020MNRAS.499.2803P} & W20 & 0 & Yes \\
    \end{tabular}
    \caption{Core-collapse model assumptions supported in v2. In addition to these different assumptions, similarly to the fiducial model assumption in v1 \citepalias{2023ApJS..264...45F}, each model further assumes ECSN according to \citet{2015MNRAS.451.2123T}, PPISN/PISN according to \citet{2023MNRAS.526.4130H}, maximum neutrino mass loss of $0.5\,\Msun$, and a maximum NS mass of $2.5\,\Msun$. 
    }
    \label{tab:models}
\end{table*}

\subsubsection{Stellar Mergers}
\label{sec:merged}

While focused attempts using hydrodynamical simulations to study stellar mergers provide some insight into mass loss during merger, final spin period, and the merger product's structure \citep{2002ApJ...568..939L, 2009A&A...497..255G, Schneider+2019, Chatzopoulos+2020}, these are difficult to account for in a general way so as to handle the broad array of stellar mergers produced by \posydon populations. Therefore, in v2 we include only a basic treatment of binary merger products by matching it to our single star grid (in v1, binaries that resulted in a merger were stopped). These mergers typically occur in \posydon when a binary entering a CE phase lacks sufficient orbital energy to unbind the donor's envelope \citepalias[see][for details]{2023ApJS..264...45F}. As in v1, HMS or HeMS donors are assumed to automatically merge with their companions upon the onset of unstable mass transfer during RLOF. 

We follow \citet{2002MNRAS.329..897H} to estimate the total mass of the merger, as well as its core masses, by summing corresponding layers of the merging components (e.g., the helium core mass of the merger will be the sum of the two helium core masses). Layers dominated by the same chemical abundance are assumed to be fully mixed, with their abundances weighted by the mass of each star's corresponding layer. While we calculate all the abundances within each layer of the merger product's structure, only the surface and central abundances are considered for the subsequent matching step. For instance, a merger involving two stars with H-rich envelopes (e.g., a MS star with another MS star or with a H-rich giant) will result in an envelope with a combined mass equal to the sum of their individual H-rich layers, and surface abundances determined as the mass-weighted average of these layers. We assume the merger has reached an equilibrium state, where layers enriched with heavier elements have settled deeper into the merger product. The central abundances of the merger become the mass-weighted average of the layers dominated by the heaviest element, which become the new core. For the treatment of formed WDs in this merging process, we maintain the central abundances of WDs during their formation, treating them as stripped carbon-oxygen cores, unless their central helium mass fraction is $1\%$ in which case we treat them as helium cores.

A schematic diagram illustrating our process for estimating the properties and the state of the newly-formed merger product for two indicative cases is shown in Figure~\ref{fig:merger_schematic}.
We match the merger product to a model in our single-star grid in order to follow its further evolution. The default matching criteria are the same as described in Section~\ref{sec:matching}, excluding the radius of the newly-formed star. We treat CE mergers considering two different cases as we match the merger product to models in our single-star grids. When the stellar cores do not mix (see left panel of Figure~\ref{fig:merger_schematic}) we include the nuclear luminosity from the core of the more evolved star to identify the nearest single-star model. In cases where the merger has cores that mix (right panel of Figure~\ref{fig:merger_schematic}), we do not use nuclear luminosity as one of the criteria identifying the closest single-star model. 

\begin{figure*}
    \centering
    \includegraphics[scale=1,angle=0]{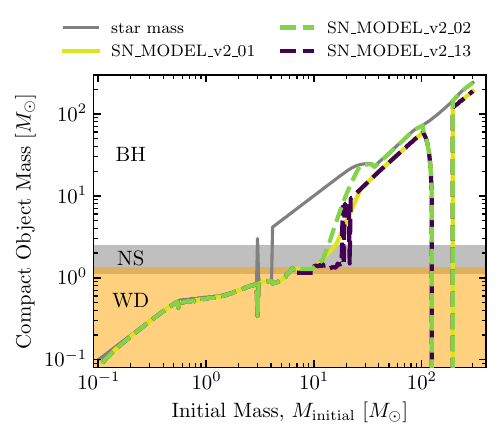}\hfill
    \includegraphics[scale=1,angle=0]{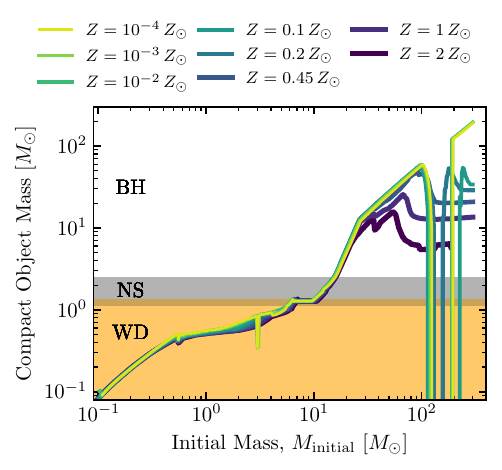}
    \caption{CO masses as a function of ZAMS mass, $M_\mathrm{ZAMS}$, for different single stellar track models. The background color defines the regime of CO type ($M_\mathrm{BH}/\Msun>2.5$, $1.1\leq M_\mathrm{NS}/\Msun\leq2.5$, $M_\mathrm{WD}/\Msun<1.37$). (\textit{Left}) We show the CO masses of single stellar models at $10^{-4}\,\Zsun$ for \texttt{SN\_MODEL\_v2\_01} (solid, yellow), \texttt{SN\_MODEL\_v2\_02} (dashed, green) and \texttt{SN\_MODEL\_v2\_13} (dotted, purple). The purple model assumes the N20 collapse mechanism from \citet{2020MNRAS.499.2803P} while the yellow and green models use the delayed prescription from \citet{2012ApJ...749...91F}. {\tt SN\_MODEL\_v2\_02} conserves the H-rich envelope in BH formation outside the PPISN/PISN regime, while {\tt SN\_MODEL\_v2\_01} assumes the H-envelope is ejected. See Table~\ref{tab:models} for a complete list of the models provided in v2. The final stellar mass at the end of the \texttt{MESA} simulations is shown as a solid, gray line. (\textit{Right}) We show the CO masses of single stellar models with metallicities in $Z\in[2\,\Zsun, 10^{-4}\,\Zsun]$ (color see legend) as a function of ZAMS mass given \texttt{SN\_MODEL\_v2\_01}.}
    \label{fig:singleHMS_co_masses}
\end{figure*}

We note that we assign a surface rotational angular velocity to the merger product, which is by default set to the ad hoc value of $\omega / \omega_\mathrm{crit} = 0.4$. We emphasize that rotation is not self-consistently incorporated into the merger products, as we are matching to our non-rotating single-star models. Instead, we track rotation as a latent parameter, approximating its evolution by following the spin-down of the star, which enables us to calculate the final spin of the CO end state. 

Caution should be exercised when interpreting results that depend on the evolution of merger products, as our matching step does not guarantee that the closest matching single-star model will provide an accurate representation of the merger product \citep[e.g., the core-to-envelope mass ratio may differ;][]{1992ApJ...391..246P, 2014ApJ...796..121J, 2024ApJ...963L..42M, 2024A&A...686A..45S}. We aim to develop single-stellar grids appropriate for the further evolution of merger products in future \posydon versions.

\subsection{Core-collapse SN}
\label{sec:CCSN}

\begin{figure*}
    \centering
    \includegraphics[width=\textwidth]{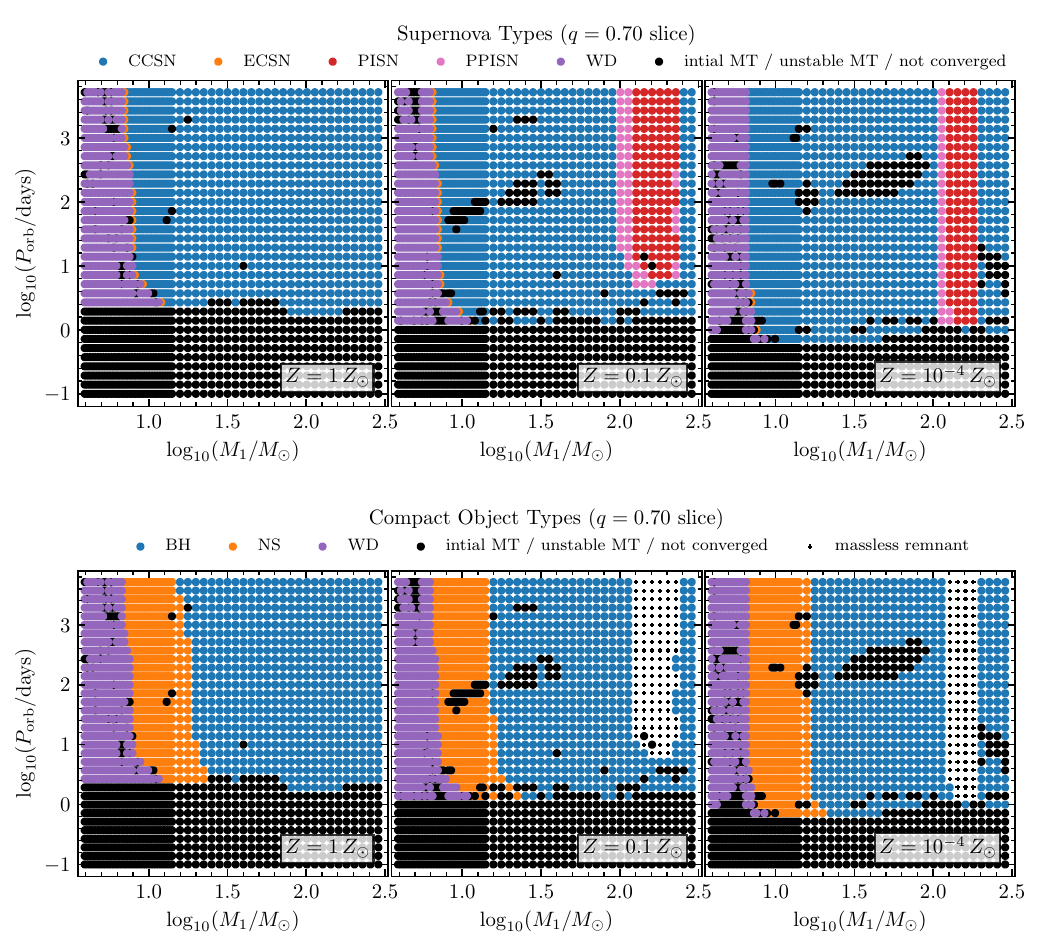}
    \caption{Example of an outcome of a core-collapse model for the primary stars in the HMS--HMS binary grids. Here, we show a comparison between the two-dimensional mass ratio slice $q=0.7$ at $\Zsun$, $0.1\,\Zsun$, and $10^{-4}\,\Zsun$ for the \texttt{SN\_MODEL\_v2\_01}. (\textit{Top}) The panels show the type of SN events occurring in the parameter space according to the legend. (\textit{Bottom}) The panels show the CO type resulting from the SN event. Note that ``islands'' of black points with $P_{\rm orb}$ $\sim10^2-10^3$ days for this $q$ slice are systems that predominantly enter a CE phase, so have not yet evolved to core collapse by the end of the simulation. }
    \label{fig:HMSHMS_CO_types}
\end{figure*}

While the overall methodology for evolving a binary through core-collapse SN (CCSNe) remains largely consistent with our implementation described by \citetalias{2023ApJS..264...45F}, the extension to a range of metallicities opens the door for exploring the PPISN and PISN regimes. Furthermore, our increased input mass range allows us to sample BH formation above the PISN mass gap. In Section~\ref{sec:PISN} we describe the conditions by which we identify which stars will undergo a PISN and PPISN. In this section we describe how we model a binary through core collapse, including PISN and PPISN, along with a set of 24 pre-computed SN models built into v2. All SN models are taken from the literature and a summary of the assumptions corresponding to each model is presented in Table~\ref{tab:models}.

All single- and binary-star model grids are post-processed to account for the outcomes of these 24 core-collapse models. Initial-to-final interpolators are trained to predict a number of parameters including the SN type, fallback mass fraction, CO state, mass, and spin for each model, following the procedure shown in Figure~\ref{fig:flow-chart} and described in Section~\ref{sec:interpolation_improvements}. Furthermore, for BH formation associated with disk accretion, we now report the amount of disk mass accreted and lost during the core-collapse process as discussed in \citetalias[Section~8.3 from][]{2023ApJS..264...45F}. 

In the left panel of Figure~\ref{fig:singleHMS_co_masses} we show the CO mass as a function of $M_\mathrm{initial}$ for three separate core-collapse models at $10^{-4}\,\Zsun$. \texttt{SN\_MODEL\_v2\_01} and \texttt{SN\_MODEL\_v2\_02} both adopt the \citet{2012ApJ...749...91F} delayed CCSN mechanism but differ in their assumption of conserving the H-rich envelope when forming a BH. Therefore, both models predict the same CO masses for WDs and NSs, but {\tt SN\_MODEL\_v2\_02} produces more massive BHs. {\tt SN\_MODEL\_v2\_13} on the other hand adopts the prescription from \citet{2020MNRAS.499.2803P} with the N20 engine, producing somewhat different boundaries separating NSs and BHs, along with different associated masses. Furthermore, the impact of PPISN and PISN is clearly evident as a sharp drop at $M_\mathrm{initial}\simeq80\,\Msun$, accompanied by a sharp rise at $M_\mathrm{initial}\simeq 200\,\Msun$, denoting the upper mass end of the PISN regime.

In the right panel of Figure~\ref{fig:singleHMS_co_masses} we show the metallicity dependence of the CO masses as a function of ZAMS mass, using {\tt SN\_MODEL\_v2\_01}. In the high-metallicity regime, winds cap the maximum BH mass. Due to changes in our adopted winds, we find a somewhat lower maximum BH mass compared with v1 \citep[see discussion in Section~\ref{sec:winds} and in][]{2023NatAs...7.1090B}. The impact of metallicity on the existence of the PISN gap is particularly striking. For single stars, the PISN gap only exists in our grids with $Z \leq0.2 \Zsun$.

In Figure~\ref{fig:HMSHMS_CO_types}, we display how these effects are realized in our binary models for three representative metallicities $\Zsun$, $0.1\,\Zsun$ and $10^{-4}\,\Zsun$ for the HMS--HMS grid using {\tt SN\_MODEL\_v2\_01}. The top panels of Figure~\ref{fig:HMSHMS_CO_types} show the primary star core-collapse event: either WD formation, electron-capture SN (ECSN), CCSN, PPISN, and, PISN, where the latter two are present only in the lower metallicity grids. In v2 we have switched our default condition for ECS to that of \citet{2015MNRAS.451.2123T} based on the CO-core mass. This region is significantly narrower than the one from \citet{2004ApJ...612.1044P} as was adopted in v1, leading to a narrow strip of models undergoing ECS in Figure~\ref{fig:HMSHMS_CO_types}. The bottom panels of Figure~\ref{fig:HMSHMS_CO_types} show the resulting CO types, either WD, NS, BH, or a massless remnant (resulting from a PISN leaving no remnant). Note that ``islands'' of black points with $P_{\rm orb}$ $\sim10^2-10^3$ days for this particular $q$ slice represent systems that predominantly enter a CE phase, and thus have not yet evolved to core collapse by the end of the simulation. In contrast, the band of black points for orbital periods below $\simeq 1-2\,\mathrm{d}$ corresponds to binaries with primary stars that initially fill their Roche lobes. Comparison between the different metallicities in Figure~\ref{fig:HMSHMS_CO_types} reveals two notable effects that align with expectations from our single-star outcomes in Figure~\ref{fig:singleHMS_co_masses}: 1) PISN occurs only in lower metallicity models, and 2) the mass boundaries between WD, NS, and BH shift toward lower initial primary masses for lower metallicity models, where winds are less effective.

\subsection{Spin--orbit misalignment due to SN kicks}
\label{sec:supernova_kicks}

We calculate the effects of mass loss and natal kicks on the binary orbit at core-collapse as in v1, with a few minor improvements. Namely, we fix a bug from v1 in the treatment of eccentric pre-SN orbits, and introduce self-consistent tracking of the spin--orbit misalignment due to natal kicks across both the core collapse of the primary and secondary masses, where applicable. 

As in v1, we calculate the binary orbital changes due to instantaneous mass-loss and additional natal kicks according to \citet{1996ApJ...471..352K}, but allow for eccentric pre-SN orbits \citep{1983ApJ...267..322H,2002MNRAS.329..897H}. As in \citet[][see Figure 1 therein]{1996ApJ...471..352K}, we work in a right-handed coordinate frame where the collapsing helium star lies on the origin and the companion lies on the negative $x$-axis at rest. The relative velocity $\vec{v}_r$ of the collapsing helium-star lies in the $x$--$y$ plane, with $v_{ry}>0$.
The orbital angular momentum vector is in the  $z$ direction, completing the right-handed coordinate system. In v2, we fix a bug wherein $\vec{v}_r$ was always taken to lie in the direction of the positive $y$-axis, which is only the case for either circular orbits or elliptical orbits exactly at very particular orientations (periapsis or apoapsis). We have corrected for this and now $\vec{v}_r$ is allowed to lie in the $x$--$y$ plane at an angle $0<\psi<\pi$ relative to the positive $x$-axis, with $\psi$ calculated as
\begin{equation}
\sin \psi = \frac{\sqrt{G (1-e^2)M_{\mathrm{tot}}a}}{rv_{r}},
\end{equation}
where $M_{\mathrm{tot}}$, $e$, $a$, and $r$ are the pre-SN total mass, eccentricity, semi-major axis, and orbital separation, respectively \citep{2012ApJ...747..111W}. 

In v2 we also track the spin--orbit misalignment due to natal kicks across multiple core collapse events. Upon each SN, the orbital plane is tilted through an angle
\begin{equation}
\cos \theta = \frac{v_{ky} + v_{ry}}{\sqrt{v_{kz}^2 + (v_{ky} + v_{ry})^2}},
\end{equation}
where $\vec{v}_r$ and $\vec{v}_k$ are the pre-SN relative velocity and natal kick velocity, respectively \citep{1996ApJ...471..352K}.
We assume that just before the first SN, the spin angular momentum of each component is aligned with the binary orbital angular momentum. Then, just after the first SN, the angles between the component spins and the binary orbital angular momentum, $\Omega_{1,\mathrm{SN1}}$ and $\Omega_{2,\mathrm{SN1}}$, are just:
\begin{equation}
\Omega_{1,\mathrm{SN1}}=\Omega_{2,\mathrm{SN1}}=\theta_{\mathrm{SN1}}.
\end{equation}
Here, the subscripts $1$ and $2$ denote the first and second component to undergo core-collapse. We assume that the spin angular momentum of the remaining non-CO binary component aligns with the new orbital angular momentum before core collapse at SN2 \citep{2018A&A...616A..28Q}, so the angle between its spin and the post-SN2 orbital angular momentum is
\begin{equation}
\Omega_{2,\mathrm{SN2}}=\theta_{\mathrm{SN2}}.
\end{equation}
We assume that MT is inefficient at realigning the spin of the first-born CO between SN1 and SN2, allowing us to calculate the cumulative misalignment between its spin and the orbital angular momentum across both SNe. However, we do account for the CO spin magnitude evolution due to accretion. In contrast, we assume that tides and MT are efficient at realigning the non-degenerate companion star to the post-SN1 orbital plane. Adopting $\hat{L}=\hat{z}$ and $\hat{L}'$ to be the unit vectors in the pre-SN1 and post-SN2 orbital angular momentum vectors respectively, we can write the post-SN2 orbital angular momentum unit vector $\hat{L}'$ as two successive rotations, one for each SN:
\begin{equation}
\hat{L}' = \mathcal{R}_\mathrm{SN2} \mathcal{R}_\mathrm{SN1} \hat{L},
\end{equation}
where $\mathcal{R} = \mathcal{R}(\theta, f)$ denotes rotation by an angle $\theta$ at the orbital position specified by the pre-SN true anomaly, $f$. Therefore, the angle between the first CO and the final binary's orbit is:
\begin{equation}
\cos(\Omega_{1,\mathrm{SN2}}) = \hat{L} \cdot \hat{L}'.
\end{equation}
Likewise, the angle between the second-born CO and the final binary's orbit is:
\begin{equation}
\cos(\Omega_{2,\mathrm{SN2}}) = \left(\mathcal{R}_{\rm SN2} \hat{L}\right) \cdot \hat{L}'.
\end{equation}
In combination, our procedure now allows for the calculation of both COs' spin vectors prior to coalescence.

\subsection{The X-ray Luminosity of Accreting COs}
\label{sec:X-ray_luminosity}

Following v1, we model accretion onto COs as fully conservative up to the Eddington limit, beyond which material is assumed to leave the binary as an isotropic wind from the vicinity of the accretor. In v2, we have added the capability to calculate the X-ray luminosity produced by XRBs in different MT regimes. For stable RLOF \mesa self-consistently calculates the MT rate within our grids. We ignore wind-fed accretion within our \mesa simulations, but account for it in a post-processing step, following the description from \citet{1944MNRAS.104..273B}, where matter lost as a wind is captured by the gravitational pull of the CO. 

As matter is accreted onto the CO, its rest mass energy is converted to radiation and the efficiency of this conversion is the CO radiative efficiency $\eta$ which is described as,
\begin{equation}
    \eta = \frac{GM_\mathrm{acc}}{R_\mathrm{acc} c^2},
\end{equation}
where $M_\mathrm{acc}$ is the CO mass and $R_{\rm acc}$ is the accretion radius and its value depends on the type of accretor. For WD accretors, the radius is calculated as $2.98\times 10^{8} (M_\mathrm{acc}/\Msun)^{-1/3}\,\mathrm{cm}$ \citep{Hansen2004}. For NS accretors, $R_{\rm acc}$ is the radius of the NS surface, which we take 12.5\,km \citep{2018PhRvL.120z1103M, 2019ApJ...887L..24M, 2019ApJ...887L..21R, 2020PhRvD.101l3007L, 2020ApJ...892L...3A, 2021A&A...650A.139K, 2021ApJ...921...63B, 2021ApJ...918L..29R}, and for BH accretors, $R_{\rm acc}$ is the spin-dependent innermost stable circular orbit \citep{2003MNRAS.341..385P}. The disc structure for the sub-Eddington regime is assumed to be geometrically thin, hence, the accretion is isotropic. 

For RLOF and wind-fed accreting binaries with sub-Eddington mass-transfer rates, we use the mass-accretion rate $\dot{M}_\mathrm{acc}$ and $\eta$ to calculate the bolometric X-ray luminosity  \citep{2002apa..book.....F},
\begin{equation}
    L^\mathrm{RLO/wind}_\mathrm{bolometric} = \eta \dot{M}_\mathrm{acc} c^2, \quad \text{if } \dot{m} \leq 1,
\end{equation}
where $\dot{m}$ is the Eddington ratio ($\equiv \dot{M}_\mathrm{transfer}/\dot{M}_\mathrm{Edd}$) and $c$ is the speed of light. Additionally, we have adopted the criterion from \citet{2021PASA...38...56H} that the donors in wind-fed BH XRBs must be filling their Roche lobes by at least 80\% to form an observable accretion disk. This criterion is included as a user option that can be activated or modified from its default value of 80\%.

For MT rates exceeding the Eddington limit ($\dot{m} > 1$) we use the super-Eddington disk model from \citet{1973A&A....24..337S} who described accretion disks for BH binaries in both sub- and super-Eddington regimes. Following this model, as the MT rate approaches the Eddington limit of the accretor, the innermost part of the accretion disk becomes geometrically thick with an outer thin part of the disk that remains geometrically thin. The mass-accretion rate at each point within this thick disk is locally Eddington limited, however, the total luminosity coming from such a bloated disk exceeds the Eddington limit by a factor of $\ln\left(1+\dot{m}\right)$. The bolometric X-ray luminosity from such an accretion disk is described as follows:
\begin{equation}\label{eq:bolo_lum}
    L^\mathrm{RLO/wind}_\mathrm{bolometric} = L_\mathrm{Edd}\left(1 + \ln\dot{m}\right), \text{ if } \dot{m} > 1,
\end{equation}
where $L_{\mathrm{Edd}}$ is the Eddington luminosity.
For $\dot{m} > 8.5$, \citet{2001ApJ...552L.109K} and \citet{2009MNRAS.393L..41K} suggest that the outgoing emission is collimated due to the thick accretion disk with a geometric beaming factor $b$, where
\begin{equation}\label{eq:beaming}
        b= 
\begin{cases}
    \left(\dfrac{8.5}{\dot{m}}\right)^2, & \text{if } \dot{m}> 8.5,\\
    1,              & \text{otherwise}.
\end{cases}
\end{equation}
The isotropic-equivalent luminosity then becomes:
\begin{equation}\label{eq:iso_lum}
    L^\mathrm{RLO/wind}_\mathrm{iso} = \frac{L^\mathrm{RLO/wind}_\mathrm{bolometric}}{b} = \frac{L_{\mathrm{Edd}}}{b}(1 + \ln{\dot{m}}).
\end{equation}
Since our \mesa simulations can evolve binaries through brief phases of exceptionally high MT rates (e.g., just prior to a CE phase), we find our XRB populations occasionally contain individual sources with extreme beaming factors leading to unphysically high X-ray luminosities ($\gtrsim 10^{41}\,\mathrm{erg}\,\mathrm{s}^{-1}$). We therefore place a lower limit on b, such that $b \geq 3.2\times 10^{-3}$ \citep{2016A&A...587A..13L, 2017ApJ...846...17W}. 
We note that since X-ray luminosities are calculated in a post-processing step, users can easily incorporate their own preferred models.

Within \posydon, Be-XRBs, characterized by rapid spin rates and luminous decretion disks, are treated separately from wind-fed XRBs. We identify Be XRBs using the criteria from \citet{2009ApJ...707..870B} and \citet{2014MNRAS.437.1187Z}: wide, detached binaries (orbital periods in the range of 10 to $300\,\mathrm{days}$) with fast-spinning hydrogen MS donors ($\gtrsim 70\%$ of critical surface velocity). However, we adopt a lower-mass limit of $6\,\Msun$ for the donor's mass \citep{2010AN....331..349H}. We also require the decretion disk radius \citep[using an approximated radius of 100 times the stellar radius;][]{2017A&A...601A..74K} to exceed the donor Roche-lobe radius at periastron. For these systems, we model X-ray luminosities using the empirical relation from \citet{2006ApJ...653.1410D} which is based on peak X-ray luminosities of observed Be XRBs,
\begin{equation}\label{eq:be_xrb}
    \log_{10}\left( \frac{L^\mathrm{Be-XRB}_\mathrm{bolometric}}{10^{35} \mathrm{erg}\,\mathrm{s}^{-1} } \right) = 4.53 - 1.5 \log_{10}\left( \frac{P_\mathrm{orb}}{\mathrm{day}} \right).
\end{equation}
Since Be-XRBs are transient systems, we assume a duty cycle of 10\% \citep[][]{2019IAUS..346..178S}. 

Recently, \citealt[][]{2024ApJ...971..133R} investigated alternative models for defining Be-XRBs based on the latest observations, changing criteria for the rotation threshold for the Be phenomenon, and adopting a more detailed X-ray luminosity calculation from \citet{2024MNRAS.527.5023L} which depends on the periastron separation, donor and CO masses, and CO type.
This experimental treatment for Be-XRBs and associated observables are also included in \posydon, where we refer the reader to \citet[][]{2024ApJ...971..133R} for a detailed discussion.

\section{Generating Populations and Population Analysis \label{sec:populations}}

The process of simulating a synthetic stellar population with \posydon, and comparing the properties of the synthetic population with observations, has been significantly streamlined in v2. In this section, we describe the new \posydon application programming interface (API) for simulating multi-metallicity, single- and binary-star populations (Section~\ref{sec:api}), how one can convolve a synthetic population with a cosmological, metallicity-specific star-formation history (SFH; Section~\ref{sec:SFH}), how to calculate rates of different transient events (Section~\ref{sec:transients}), and how to model the selection effects of GW detectors (Section~\ref{sec:gw_selection}).

\subsection{API to Run Binary Populations}\label{sec:api}

The v2 software infrastructure includes a modular API for evolving binary populations locally and on high-performance computing clusters using {\tt slurm} with optional parallelization. Using inifiles, \posydon{} users can easily specify initial conditions (e.g., initial-mass function, $q$ distribution, etc.), rapid modeling choices (e.g., SNe prescriptions, CE efficiency, etc.), and detailed model grids (e.g., metallicity) facilitating the creation of customized BPS models for different science cases. While some changes to \posydon{} can be made instantaneously (e.g., changing CE efficiency), more complex changes to the stellar or binary physics require re-running grids (e.g. altering stellar winds). Users who wish to develop their own \mesa model grids must first go through the post-processing pipeline (see Section~\ref{sec:postprocessing}) before entering the \posydon{} BPS framework. We refer the reader to the \posydon{} documentation for further details and in-depth tutorials.\footnote{\url{https://posydon.org/POSYDON/v2.0.0/index.html}\label{fn:docsurl}}

\subsection{Redshift- and Metallicity-dependent Star-Formation History of the Universe}\label{sec:SFH}

\begin{figure*}
    \centering
    \includegraphics{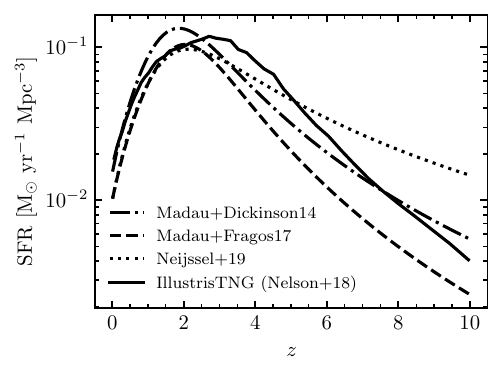}\hfill
    \includegraphics{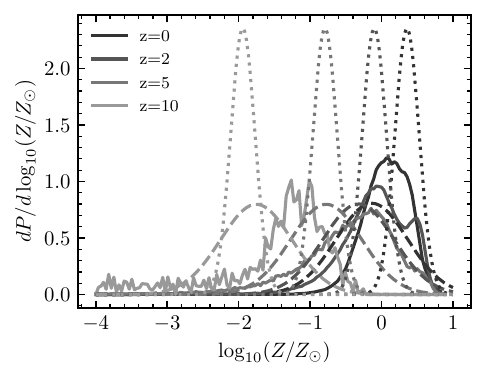}
    \caption{ We show the SFR density as a function of redshift for four separate models (left panel) and the metallicity distribution of star formation at four separate redshifts for each of our star-formation models (right panel). To produce their own cosmological populations accounting for both metallicity and SFR evolution as a function of redshift, a user can specify which of these models to use or optionally include their own. Note we do not show the \citet{2014ARA&A..52..415M} model in the right panel.}
    \label{fig:SFRH}
\end{figure*}

The evolution of galaxies and, in turn, the Universe, is influenced by a complex interplay of processes, among which star formation and chemical enrichment are critical \citep{2014ARA&A..52..415M}. The SFH is, thus, a fundamental distribution describing the cosmic evolution of stellar populations. Here, we define the redshift- and metallicity-dependent SFH, 
\begin{equation}
\mathrm{SFH}(z,Z) \equiv \frac{\mathrm{d}^3M}{\mathrm{d}V_\mathrm{c} \, \mathrm{d}t \, \mathrm{d}\log_{10}(Z)}(z,Z),
\end{equation}
as the total stellar mass formed, $\mathrm{d}M$, per comoving volume interval, $\mathrm{d}V_\mathrm{c}$, per unit of cosmic time, $\mathrm{d}t$, and per log-metallicity range $\mathrm{d}\log_{10}(Z)$.
The remaining part of this section presents different options for calculating the SFH within \posydon{}, which includes but is not limited to the IllustrisTNG large-scale cosmological simulation \citep{2018MNRAS.477.1206N, 2018MNRAS.475..676S, 2018MNRAS.480.5113M, 2018MNRAS.475..648P, 2018MNRAS.475..624N} and the empirical models from \citet{2014ARA&A..52..415M}, \citet{2017ApJ...840...39M}, and \citet{2019MNRAS.490.3740N}.

The Next Generation Illustris (IllustrisTNG) cosmological simulation \citep{2019ComAC...6....2N} provides a powerful tool for understanding the SFH in the context of a self-consistent physical model of the Universe through magnetohydrodynamical simulations of dark and baryonic matter. In this model, the metallicity distribution of the SFH emerges naturally from the underlying physical assumptions, such as stellar evolution and SN feedback \citep[for more details, see][]{2018MNRAS.477.1206N}. The TNG100-1 model included as an option in \posydon{} simulated a comoving volume of $75\,(\mathrm{Mpc}/h)^3$ containing tens of thousands of galaxies captured in high detail.

As an alternative, empirical models provide prescriptions for incorporating metallicity into the SFH. In these models, the SFH is constructed from observational data, and assumptions are made about the metallicity distribution and its redshift evolution. 

The \citet{2014ARA&A..52..415M} model, for instance, uses a compilation of observational data to estimate the cosmic star formation rate, $\mathrm{SFR}(z)$, and the measured mean metallicity of the Universe, $\bar{Z}(z)$, while the \citet{2017ApJ...840...39M} model extends this data set to include more recent surveys. Meanwhile, the \citet{2019MNRAS.490.3740N} model estimates the free-parameters of the $\mathrm{SFR}(z)$ functional form of the \citet{2014ARA&A..52..415M} model, the $\bar{Z}(z)$ parameters of the \citet{2006ApJ...638L..63L} functional form and the log-normal metallicity distribution standard deviation $\sigma$ using best-fit parameters based on compact binary coalescence rates from GW observations.

Typically, the empirical models are implemented by assuming metallicities are log-normally distributed around $\bar{Z}(z)$. These models have the advantage of being directly tied to observed data, and the log-normal assumption is a reasonable approximation given observational constraints. For these models, we can decouple the SFR and metallicity evolution:
\begin{equation} \label{eq:SFH_split}
\begin{split}
\mathrm{SFH}(z,Z) &= \frac{\mathrm{d}^2M}{\mathrm{d}V_c \, \mathrm{d}t}(z) \times \frac{\mathrm{d}P}{\mathrm{d}\log_{10}(Z)}(z,Z) \\ 
&= \mathrm{SFR}(z) \times \mathcal{N}(\log_{10}(Z) | \mu(z), \sigma),
\end{split}
\end{equation}
where $\mathrm{d}P/\mathrm{d}\log_{10}(Z)(z,Z)$ is the probability per decade of metallicity of a star being formed with a metallicity $Z$ at a redshift $z$, and $\mathcal{N}$ stands for a normal distribution with mean $\mu(z)$ and standard deviation $\sigma$, evaluated at $\log_{10}(Z)$. We define the SFR density as $\mathrm{SFR}(z) \equiv \mathrm{d}^2M/\mathrm{d}V_c\,\mathrm{d}t(z)$, $\sigma$ is the log-normal distribution standard deviation, and $\mu = \log_{10}(\bar{Z}(z)) - \ln(10) \sigma^2/2$ is the mean metallicity for \citet{2014ARA&A..52..415M} and \citet{2017ApJ...840...39M}, as implemented in \posydon{} \citep[see, e.g., Appendix B in][]{2020A&A...635A..97B}.
However, for the \citet{2019MNRAS.490.3740N} distribution we adopt a mean of $\mu = \ln(\bar{Z}(z)) - \sigma^2 / 2$ instead.
Depending on the arbitrary choice of $\sigma$, one might need to truncate the log-normal distribution to ensure $Z \in [0,1]$. 
By default, \posydon{} assumes a log-normal distribution truncated at $Z_\mathrm{max} = 1$.

In Figure~\ref{fig:SFRH}, we present a comparison of the different SFR density models in \posydon{}, as well as the metallicity distribution evolution as a function of redshift. To illustrate the redshift evolution of the \citet{2017ApJ...840...39M} metallicity distribution, we assume $\sigma = 0.5$ according to \citet{2020A&A...635A..97B}; however, \posydon{} allows the user to specify their preferred value.
All considered SFR models tend to agree at low redshift, but show some variation at $z>2$. Figure~\ref{fig:SFRH} also shows that the Illustris simulation predicts a larger mean metallicity with respect to empirical models. The variation of different models in Figure~\ref{fig:SFRH} can be interpreted as an indirect measure of current uncertainties in the SFH \citep[for the effect on GW merger rates, see for example:][]{2022MNRAS.516.5737B, 2022MNRAS.514.1315B, 2024AnP...53600170C}.

\subsection{Cosmological Rates for Transient Event Populations}
\label{sec:transients}

DCO mergers, CCSNe, and long gamma-ray bursts are all examples of non-repeating, nearly instantaneous events in the evolution of a single star or a binary system. In this section, we detail our methodology for calculating the cosmological rates of such events.

From a \posydon{} binary population simulation with a burst of star formation at a discrete metallicity $Z_j$, we can extract the time between stellar birth and the transient event (the delay time, $t_\mathrm{delay}$). We then convolve these times with a SFH, which artificially distributes each transient event across the cosmic history of the Universe every $\Delta t_i$ interval with a corresponding redshift $z_i$ at the center of each time interval.
The default \posydon{} assumption is $\Delta t_i = 100\,\mathrm{Myr}$, although different values can be specified by the user.
Hence, we can assign a redshift of formation $z_{\mathrm{form},i}$ corresponding to the center of the $\Delta t_i$ bin and compute the redshift of the event, $z_{\mathrm{event},i,k}$, given its $t_\mathrm{delay}$. Note that this calculation requires a population generated with a single burst, which is then post-processed with a particular choice of star formation history, allowing one to test the impact of that choice without rerunning the population model in its entirety.

We calculate the event rate density, which quantifies the number of transient events per unit of comoving volume per unit of cosmic time at a given redshift, similar to  \citet{2020A&A...635A..97B,2021A&A...647A.153B,2022A&A...657L...8B}, by approximating it as a Monte Carlo sum:

\begin{equation}
\mathcal{R}(z) = \frac{1}{\Delta V_\mathrm{c}(z)}\sum_{\Delta Z_j} \sum_{k} \sum_{z_{\mathrm{event},i,k} \text{ in }\Delta t}  w_{k,i,j},
\label{eq:R_DCOs}
\end{equation}
where $w_{k,i,j}$ is a weight representing the rate contribution of a single event per year given its metallicity and its formation redshift. For each metallicity bin ($\Delta Z_j$), we sum over all $k$ events in that bin that have their $z_{\mathrm{event},i,k}$ occurring in an interval ($\Delta t$) around $z$. This interval is the same $\Delta t_i$ used to distribute the birth redshifts across cosmic history. We finally sum over all the metallicity bins and normalize using the comoving volume shell, $\Delta V_\mathrm{c} (z)$, corresponding to the interval of $\Delta t$ around $z$:

\begin{equation} \label{eq:event_rate}
\begin{split}
    \Delta V_\mathrm{c} (z) & \equiv \int_{\Delta t} \frac{1}{1+z} \frac{\mathrm{d}V_\mathrm{c}}{\mathrm{d}z} \mathrm{d}z \\ &= \frac{4\pi c}{H_0} \int_{\Delta z} \frac{D^2_\mathrm{c}(z)}{E(z)(1+z)} \mathrm{d}z ,
\end{split}
\end{equation}
where $\Delta z$ is the redshift interval corresponding to $\Delta t$ and $D_\mathrm{c} (z) = ({c}/{H_0}) \int_0^z E(z')^{-1} \mathrm{d}z'$ is the comoving distance. $E(z)=\sqrt{\Omega_\mathrm{m}(1+z)^3+\Omega_\Lambda}$ and $\Omega_\Lambda = 1 - \Omega_\mathrm{m}$ assuming a flat $\Lambda$CDM cosmology \citep{Planck2015}.

The weights $w_{k,i,j}$ in Eq.~\ref{eq:R_DCOs} can be calculated at the specific metallicity of the event for each $z_{\mathrm{form},i}$ following \citet{2020A&A...635A..97B}:
\begin{equation} \label{eq:event_weight}
\begin{split}
    w_{k,i,j} = \frac{f_\mathrm{corr}}{M_{\mathrm{sim}, \Delta Z_j}} 4 \pi c \left[D_c(z_{\mathrm{event},i,k})\right]^2 \\ 
    \times \ \mathrm{SFH}(z_{\mathrm{form},i},\Delta Z_j) \Delta t_i\quad \mathrm{yr}^{-1},
\end{split}
\end{equation}
where $f_\mathrm{corr}/M_{\mathrm{sim}, \Delta Z_j}$ is the simulation mass correction with $f_\mathrm{corr}$ accounting for the mass contribution of stars not considered in the population model, $\mathrm{SFH}(z_{\mathrm{form},i}, \Delta Z_j)$ is the SFR at the formation redshift and the metallicity of the event, and $4\pi c\left[D_c(z_{\mathrm{event},i,k})\right]^2$ arises from the conversion of the integration over the comoving volume to cosmic time and accounting for the difference in clock times between the source and detector frame \citep[for more details, see][]{2015ApJ...806..263D}.

The underlying stellar mass of the population at each metallicity is determined from the simulation mass $M_{\mathrm{sim}, \Delta Z_j}$, and the correction factor $f_\mathrm{corr}$ for the sampling limits of the initial binary parameters \citep[see, e.g., Appendix~A in][]{2020A&A...635A..97B}. The default \posydon assumption for this normalization is a binary fraction of $f_\mathrm{b} = 0.7$, although the user can choose different values. We provide routines to calculate the metallicity-specific event rate efficiency, which is the number of transient events per unit of stellar mass formed at a given metallicity, for a \posydon{} transient event population.

The last contribution to $w_{k,i,j}$ is from the SFH, which as discussed in Section~\ref{sec:SFH}, can be decoupled into a SFR and metallicity evolution component. Since a population in v2 is obtained by evolving ZAMS binary systems at each of the eight supported discrete metallicities, $Z_j$, without interpolation between these metallicities, we divide the continuous metallicity interval $Z\in[0,1]$ into uniform-in-log intervals, $\Delta Z_j$, centered around $Z_j$. This provides a discretized weighting of each metallicity across redshift, $f_\mathrm{SFR}(z| \Delta Z_j)$, but excludes part of the metallicity distribution at high and low $Z$. By default, \posydon{} does not include these parts of the metallicity distribution outside the bin range. However, different boundary conditions, $Z_\mathrm{min}$ and $Z_\mathrm{max}$, can be given to alter the considered metallicity range. For example, if $Z_\mathrm{max}$ is higher than the upper bin edge, the remaining part of the metallicity distribution is added to the upper bin, and similarly for $Z_\mathrm{min}$ and the lowest bin. Additionally, if either boundary is within the bin range, only the part of the distribution, where $Z\in[Z_\mathrm{min}, Z_\mathrm{max}]$, is included in the metallicity calculation. \posydon also includes the option to renormalize the metallicity fraction to 1 for every redshift, such when convolved with the star formation rates, the full cosmic star formation rate is considered. Combining this with $\mathrm{SFR}(z)$ gives the SFH weight of the event:
\begin{equation}
\begin{split}
\mathrm{SFH}(z_{\mathrm{form},i}, \Delta Z_j) = & \\
f_\mathrm{SFR}&(z_{\mathrm{form},i}|\Delta Z_j)\ \mathrm{SFR}(z_{\mathrm{form},i}).      
\end{split}
\end{equation}

All these components come together to create the weight per event in Equation~\ref{eq:event_weight}, which can be combined to create the intrinsic rate density calculation in Equation~\ref{eq:R_DCOs}. This calculation can be applied to any transient population with a short duration relative to a stellar evolution timescale. \posydon{} allows for the generation of arbitrary transient event populations from its burst stellar populations, as long as an instantaneous moment in time can be defined for the event, providing flexibility for many use cases.

\subsection{Detection Probabilities for DCO Mergers}\label{sec:gw_selection}

\begin{figure*}[t]
    \center
    \includegraphics[width=0.97\textwidth]{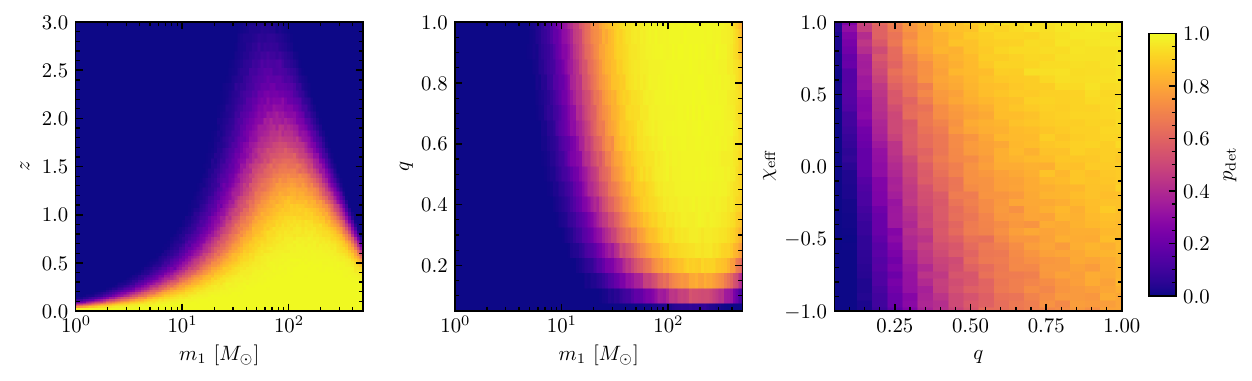}\\
    \caption{Two-dimensional slices of the grid used for estimating detection probabilities. For each panel, the two parameters not shown are held fixed at $m_1=30\,\Msun$, $q=1$, $\chi_\mathrm{eff}=0$, or $z=0.5$. The detection probabilities $p_\mathrm{det}$ shown are for a 3-detector network consisting of LIGO-Hanford, LIGO-Livingston, and Virgo operating at design sensitivity \citep{2020LRR....23....3A}. 
    }
    \label{fig:pdet_grid}
\end{figure*}

One can use routines within \posydon{} described in the previous section to calculate the rate of DCO mergers due to GW radiation. However, in v2 we additionally provide functionality to fold in GW detector selection effects and observability and calculate projected detection DCO merger rates. In this section, we describe how we compute the selection effects of a network of GW detectors given the redshift of the DCO merger, its masses, and the effective spin parameter.

One can approximate the DCO detection rate in events per year with a Monte Carlo sum given by
\begin{equation}
R^\mathrm{det}_\mathrm{DCO} = \sum_{\Delta t_i} \sum_{\Delta Z_j} \sum_{k} w_{k,i,j}\ p_{\mathrm{det}, i,k},
\end{equation}
where $w_{k,i,j}$ is the contribution of the DCO binary $k$ to the detection rate, as calculated in Equation \ref{eq:event_weight}. Here, $p_{\mathrm{det},i,k} \equiv p_\mathrm{det}(z_{\mathrm{m},i,k},M_{1,k},M_{2,k},\chi_{\mathrm{eff},k})$ is the probability of detecting a DCO merger at redshift $z_{\mathrm{m},k}$ with source-frame masses $m_{1,k}$ and $m_{2,k}$ and effective spin $\chi_{\mathrm{eff},k}$ for a given network of GW detectors. Although these are the only three intrinsic parameters of the system we currently account for in calculating detectability, one could expand the formalism in this section to include additional parameters characterizing the binary (e.g., $\chi_{\rm p}$). With $p_\mathrm{det}$, we transform the intrinsic weight $w_{k,i,j}$ of the event into a detection weight $\tilde{w}_{k,i,j}$ for a specific GW detector configuration. This allows us to calculate observable rates and populations of merging DCOs that can be directly compared against observations \citep[e.g.,][]{2024A&A...683A.144X,2024arXiv241020415X}.

To determine the detection probability $p_{\mathrm{det},i,k}$ for the $k$th system, we use precalculated grids of detection probabilities in the 4-dimensional space of primary mass $m_1$, mass ratio $q \equiv M_2/M_1$, effective inspiral spin $\chi_{\mathrm{eff}}$, and redshift $z$. This grid has $100$ log-spaced gridpoints for $M_1/\Msun \in [1,500]$, $20$ gridpoints for $q \in [0.05, 1]$, $41$ gridpoints for $\chi_{\mathrm{eff}} \in [-1, 1]$, and $100$ gridpoints for $z$ that are linearly spaced in luminosity distance between $D_\mathrm{L}(z=0.001)$ and an upper bound determined by the maximum luminosity distance that systems in the grid can be observed above a threshold signal-to-noise ratio (SNR) for a given network sensitivity. This totals in $8.2 \times 10^6$ SNR calculations for each network sensitivity considered. 

We generate detection-probability grids assuming a three-detector network consisting of LIGO-Hanford, LIGO-Livingston, and Virgo operating at sensitivities representative of four different observing eras \citep[\texttt{O3}, \texttt{O4-low}, \texttt{O4-high}, and \texttt{design}, see][]{2020LRR....23....3A}. We use the waveform approximant \texttt{IMRPhenomXHM} for all SNR calculations \citep{2020PhRvD.102f4002G}. We first calculate the optimal matched-filter network SNR assuming the system is face-on and directly overhead of each detector, which is an overly optimistic SNR since the system cannot be directly overhead of all detectors simultaneously \citep{1993PhRvD..47.2198F}. If the system has an optimal network SNR less than $\rho_\mathrm{thresh}=10$ (see \citealt{2023PhRvD.108d3011E} for a comparison of SNR threshold choices for semi-analytic sensitivity estimates), we assign a detection probability of $0$. Otherwise, we randomly vary the extrinsic parameters of the system $\psi_j$ (right ascension, declination, inclination) $N=10^3$ times and calculate the detection probability $p_\mathrm{det}$ as the fraction of trials that have a network SNR $\rho_\mathrm{net} \geq \rho_\mathrm{thresh}$, using the proper response function for each detector: 
\begin{equation}
    p_\mathrm{det} =  \frac{1}{N} \sum^{N}_{j=1} \mathcal{H}\left[ \sqrt{\sum_i (\rho_i(\psi_j))^2} - \rho_\mathrm{thresh} \right],
\end{equation}
where $\rho_i$ is the single-detector SNR for each of the $i$ detectors in the network given $\psi_j$ and $\mathcal{H}$ is the Heaviside step function. 
In Figure~\ref{fig:pdet_grid} we show two-dimensional slices of detection probabilities for the grid assuming a three-detector network operating at design sensitivity.

With the precomputed grids of detection probabilities in hand, we estimate detection probabilities of new data by normalizing the grids and data to a unit cube. We then train the \texttt{scikit-learn} implementation of a $k$-NN regressor with $10$ neighbors measured using Euclidean distance. Using a population resampled according to the procedures above and a three-detector network operating at design sensitivity, we compare the predicted detection probabilities to directly calculated detection probabilities and find a mean residual of $0.8\%$ ($<4.3\%$ at the $99^\mathrm{th}$ percentile). 

The functionality described in this section is all wrapped into pre-packaged routines, allowing for the GW merger rate to be calculated for any particular binary population model. The user only needs to generate statistically significant binary populations ($\gtrsim10^6$) across one or multiple of the eight metallicities as input for these functions to calculate the observed merger rate.

\section{Caveats, Considerations, and Future Work}
\label{sec:caveats}

\subsection{Overview and Caveats}

\posydon{} is designed to be a software package that can efficiently and accurately evolve statistically large populations of single and binary stars in a reasonable computational time, while incorporating the best available physics treatment. Our aim is to evolve a binary in $\mathcal{O}$(1 s), which we achieve in v2 using the extensive grids of pre-calculated detailed single- and binary-star models presented here, a rate which allows a population of $10^6$ binaries to be evolved in $\sim 300\,\mathrm{h}$. While this computation rate is $\sim 3$ orders of magnitude slower per binary than rapid BPS codes such as {\tt COSMIC}, {\tt COMPAS}, or {\tt BSE}, the runtime is still manageable with access to even modest CPU resources. In particular, \posydon{} is designed to evolve large populations of binaries in parallel, making it optimal for high-performance computing environments. However, the sheer size of our pre-computed simulation grids demands considerable memory and disk resources: for synthesizing populations using our publicly available v2 grids, our benchmarking tests find that $\sim 9\,\mathrm{GB}$ of RAM per core and $150\,\mathrm{GB}$ of disk space are required to run \posydon{}. Alternatively, if one wishes to generate their own single- or binary-star grids with \posydon{}, the computation requirements mirror those of \mesa. To improve runtime, users of \posydon{} can optimize their population runs by focusing on a particular regime of interest (e.g., restrict for binaries within a specific mass or orbital period range, whether or not to include initially single stars along with binaries in their populations, etc.). Our online documentation and tutorials\footref{fn:docsurl} provide a guide for setting up population runs.

The sheer size and runtime make computing individual grids a potentially expensive endeavor. Therefore, parameters must be carefully chosen. While \posydon{} is written in a modular way, so that individual grids can be easily swapped for evolutionary phases of interest, the expense of running new grids makes \posydon{} inflexible in other ways. In generating our grids, we made specific choices about the stellar winds, overshooting, and accretion efficiency, among other parameters. Altering any one of these prescriptions would require the entire grid to be re-run. We have therefore updated for v2---and will continue to improve in future versions---the physical treatments within \mesa{} to incorporate our best understanding of stellar and binary evolution. 

Our underlying single and binary stellar evolution model grids represent a substantial augmentation in v2 compared to our v1 grids. The addition of seven new metallicities effectively expands our grids from three dimensions to four. However, as discussed in Section~\ref{sec:grids} we do not train any interpolation methods to calculate the full time-evolution of stars in-between metallicities. Given the computational expense of our \mesa{} grids, this expansion would be unfeasible for our available resources (without major machine-learning advances, e.g., active learning and emulators, planned for future \posydon versions); therefore, our v2 grids have a somewhat lower resolution than those simulated in v1. The result is that while our v1 grids contain $\simeq 120,000$ individual \mesa simulations at one metallicity, our v2 grids contain $\simeq 500,000$, but with the additional dimension of metallicity.
Furthermore, our single- and binary-star model grids in v2 (as was the case for v1) are motivated by phenomena produced by relatively massive stars. However, compared with v1 we have extended the mass range from $4.0\,\Msun$ going up to $286\,\Msun$ for the HMS--HMS grid.

\subsection{Benefits of the \posydon{} Approach to BPS}

Binary population synthesis involves a complex combination of different prescriptions intended to represent the evolution of binary systems through a range of interactions. As \posydon takes a substantially different approach to binary population synthesis compared to other rapid BPS codes, as well as codes such as {\tt BPASS}, it is nontrivial and, in some cases not feasible, to discern how its implementation differences influence the predicted properties of binary populations relative to other codes. Such a comparison is beyond the scope of this work. However, it is possible to identify physical processes that {\tt POSYDON} models with an improved accuracy. Thus, the predictions of population characteristics made by {\tt POSYDON} are expected to be more robust, as they are governed by these accurately modeled processes.

First, the stability of Roche-lobe overflow mass-transfer is assessed self-consistently by the response itself of the stellar structure of the two binary components, and it is not based on any fixed boundaries on binary mass-ratios, or analytical approximations of the derivative of the mass-radius relation ($\zeta=d\ln R / d\ln M$) of stars. A characteristic example of this is the delayed onset of dynamical instability \citep{1987ApJ...318..794H} due to the exposure of an inner convective layer of the donor after the binary has transferred several solar masses during a stable mass-transfer phase \citep[e.g.,][]{2021A&A...645A..54K}. These binaries cannot be accurately modeled in rapid binary population synthesis (BPS) codes, as the structure of the star at the onset of RLOF is substantially different from its structure at the onset of CE evolution.

For the same reason, namely the self-consistent modeling of the response of a star's stellar structure to mass loss, we expect that \posydon populations will calculate the MT rates more accurately during stable MT phases. This assertion is particularly relevant for thermal timescale MT, as the outer envelopes of donor stars in this regime are, by definition, out of thermal equilibrium. In contrast, rapid BPS codes implicitly assume that stars maintain a structure akin to single stars, even during MT phases. Similarly, in Case A MT, as the donor star is stripped, its internal layers that were previously part of the receding convective core, and thus enriched with heavier elements, are now exposed. This changes once again the response of the donor star to mass loss, and affects the duration, MT rate, and point of detachment of Case A MT phases. While all these effects are self-consistently taken into account in \posydon, rapid BPS codes are known to poorly model Case A MT \citep[e.g.,][]{2024MNRAS.530.3706D}.

Third, the rotation of both stars in a binary system is tracked and evolved. Within individual \mesa{} grids, angular momentum transport mechanisms are self-consistently included, both externally (e.g., tides and magnetic braking) and internally (e.g., transport through the Taylor-Spruit mechanism). Previous studies with \posydon{} have determined that carefully accounting for this transport is critical for predicting the characteristics of a range of binary populations including ultra-compact X-ray binaries \citep{gossage2023}, the spins of BH mergers \citep{2018A&A...616A..28Q, 2020A&A...635A..97B,2023NatAs...7.1090B}, and the production of long gamma-ray bursts \citep{2022A&A...657L...8B,2025arXiv250209187B}. 

Additionally, \posydon{} provides the internal structure of stars at two key evolutionary phases: the onset of a common envelope and prior to core collapse. The former allows for a detailed understanding of the core-envelope boundary, avoiding the need for fitting formula approximations to calculate envelope binding energies \citep[e.g.,][]{2010ApJ...716..114X}. This feature of \posydon{} has been used to account for the wide orbital period range of observed double neutron stars within the Milky Way \citep{Chattaraj2024inprep}, as well as the BH mass spectrum in NS-BH mergers \citep{2024arXiv241020415X}. We include multiple parameterizations as built-in options for evolving a binary through a common envelope; however, the modular structure of \posydon{} additionally allows one to incorporate their own prescription that is informed by the complete donor's structure, information unavailable to rapid BPS codes. At the same time, knowledge of a star's structure at core collapse in principle allows for an improved prediction of the remnant's characteristics. While the \mesa{} simulations forming our binary grids are stopped at central carbon depletion rather than iron core collapse, the evolution ought to be sufficiently advanced that subsequent binary interactions are ignorable. Our 24 pre-calculated supernova models include prescriptions widely used in the binary population synthesis literature. However, here again, the modularity of \posydon{} allows a user to synthesize binary populations integrating alternative core-collapse prescriptions. As an example, \citet{2021A&A...656L..19Z} revisited the explodability of single massive star progenitors of stripped-envelope supernovae, using core-collapse prescriptions that are based on the final structure of our stellar models.

Finally, the Monte Carlo approach adopted within \posydon{} represents an innovation compared with other binary population studies using grid-based codes. Previous grid-based BPS codes calculate population characteristics by combining each simulated binary in a grid using a weighting factor corresponding to its likelihood of generation based on our understanding of the initial binary properties \citep[e.g.,][]{2017PASA...34...58E}. \posydon{} veers from that approach, instead randomly generating initial binaries and using classification and interpolation mechanisms trained on our \mesa{} grids to evolve a binary forward. The comparisons in Figures~\ref{fig:nearest_neighbor_HMS-HMS}, \ref{fig:nearest_neighbor_CO-HMS}, and \ref{fig:nearest_neighbor_CO-HeMS} and the associated discussion in Appendix~\ref{sec:nearest_neighbor} demonstrate that the approach taken by \posydon{} allows for a significantly improved model accuracy compared with nearest neighbor matching which ought to approximate the weighted-grid approach adopted by other codes. Although catastrophic failures can occur when interpolating near grid boundaries, we have generally found large errors to be rare. Further, our approach is more amenable to irregularly spaced model grids that we expect to characterize future \posydon{} versions. While, ultimately, the accuracy of population generation is dependent upon a number of factors including grid density and the particular type of binary being evolved, we have found the classification and interpolation methods adopted here to be a significant step forward.

\subsection{Looking Forward}

With this v2 release of \posydon it is possible to model cosmological binary star populations involving NS and/or BH with low- or high-mass companions and their progenitors throughout their evolutionary history. Considering future \posydon versions our focus includes: extending to binaries with low-mass primaries and WD populations, accounting for pulsar treatment for NS, XRB evolution with their persistent and transient behavior and associated selection effects, improved treatments of the CE phase, alternative models for estimating the accretion efficiency on both compact objects and non-degenerate stars, mass-transfer in eccentric orbits (without assuming instant circularization upon RLOF), modeling of stellar spectra, models with $\alpha$-element enhancement for low-metallicity stars, and incorporating single-star grids at multiple initial rotational velocities. Several of these improvements will be aided by technical advacements involving binary-evolution track interpolation \citep{2025ApJ...984..154S} and incorporation of active learning for the development of \mesa model grids \citep{2022ApJ...938...64R}.

\acknowledgments

We thank the anonymous referee whose detailed comments greatly improved the manuscript.

We thank Christopher Berry for useful discussions and detailed comments on the draft manuscript. 
The \posydon{} project is supported primarily by two sources: the Gordon and Betty Moore Foundation (PI Kalogera, grant awards GBMF8477 and GBMF12341) and a Swiss National Science Foundation Professorship grant (PI Fragos, project number CRSII5\_213497). 

The collaboration was also supported by the European Union's Horizon 2020 research and innovation program under the Marie Sklodowska-Curie RISE action, grant agreements No 691164 (ASTROSTAT) and No 873089 (ASTROSTAT-II). Individual team members were supported by additional sources:  
JJA acknowledges support for Program number (JWST-AR-04369.001-A) provided through a grant from the STScI under NASA contract NAS5-03127.
MMB was supported by the Boninchi Foundation, the project number CRSII5\_21349, and the Swiss Government Excellence Scholarship.
VK was partially supported through the D.I.Linzer Distinguished University Professorship fund. 
PMS, AK, VK, SG, ET, KAR, and MS were supported by the project numbers GBMF8477 and GBMF12341.
SG, VK, AK, PMS thank SkAI for support and hospitality for project operations.
KAR is also supported by the NASA grant awarded to the Illinois/NASA Space Grant Consortium, and any opinions, findings, conclusions, or recommendations expressed in this material are those of the author and do not necessarily reflect the views of NASA.
KK and EZ were partially supported by the Federal Commission for Scholarships for Foreign Students for the Swiss Government Excellence Scholarship (ESKAS No.~2021.0277 and ESKAS No.~2019.0091, respectively).
KK is supported by a fellowship program at the Institute of Space Sciences (ICE-CSIC) funded by the program Unidad de Excelencia Mar\'ia de Maeztu CEX2020-001058-M.
EZ and DS acknowledge support from the Hellenic Foundation for Research and Innovation (H.F.R.I.) under the “3rd Call for H.F.R.I. Research Projects to support Post-Doctoral Researchers” (Project No: 7933). 
DM and KAR thank the LSSTC Data Science Fellowship Program, which is funded by LSSTCorporation, NSF Cybertraining Grant No. 1829740, the Brinson Foundation, and the Gordon and Betty Moore Foundation; their participation in the program has benefited this work.
ZX was supported by the Chinese Scholarship Council (CSC). 
Support for M.Z. was provided by NASA through the NASA Hubble Fellowship grant HST-HF2-51474.001-A awarded by the Space Telescope Science Institute, which is operated by the Association of Universities for Research in Astronomy, Incorporated, under NASA contract NAS5-26555. 

The computations were performed at Northwestern University on the Trident computer cluster (funded by the GBMF8477 award) and at the University of Geneva on the Yggdrasil computer cluster. This research was supported in part through the computational resources and staff contributions provided for the Quest high performance computing facility at Northwestern University which is jointly supported by the Office of the Provost, the Office for Research, and Northwestern University Information Technology.

\software{This manuscript has made use of the following Python modules: 
\texttt{numpy} \citep{2020NumPy-Array},
\texttt{scipy} \citep{2020SciPy-NMeth},
\texttt{pandas} \citep{mckinney2010data},
\texttt{matplotlib} \citep{hunter2007matplotlib},
\texttt{astropy} \citep{2013A&A...558A..33A,2018AJ....156..123A, astropy_v3},
\texttt{scikit-learn} \citep{scikit-learn}. 
}

\newpage

\bibliographystyle{aasjournal}
\bibliography{references}

\appendix
\section{Comparing Interpolation to Nearest Neighbor Matching}
\label{sec:nearest_neighbor}

\begin{figure*}[t]
    \centering
    \includegraphics[width=0.9\textwidth,angle=0]{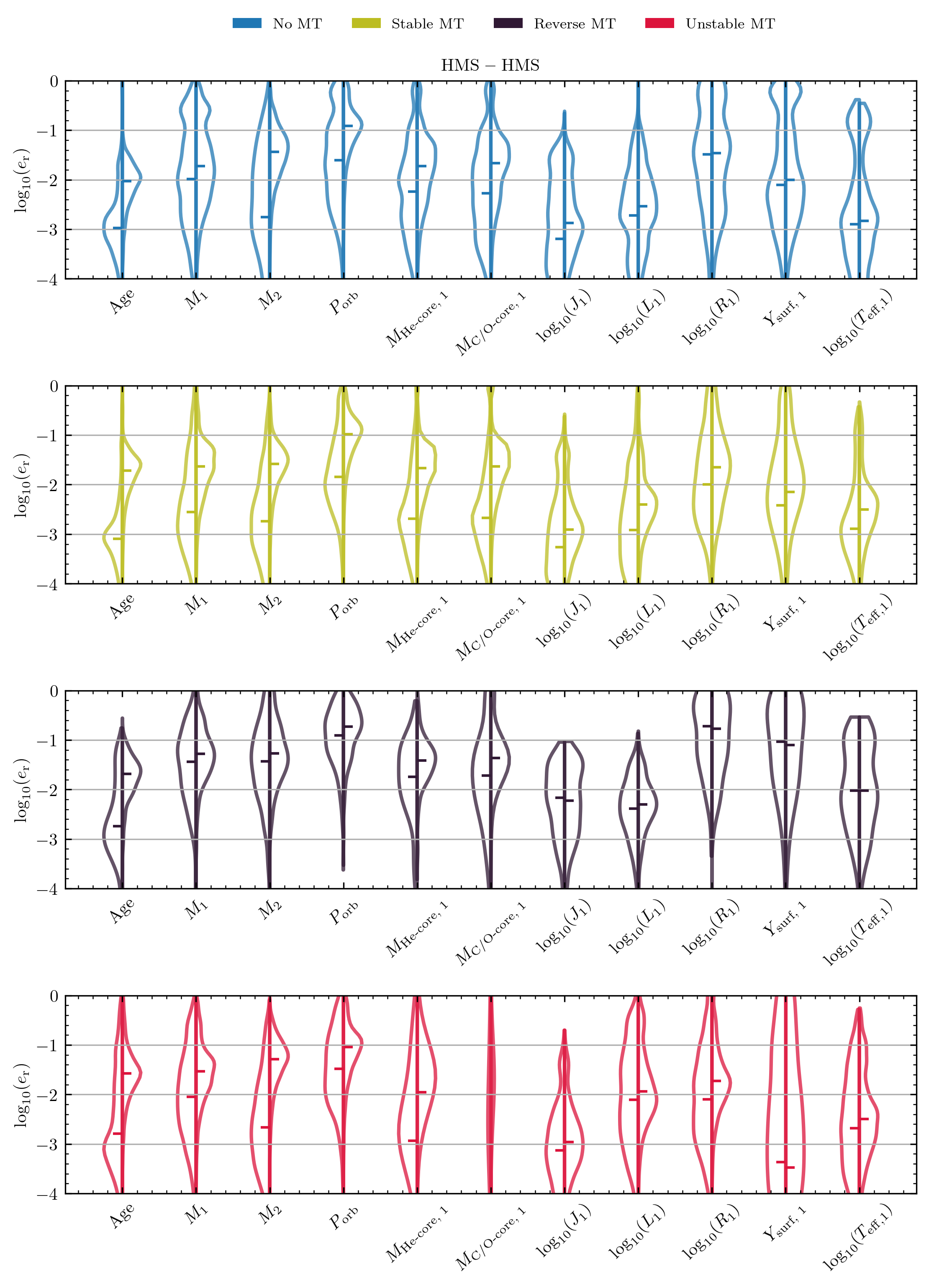}\\
    \caption{Comparison between the accuracy of our methods to interpolate between our grid points (left side of violins) and the alternative where we match to the nearest neighbor point in our regularly spaced grid (right side of violins) for our HMS--HMS grid. For ease of interpretation, we have separated out the comparison so each panel shows one class. The parameters and format is consistent with Figure~\ref{fig:error_distributions}. For nearly every parameter in every class, our interpolation methods are an improvement over nearest neighbor matching. }
    \label{fig:nearest_neighbor_HMS-HMS}
\end{figure*}

\begin{figure*}[t]
    \center
    \includegraphics[width=\textwidth,angle=0]{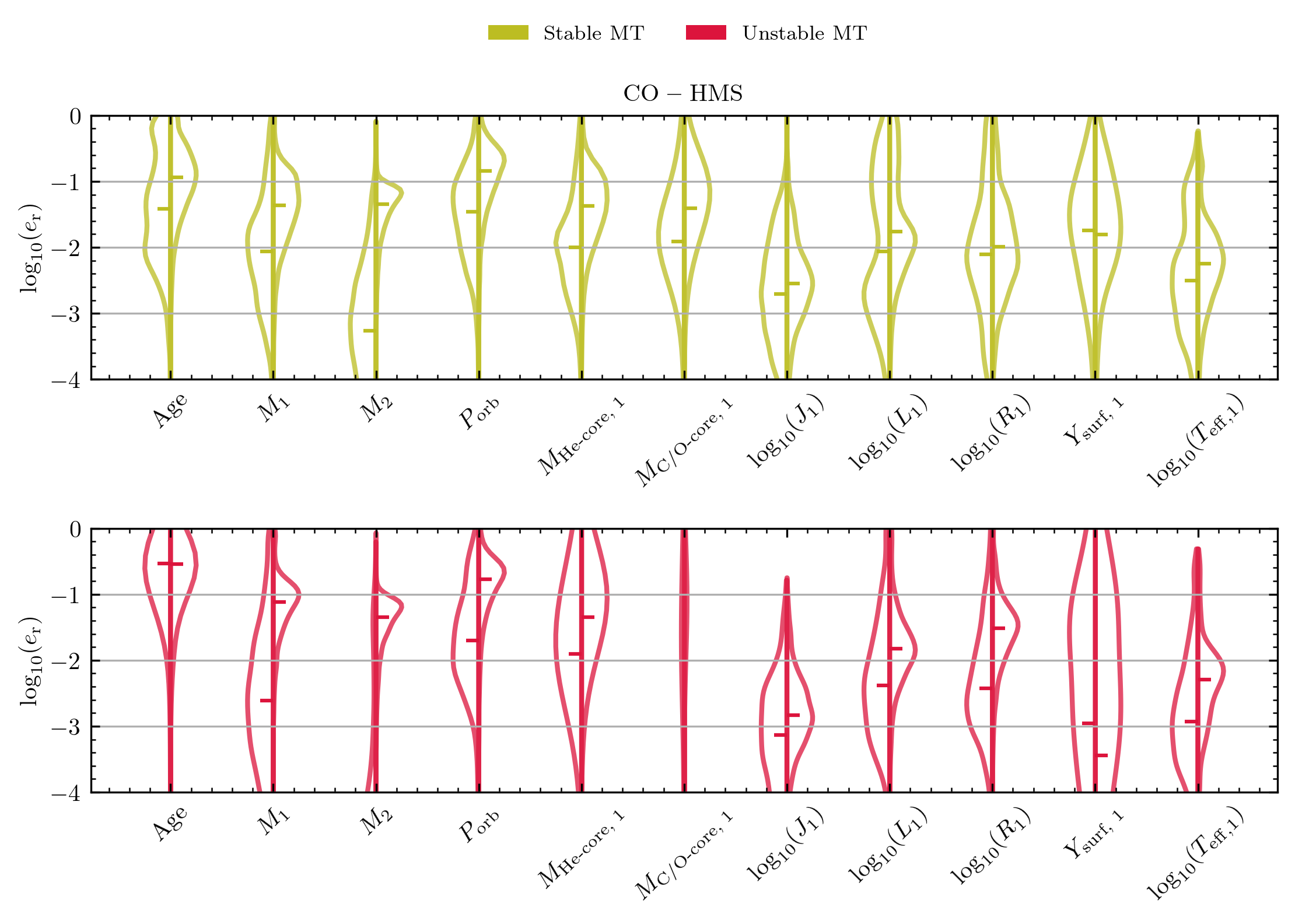}\\
    \caption{Comparison between the accuracy of our methods to interpolate between our grid points (left side of violins) and the alternative where we match to the nearest neighbor point in our regularly spaced grid (right side of violins) for our CO--HMS grid. As is the case for the HMS--HMS grid shown in Figure~\ref{fig:nearest_neighbor_HMS-HMS}, interpolation is an improvement over nearest neighbor matching. }
    \label{fig:nearest_neighbor_CO-HMS}
\end{figure*}

\begin{figure*}[t]
    \center
    \includegraphics[width=\textwidth,angle=0]{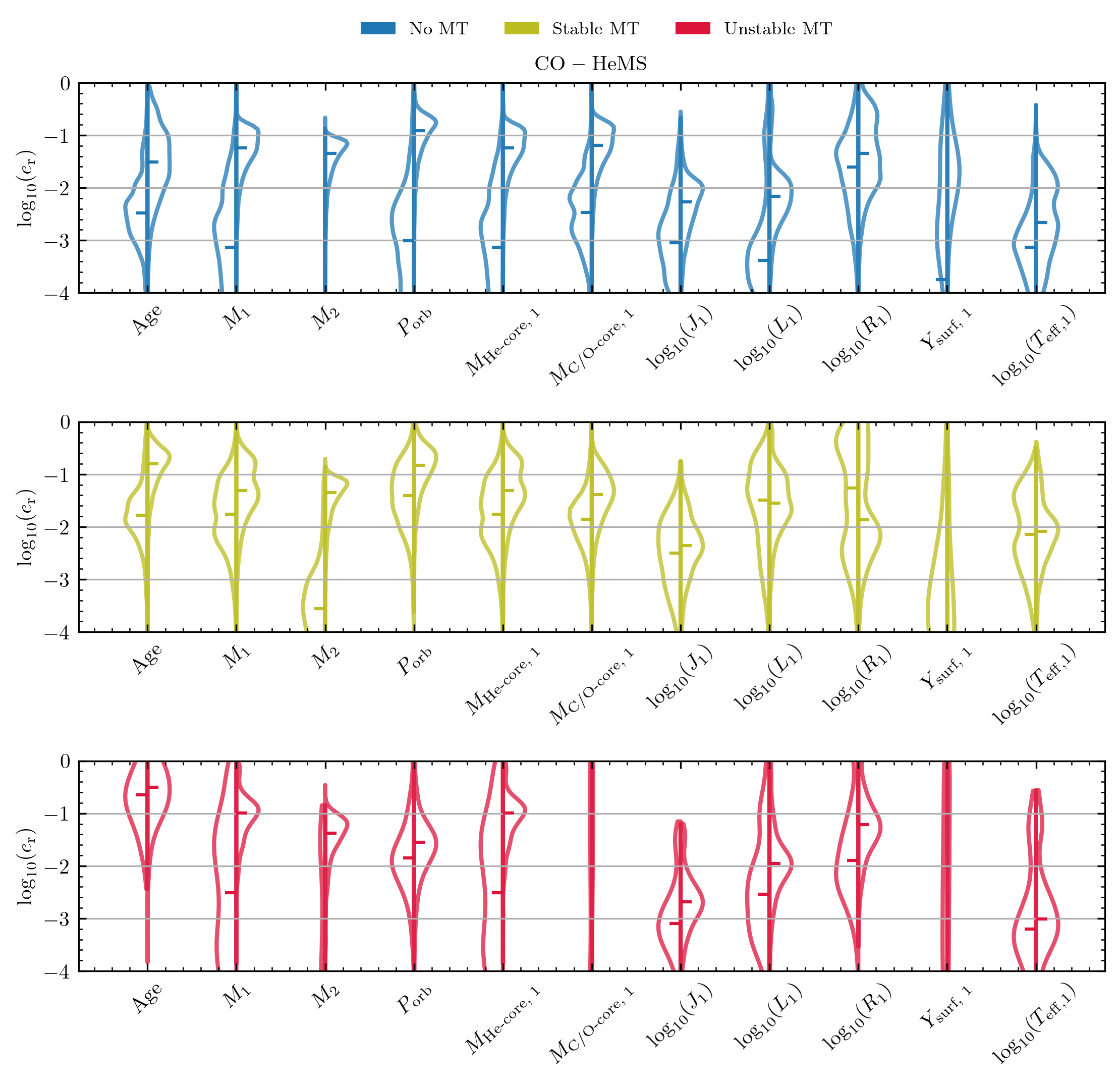}\\
    \caption{Comparison between the accuracy of our methods to interpolate between our grid points (left side of violins) and the alternative where we match to the nearest neighbor point in our regularly spaced grid (right side of violins) for our CO--HeMS grid. As is the case for the HMS--HMS and CO--HMS grids shown in Figures~\ref{fig:nearest_neighbor_HMS-HMS} and \ref{fig:nearest_neighbor_CO-HMS}, interpolation is an improvement over nearest neighbor matching. }
    \label{fig:nearest_neighbor_CO-HeMS}
\end{figure*}

To evolve binary populations within \posydon, our default approach is to use the classification and interpolation methods that are described in Section~\ref{sec:machine_learning} and trained on our grids of binary models computed with \mesa. However, as an alternative option to users, we provide the ability to evolve binaries using nearest neighbor matching; rather than interpolate our grids to find the result of a binary, we determine its parameters by adopting those of the nearest neighbor. Matching is performed as described in Section~\ref{sec:matching}. We compare the errors for 11 single- and binary-star parameters, one panel per class, for the HMS--HMS, CO--HMS, and CO--HeMS grids in Figures~\ref{fig:nearest_neighbor_HMS-HMS}, \ref{fig:nearest_neighbor_CO-HMS}, and \ref{fig:nearest_neighbor_CO-HeMS}, respectively. For nearly every parameter in every class and each of the three binary grids, the initial-final interpolation values are an improvement (often a significant one) over the nearest neighbor matching scheme, and are never significantly worse. 

\section{HMS--HMS RLOF Grid}
\label{sec:HMS-HMS_RLO}

In v1, we assumed our binary populations are initialized as a circular binary. While that remains the default prescription, in v2, we allow users to initialize binaries with non-zero eccentricity at ZAMS. Although a minor change in concept, in practice this requires a substantial shift in how binaries are evolved within \posydon. When binaries are formed in circular orbits, we can directly refer to our grids of \mesa models. Initially eccentric binaries, on the other hand, require a different treatment. These are initialized as detached binaries where the individual stars are assumed to evolve independently as essentially single stars but with orbital evolution due to mass loss and tides calculated simultaneously (using our detached step, described in Section 8.1 of \citetalias{2023ApJS..264...45F}). If one of the stars overfills its Roche lobe at periastron, then we assume the binary instantaneously circularizes at the periastron separation, and we map this binary onto our HMS--HMS \mesa grid. 

\begin{figure}
    \centering
    \includegraphics[width=\linewidth]{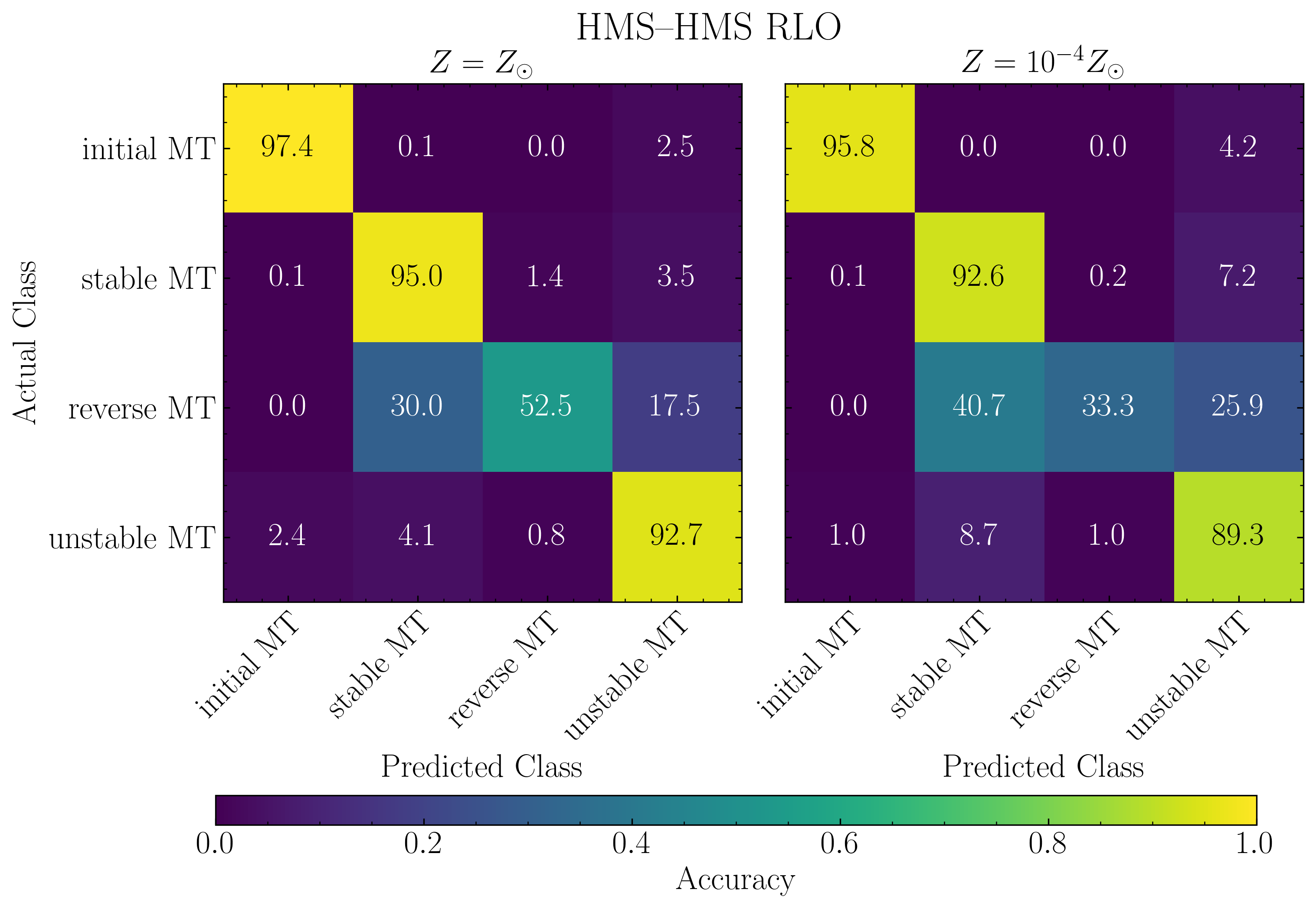}
    \caption{Confusion matrices showing the classification error exhibited by the classifier trained on the HMS--HMS RLOF grid for two different metallicities selected. The left column of confusion matrices corresponds to $\Zsun$ while the right column corresponds to $Z = 10^{-4}\,\Zsun$. The horizontal axis of each matrix corresponds to the predicted class while the vertical axis corresponds to the ground truth class. Each value in cell $C_{ij}$ corresponds to the fraction of samples that belong to class $i$ (horizontal axis) while being classified as class $j$ (vertical axis). The color represents the magnitude of the value $C_{ij}$.}
    \label{fig:HMS-HMS_RLO_confusion_matrix}
\end{figure}

\begin{figure}
    \centering
    \includegraphics[width=\linewidth]{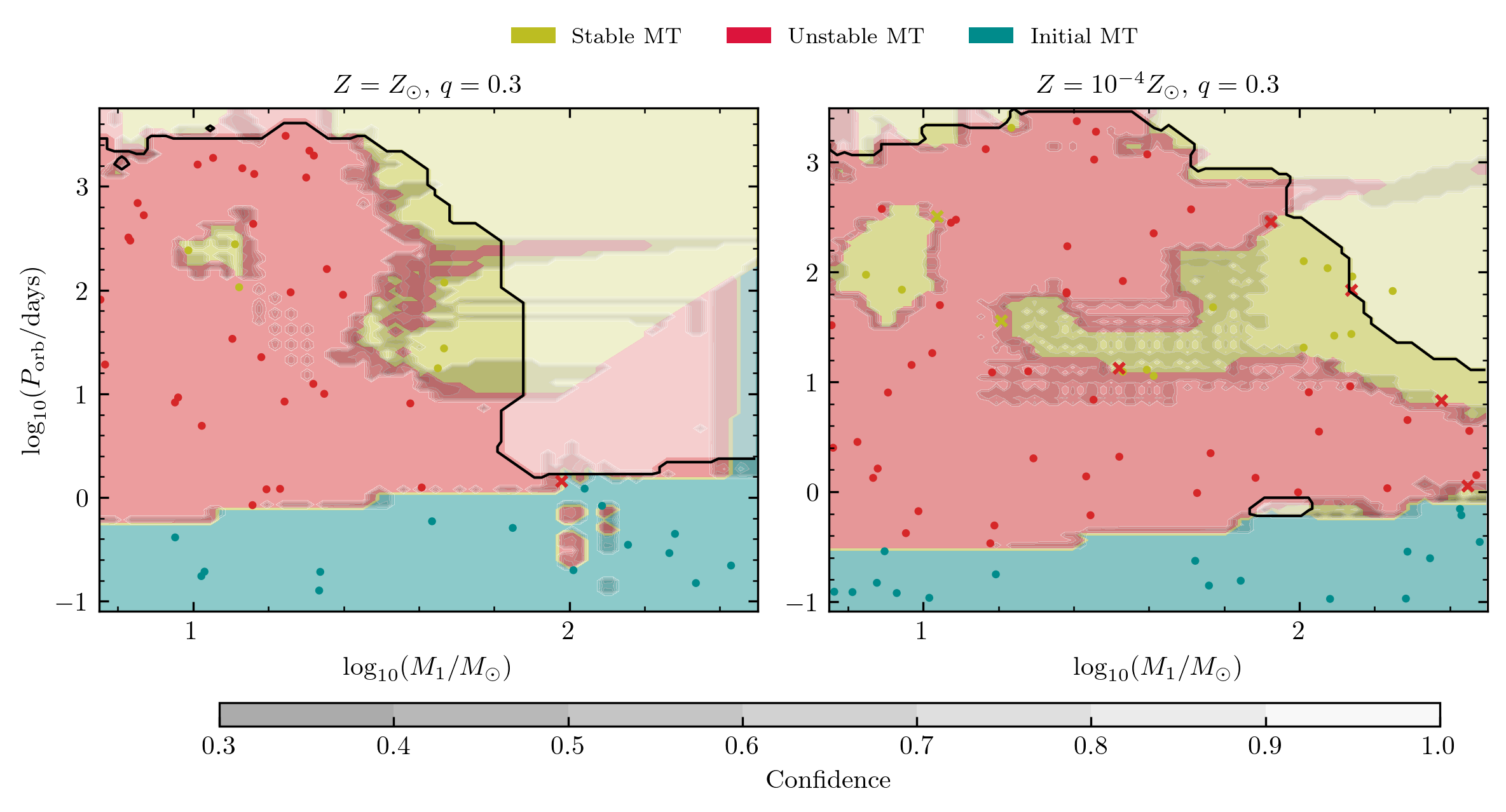}
    \caption{Decision boundaries of our classifier for two sample metallicities (\Zsun, left panel; $10^{-4}\,\Zsun$, right panel). Each panel shows the logarithm of the primary's initial mass on the horizontal axis and logarithm of the initial orbital period on the vertical axis. The constant (mass ratio) at which the slice is fixed is indicated in the titles of the plots in the top row. Each decision boundary has a different color as indicated by the legend, with grayscale as a transparency layer indicating classification confidence. Colored points indicate the positions and classes of our validation set, where predicted correctly (filled circle) and incorrectly (crosses). The white overlay, marked by the black boundary corresponds to regions of the grid not exhibiting mass transfer.}
    \label{fig:HMS-HMS_RLO_decision_boundary}
\end{figure}

\begin{figure}
    \centering
    \includegraphics[width=\linewidth]{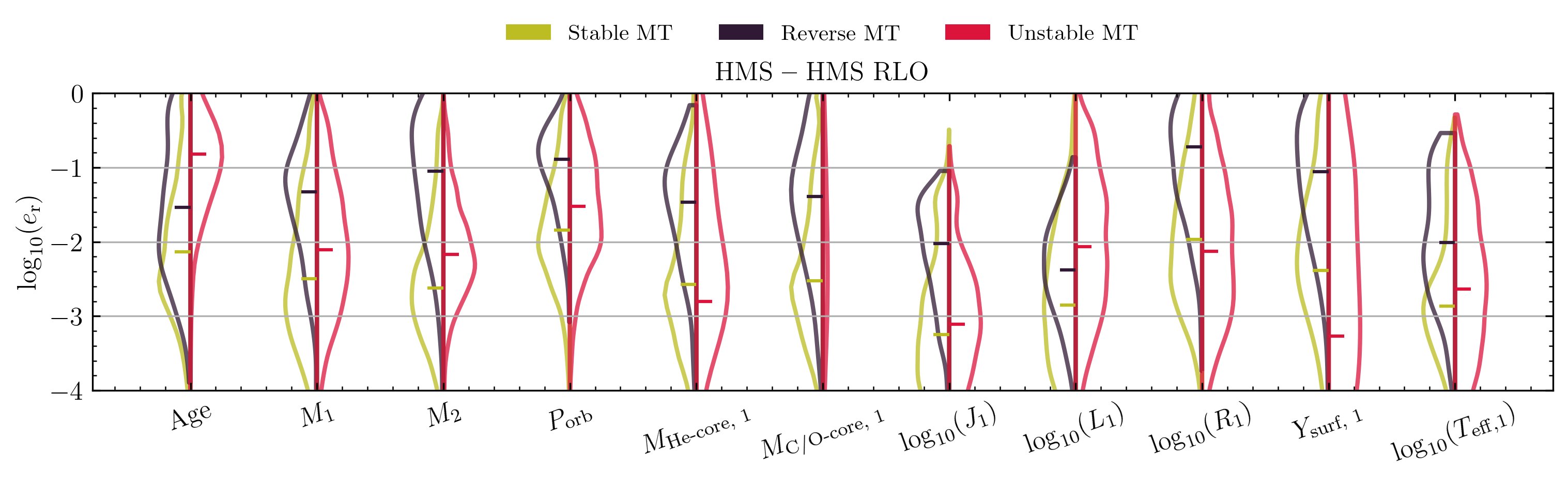}
    \caption{Violin plots showing relative interpolation error for the HMS--HMS RLOF for all metallicities. Each of the different colored curves corresponds to a different interpolation class. The horizontal axis shows a select 11 interpolation fields while the vertical axis shows the relative error in $\log_{10}$ scale. Tick marks indicate median values. The bulk of the distributions for most parameters are almost all within 10\%, typically under 1\%.}
    \label{fig:HMS-HMS_RLO_interpolation_error}
\end{figure}

Our mapping procedure requires that we re-post-process our HMS--HMS grid to ignore the evolution of binaries prior to RLOF so that the ``initial'' values of each \mesa binary simulation are not taken to from the first timestep of their evolution in \mesa, but the timestep where RLOF occurs. This procedure is similar to our treatment of the CO--HMS grid described in Section~\ref{sec:CO-HMS}. Since the initial values of this grid differ from those of our original HMS--HMS, we train new classification and interpolation models, the summary of which we show in Figures~\ref{fig:HMS-HMS_RLO_confusion_matrix}, \ref{fig:HMS-HMS_RLO_decision_boundary}, and \ref{fig:HMS-HMS_RLO_interpolation_error}. These figures demonstrate that our reprocessed form of the HMS--HMS grid has similar metrics when compared with the original HMS--HMS grid.

\section{CO--HeMS RLOF Grid}
\label{sec:CO-HeMS_RLO}

When a primary star undergoes a supernova with an evolved, stripped companion, the resulting binary (assuming it is not disrupted during collapse) is a compact object with a helium star in an eccentric orbit. Since our CO--HeMS grid described in Section~\ref{sec:CO-HeMS} assumes circular orbits, we could not properly handle these systems. Although rare, they represented a failure mode in v1 that we have resolved in v2 by allowing these binaries to enter the detached step (described in Section 8.1 of \citetalias{2023ApJS..264...45F}). Upon RLOF, we then map them to a new form of our CO--HeMS grids.

\begin{figure}
    \centering
    \includegraphics[width=\linewidth]{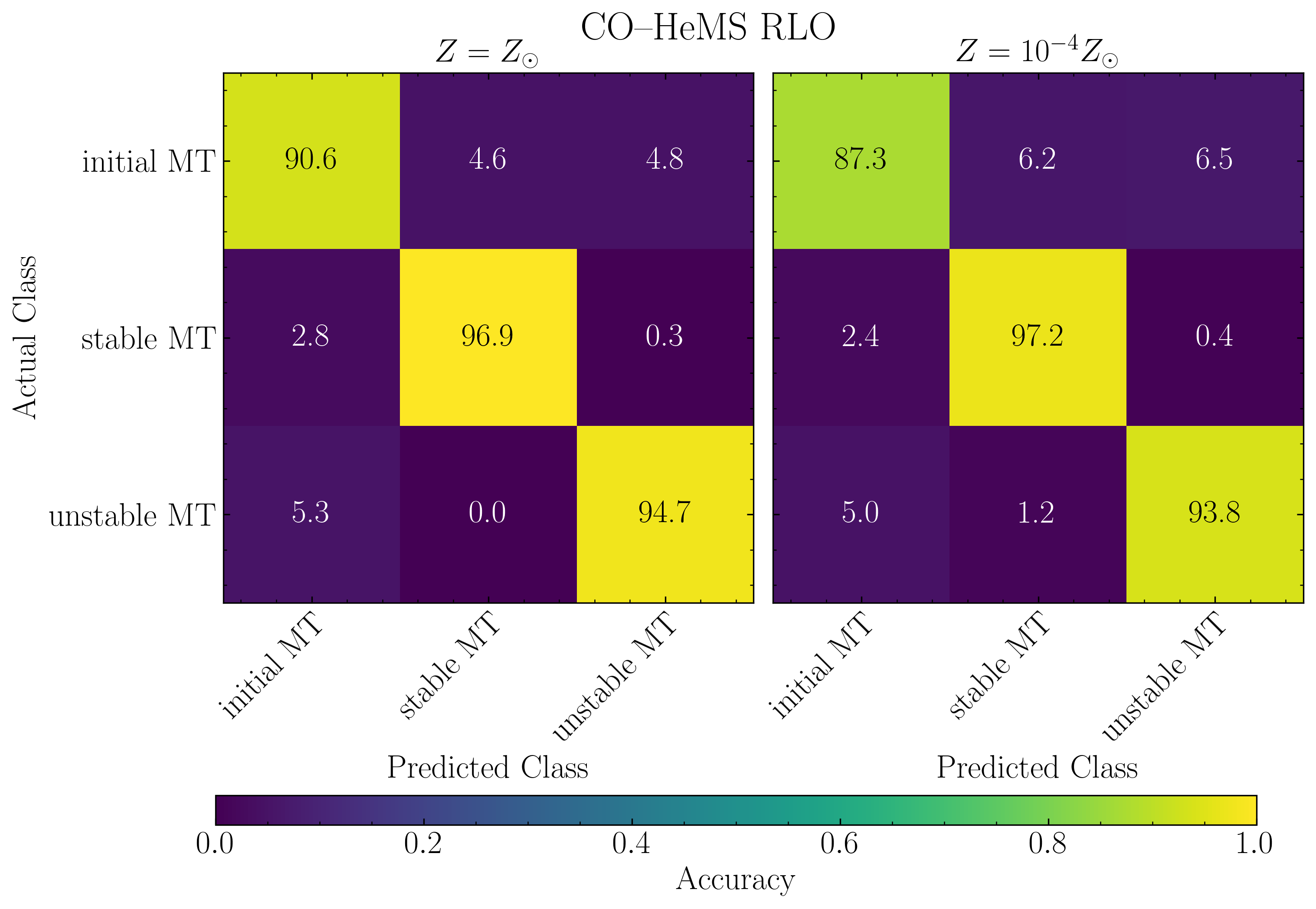}
    \caption{Confusion matrices showing the classification error exhibited by the classifier trained on the CO--HeMS RLOF grid for two different metallicities selected. The left column of confusion matrices corresponds to $\Zsun$ while the right column corresponds to $Z = 10^{-4}\,\Zsun$. The horizontal axis of each matrix corresponds to the predicted class while the vertical axis corresponds to the ground truth class. Each value in cell $C_{ij}$ corresponds to the fraction of samples that belong to class $i$ (horizontal axis) while being classified as class $j$ (vertical axis). The color represents the magnitude of the value $C_{ij}$.}
    \label{fig:CO-HeMS_RLO_confusion_matrix}
\end{figure}

\begin{figure}
    \centering
    \includegraphics[width=\linewidth]{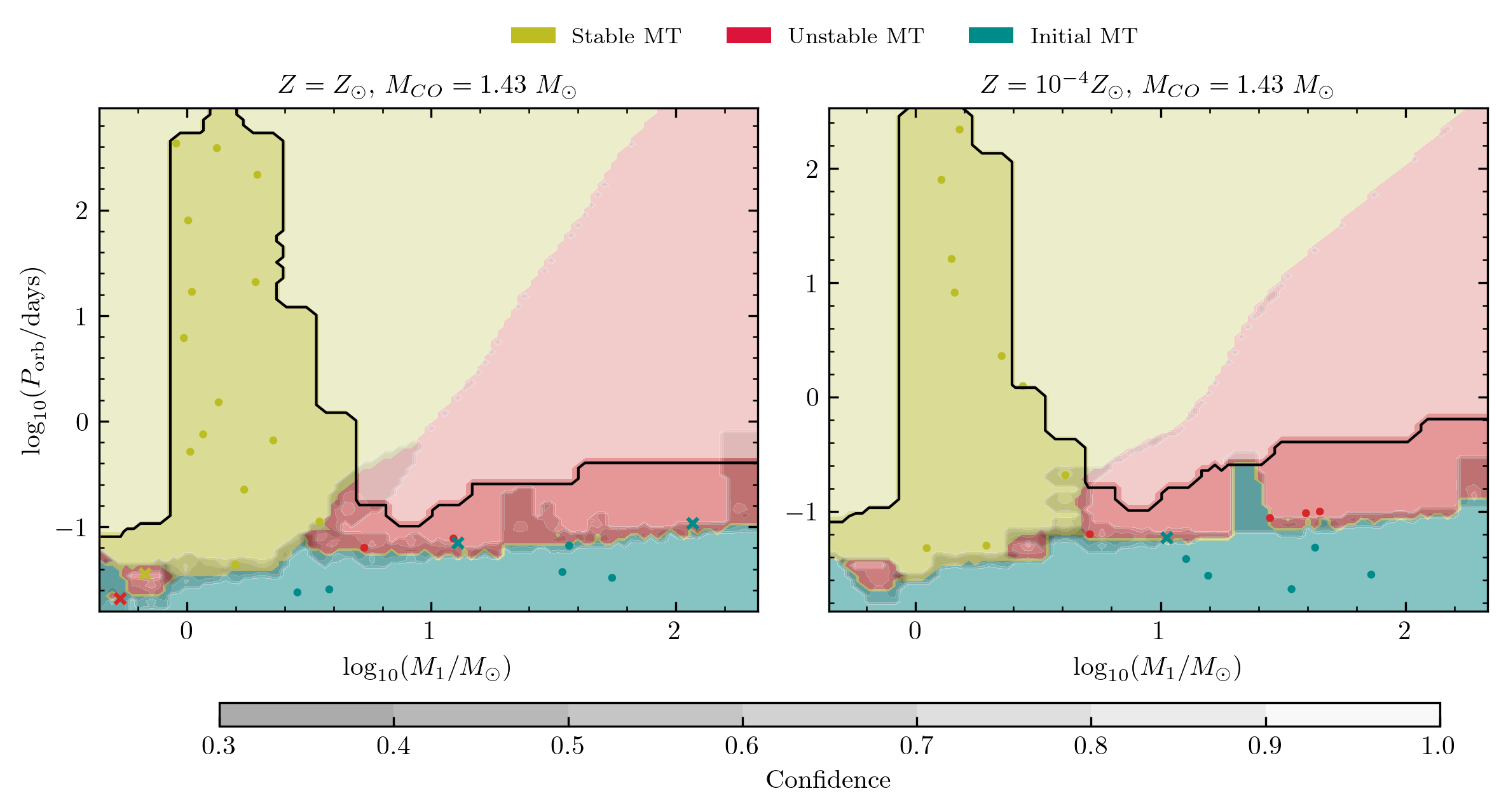}
    \caption{Decision boundaries of our classifier for two sample metallicities (\Zsun, left panel; $10^{-4}\,\Zsun$, right panel). Each panel shows the logarithm of the primary's initial mass on the horizontal axis and logarithm of the initial orbital period on the vertical axis. The constant (CO mass) at which the slice is fixed is indicated in the titles of the plots in the top row. Each decision boundary has a different color as indicated by the legend, with grayscale as a transparency layer indicating classification confidence. Colored points indicate the positions and classes of our validation set, where predicted correctly (filled circle) and incorrectly (crosses). The white overlay, marked by the black boundary corresponds to regions of the grid not exhibiting mass transfer.}
    \label{fig:CO-HeMS_RLO_decision_boundary}
\end{figure}

\begin{figure}
    \centering
    \includegraphics[width=\linewidth]{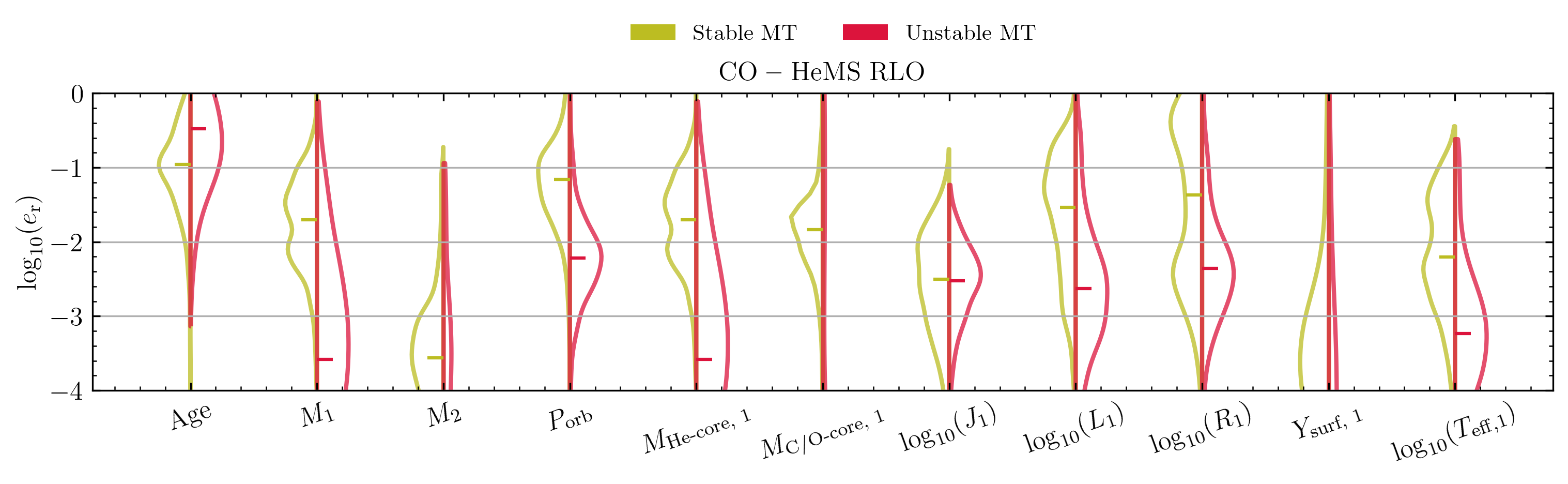}
    \caption{Violin plots showing relative interpolation error for the CO--HeMS RLOF grid for all metallicities. Each of the different colored curves corresponds to a different interpolation class. The horizontal axis shows a select 11 interpolation fields while the vertical axis shows the relative error in $\log_{10}$ scale. Tick marks indicate median values. The bulk of the distributions for most parameters are almost all within 10\%, typically under 1\%.}
    \label{fig:CO-HeMS_RLO_interpolation_error}
\end{figure}

Similar to our HMS--HMS RLOF grid described in Appendix~\ref{sec:HMS-HMS_RLO}, we generate a re-post-processed form of our CO--HeMS grid described in Section~\ref{sec:CO-HeMS}; for each of our \mesa simulations any evolution prior to RLOF is ignored. We present our classification and interpolation metrics for this new form of the CO--HeMS grid in Figures~\ref{fig:CO-HeMS_RLO_confusion_matrix}, \ref{fig:CO-HeMS_RLO_decision_boundary}, and \ref{fig:CO-HeMS_RLO_interpolation_error}. These figures demonstrate metrics comparable to our CO--HeMS grid described in Section~\ref{sec:CO-HeMS}.

\end{document}